\documentclass[fleqn,usenatbib]{mnras}

\usepackage{newtxtext,newtxmath}
\usepackage[T1]{fontenc}

\DeclareRobustCommand{\VAN}[3]{#2}
\let\VANthebibliography\thebibliography
\def\thebibliography{\DeclareRobustCommand{\VAN}[3]{##3}\VANthebibliography}

\usepackage{graphicx}	% Including figure files
\usepackage{amsfonts}
\usepackage{mathtools}
\usepackage{dsfont}
\usepackage{amsmath}	% Advanced maths commands
\usepackage{multicol}        % Multi-column entries in tables
\usepackage{bm}		% Bold maths symbols, including upright Greek
\usepackage{pdflscape}	% Landscape pages
\usepackage{dcolumn}   % Align table columns on decimal point
\usepackage[]{algorithm2e} % display algorithms
\usepackage{grffile}
\usepackage{color}
\usepackage{simplewick}
\usepackage{cancel}
\usepackage{subcaption}

\allowdisplaybreaks % Makes evertithing 'breakable'

\def\bthet{{\bm\theta}}
\def\dthet{{{\rm d}^2\bm\theta}}

\newcommand{\dellnorm}[1]{\frac{{\rm d}^2\bm\ell_{#1}}{(2\pi)^2}}
\newcommand{\ellvec}[1]{\vec{\ell}_{#1}}

\def\Map{M_{\rm ap}}

\def\MapEst{\widehat{M_{\rm ap}}}

\newcommand{\MapStat}[1]{M_{\rm ap}^{#1}}

\newcommand{\MgStat}[1]{M_{g,2}^{#1}}
\newcommand{\MapStatEns}[1]{\left\langle \mathcal{M}_{\rm ap}^{#1} \right\rangle}

\newcommand{\MsMapStatEns}[2]{\langle \mathcal{M}_{s,2}^{#1} \mathcal{M}_{\text{ap}}^{#2} \rangle}

\newcommand{\MapStatEst}[1]{\widehat{\MapStat{#1}}}

\newcommand{\MsEst}[1]{M_{{\rm s}, #1}}

\newcommand{\MsEstPoly}[2]{{M_{\rm s, ({#2})}^{({#1})}}}
\newcommand{\SEst}[1]{S_{#1}}
\newcommand{\SEstPow}[2]{\left(S_{{#1}}\right)^{#2}}
\newcommand{\SEstPoly}[2]{{S_{ ({#2})}^{({#1})}}}
\newcommand{\sigsq}[1]{\left(\frac{\sigma_\epsilon^2}{2}\right)^{#1}}

\def\cov{{\rm cov}}

\newcommand\Eqn[1]     {Eq.\,(\ref{#1})}
\newcommand\Eqns[2]    {Eqs\,(\ref{#1}) and~(\ref{#2})}

\newcommand\Figure[1]     {Figure~\,{\ref{#1}}}
\newcommand\Fig[1]     {Fig.\,{\ref{#1}}}

\newcommand\Sect[1]    {\S\ref{#1}}
\newcommand\App[1]     {Appendix~\ref{#1}}

\newcommand\nn         {\nonumber}

\def\cfhtl{\mbox{CFHTLenS}}

\def\euclid{\emph{Euclid}}
\def\kids{\mbox{KiDS}}

\def\pasj{Publications of the ASJ}
\def\mnras{{Mon.~ Not.~ R.~ Astron.~ Soc.~}}

\def\prd{{Phys.~ Rev.~ D.~}}
\def\prl{{Phys.~ Rev.~ Lett.~}}
\def\apj{{Astrophys.~ J.~}}
\def\apjs{{Astrophys.~ J.~ Suppl.~}}
\def\apjl{{Astrophys.~ J.~ Lett.~}}

\def\nat{{Nature (London)~}}

\def\mnras{{MNRAS}}

\def\prd{{PRD}}
\def\prl{{PRL}}
\def\apj{{ApJ}}
\def\apjs{{ApJS}}
\def\apjl{{ApJL}}
\def\aap{{A\&A}}
\def\nat{{Nature}}

\def\physrep{{Phys.~ Rep.~}}
\def\jcap{Journal of Cosmology and Astro-Particle Physics}

\newcommand{\be}{\begin{equation}}
\newcommand{\ee}{\end{equation}}

\newcommand{\ba}{\begin{align}}
\newcommand{\ea}{\end{align}}

\newcommand{\bes}{\begin{subequations}}
\newcommand{\ees}{\end{subequations}}

\def\ex{{\rm e}}

\def\RR{{\mathbb{R}^2}}

\def\chiH{\chi_{\rm H}}

\def\omr{\Omega_{\rm r}}
\def\oml{\Omega_{\Lambda}}
\def\omk{\Omega_{\rm k}}

\def\omm0{\Omega_{\rm m,0}}
\def\omr0{\Omega_{\rm r,0}}
\def\oml0{\Omega_{\Lambda,0}}
\def\omk0{\Omega_{\rm k,0}}

\def\Mpc{\, h^{-1}{\rm Mpc}}

\def\1D{{\rm 1D}}
\def\2D{{\rm 2D}}
\def\3D{{\rm 3D}}

\def\d{{\rm d}}

\def\1Loop{{\rm 1Loop}}

\def\wk{U}
\def\wg{Q}
\def\Map{{\mathcal M}_{\rm ap}}
\def\thetc{{\vartheta}}

\usepackage{tikz}
\newcommand*\circled[1]{\tikz[baseline=(char.base)]{
            \node[shape=circle,draw,inner sep=2pt] (char) {#1};}}
\title[Fast estimation of aperture-mass statistics II] {Fast estimation
  of aperture-mass statistics II: Detectability of higher order statistics in current and future surveys}

      \author[Porth \& Smith]{
        Lucas Porth$^{1,2}$\thanks{lporth@uni-bonn.de} and 
        Robert E. Smith$^{1}$\thanks{r.e.smith@sussex.ac.uk}.
\\
$^{1}$Astronomy Centre, Department of Physics \& Astronomy, University of Sussex, Brighton, BN1 9RH, UK\\
$^{1}$Argelander-Institut f\"ur Astronomie, Universit\"at Bonn, Auf dem H\"ugel 71, 53121 Bonn, Germany\\
}

\date{Accepted XXX. Received YYY; in original form ZZZ}

\pubyear{2021}

\begin{document}
\label{firstpage}
\pagerange{\pageref{firstpage}--\pageref{lastpage}}
\maketitle

%%%%%%%%%%%%%%%%%%%%%%%%%%%%%%%%%%%%%%%%%%%%%%%%%%%%%%%

\begin{abstract}
We explore an alternative method to the usual shear correlation
function approach for the estimation of aperture mass statistics in
weak lensing survey data. Our approach builds on the direct estimator
method. In this paper, we extend our analysis to statistics of
arbitrary order and to the multiscale aperture mass
statistics. We show that there always exists a linear order algorithm
to retrieve any of these generalised aperture mass statistics from
shape catalogs when the direct estimator approach is adopted. We
validate our approach through application to a large number of Gaussian
mock lensing surveys where the true answer is known and we do this up
to 10th order statistics. We then apply our estimators to an ensemble
of real-world mock catalogs obtained from $N$-body simulations -- the
SLICS mocks, and show that one can expect to retrieve detections of
higher order clustering up to fourth order in a KiDS-1000 like
survey. We expect that these methods will be of most utility for
future wide-field surveys like Euclid and the Rubin Telescope.
\end{abstract}

%%%%%%%%%%%%%%%%%%%%%%%%%%%%%%%%%%%%%%%%%%%%%%%%%%%%%%%


%%%%%%%%%%%%%%%%%%%%%%%%%%%%%%%%%%%%%%%%%%%%%%%%%%
%%%%%%%%%%%%%%%%% BODY OF PAPER %%%%%%%%%%%%%%%%%%

\section{Introduction}
Weak gravitational lensing by large-scale structure of the light from
distant galaxies is a powerful probe for constraining the cosmological
parameters and distinguishing between competing models of the Universe
\citep{Blandfordetal1991,Seitzetal1994,JainSeljak1997,Kaiser1998,Schneideretal1998,ZhangPetal2007}.
The first measurements of the correlations in the shapes of distant
background galaxies date back more than two decades
\citep{Baconetal2000,Kaiseretal2000,VanWaerbekeetal2000,Wittmanetal2000}. Since
then, cosmic shear observations have become ever more precise as the
coupling of techological advancements and algorithmic developments
have enabled us to conduct unprecedented deep optical imaging surveys
of the cosmos \kids\footnote{{\tt kids.strw.leidenuniv.nl}},
DES\footnote{{\tt www.darkenergysurvey.org}} and HSC\footnote{{\tt
  hsc.mtk.nao.ac.jp/ssp/}}, with current state-of-the-art surveys now
mapping thousands of square degrees
\citep{Hildebrandtetal2017,Troxeletal2018,HSC2018,Hikageetal2019,
  Asgarietal2021}. By the end of the decade planned experiments like
\euclid\footnote{{\tt www.cosmos.esa.int/web/euclid}} and the Rubin
Telescope\footnote{{\tt www.lsst.org}} \citep{Euclid2011,LSST2009}
will map volumes close to the entire physical volume of our observable
Universe. In order to make optimal use of these rich data sets we will
need to push forward our understanding and modelling of various
physical and measurement effects. In particular: accurate modelling of
the nonlinear evolution of large-scale structure, including the
baryonic physics effects; accurate modelling and correction of the
point-spread function of the telescope; correcting the bias in the
weak lensing shape estimation algorithms; and accounting for the
intrinsic alignments, to name but a few of the main systematics 
\citep[see][for a more detailed discussion of these effects]{Schneider2006p3,Masseyetal2013,TroxelIshak2015}.

If the underlying matter density field were a Gaussian random field,
then all of the information in a weak lensing survey would be
contained in the shear two-point correlation function.  However,
physical effects like: the nonlinear growth of structure
\citep{Bernardeauetal2002}, the mapping between cosmic shear and
galaxy ellipticities \citep{Miralda-Escude1991}, and lensing beyond
the Born approximation
\citep{Hilbertetal2009,PrattenLewis2016,Fabbianetal2018}, all
introduce non-Gaussianity in the maps. Furthermore, the nonlinear
evolution also induces correlations in the convergence power spectrum
multipoles, which grow stronger on small scales. This means that the
information content of the second order statistics becomes saturated
after a given multipole
\citep{Satoetal2011,Hilbertetal2012,Kayoetal2013,Marianetal2013,Byunetal2017}. Thus
in order to capture all of the cosmological information available in
lensing surveys one must look to the higher order statistics of the
shear field
\citep{Schneideretal1998,Bernardeauetal2002,SchneiderLombardi2003}. Furthermore,
owing to the different ways in which the cosmological parameters and
nuisance parameters depend on the higher-order statistics, the
inclusion of such measurements brings with it the further virtue of
being able to break parameter degeneracies, e.g. by combining second
and third order statistics
\citep{KilbingerSchneider2005,Sembolonietal2011,Fuetal2014}, or by
incorporating the information found in the statistical properties of
the peaks in the shear field \citep{Marianetal2013,Kacprzaketal2016}.

A powerful method to disentangle systematic effects from cosmic shear
signals is the E/B decomposition
\citep{Crittendenetal2001,Schneideretal2002a}. At leading order, pure
weak lensing signals are sourced by a scalar lensing potential, which
means that their deflection fields are curl free.  Equivalently, the
ring-averaged cross component of the shear is expected to be zero (the
B mode), while the tangential one contains all the lensing signal (the
E mode). Thus B modes enable a robust test for the presence of
systematic errors. One method to take advantage of this E/B
decomposition is the so-called `aperture mass statistics'
\citep{Kaiser1995,Schneider1996,Schneideretal1998}. `Aperture mass'
($\Map$) and `Map-Cross' $(\mathcal{M}_{\times})$ are obtained by
convolving the tangential and cross shear with an isotropic filter
function. Therefore by construction they are E/B-decomposed. Taking
the second moment leads to the variance of aperture mass, the third to
the skewness, the fourth to the kurtosis, etc.

The standard approach for measuring the aperture mass statistics in
data utilises the fact that, for the flat sky, any $n$-point moment
can be expressed in terms of integrals over the $n$-point shear
correlation functions, modulo a kernel function
\citep{Schneideretal2002a,Jarvisetal2004}. The reason for adopting
this strategy stems from the fact that the correlation functions can
reliably be estimated in the presence of a nontrivial survey
mask. However, for these estimators to be accurate and E/B decomposed,
one requires three conditions to be satisfied: (i) the $\xi_+$/$\xi_-$
correlations need to be measured down to zero separation; (ii) they
also need to be measured up to a maximum angular scale, set by the
exact form of the aperture mass filter and its angular scale; (iii)
the angular bins must be sufficiently fine for the discretisation of
the integrals to be reliable
\citep{KilbingerSchneider2005,Fuetal2014}. Owing to galaxy image
blending, signal-to-noise issues and the finite size of the survey,
the lower bound is never possible and the upper bound means that
biases can occur due to edge effects. In addition, while the mean
estimate is unbiased, the covariance matrix does require one to
carefully account for the mask
\citep{Schneideretal2002b,Friedrichetal2016}. More recent developments
that also make use of the shear correlation functions, while
circumventing the issues of E/B leakage on small scales are the ring
statistics and COSEBIs
\citep{Schneideretal2007,Schneideretal2010}. While those approaches
can in principle be extended to higher order statistics, the
estimation of the $n$-point correlation functions turns out to be
notoriously time consuming \citep{Schneideretal2005a,
  Jarvisetal2003}. Further methods to extract non-Gaussian information
from the aperture mass look at its probability density function as a
whole \citep{Bernardeauetal2000, Munshietal2004, Barthelemyetal2020}
or at the distribution of its signal-to-noise peaks
\citep{Marianetal2012, Heydenreichetal2020, Martinetetal2021}.

In \citet{Porthetal2020} we took a different approach and explored a
computationally efficient (accelerated) implementation of the original
direct estimator of the aperture mass dispersion
\citep{Schneider1998}. Rather than measuring the correlation functions
of the shear polar, in this formulation one instead directly measures
cumulants of $\mathcal{M}_{\rm ap}$ on a set of apertures and then
uses an optimised weighting scheme to average the estimates, along
with a restriction on the types of apertures that are acceptable. The
present work extends our previous investigation in a number of
important ways. First, we construct accelerated direct estimators for
the higher-order aperture mass moments, including the skewness,
kurtosis, etc. Second, we also develop further the multiscale
aperture moments \citep{Jarvisetal2003,Schneideretal2005a}. These two
improvements enable us to better trace the full, harmonic mode,
configuration dependence of the convergence polyspectra.

This paper is organised as follows: In \Sect{sec:Theory} we introduce
key concepts of weak lensing, define the aperture mass and show how
its connected cumulants are related to the convergence polyspectra. In \Sect{sec:Estimators} we
revisit the direct estimators for higher order aperture mass measures
and construct suitable bases, in which each statistic can be computed
in linear time complexity. After investigating the variance of the
direct estimators, we give details of our updated algorithm used to
perform the measurements. In \Sect{sec:Measurements} we
empirically verify the linear scaling and the measurements of our implementation of the direct estimator on Gaussian mocks. In \Sect{sec:DetectionSignificance} we then apply the estimator to the SLICS simulation suite in
order to assess up to which order one can expect to extract
information from the aperture mass statistics on a KiDS-1000 like
survey. Finally, in \Sect{sec:Conclusions} we summarise our
findings, conclude and discuss future work.

%%%%%%%%%%%%%%%%%%%%%%%%%%%%%%%%%%%%%%%%%%%%%%%%%%%%%%%

\section{Higher order aperture mass measures for cosmic shear}\label{sec:Theory}

%%%%%%%%%%%%%%%%%%%%%%%%%%%%%%%%%%%%%%%%%%%%%%%%%%%%%%%

\subsection{Weak gravitational lensing and aperture mass}\label{ssec:apmass}

In this paper we are mainly concerned with the weak lensing of distant
background (source) galaxy shapes by the intervening large-scale
structure
\citep[for detailed reviews of the topic
  see][]{BartelmannSchneider2001,Schneider2006p1,Schneider2006p3,Dodelson2003,Dodelson2017,Kilbinger2015,Mandelbaum2018}.
The two fundamental quantities describing this mapping from true to
observed galaxy images are the convergence $\kappa$ and the complex
shear $\gamma=\gamma_1+i\gamma_2$, which, assuming a metric theory of
gravity, are all derived from an underlying scalar lensing
potential. In a galaxy survey the effective convergence at angular
position $\bm \theta$ and radial comoving distance $\chi$ can be
connected to the density contrast $\delta(\chi\bm\theta,\chi)$ through:
\begin{align}
  \kappa(\bthet) & =  \frac{3}{2}\Omega_{\rm m,0}\left(\frac{H_0}{c}\right)^2
  \int_0^{\chiH} \d\chi' \frac{\chi'}{a(\chi')} g(\chi') \delta(\chi'\bthet,\chi') \ ,\label{eq:kappaeff}
\end{align}
where $\Omega_{\rm m,0}$ is the total matter density, $H_0$ denotes
the Hubble constant, $a$ is the scale factor, $c$ is the speed of
light, $\chi_H$ is the comoving distance to the horizon and $g(\chi)$
is a weight function related to the normalized redshift distribution
$\d n(z)/\d z$ of the source galaxies as
\begin{align}
  g(\chi') &\equiv \int_{z(\chi')}^{z_{\rm H}} \d z \ 
   \frac{\d n(z)}{\d z} 
  \frac{\left[\chi(z)-\chi'\right]}{\chi(z)} \ . \label{eq:gchi}
\end{align} 
Aperture mass was developed by \citet{Schneider1996} as a technique to
estimate projected mass overdensities enclosed within a circular
region:
\be 
\Map(\bthet_0;\thetc) = \int_{\RR} \dthet_1 \kappa(\bthet_1) 
\wk(\left|\bthet_1-\bthet_0\right|;\thetc) \ , \label{eq:MapKappa} 
\ee
where $U$ is a compensated filter function.  In the flat sky limit the
(cross) aperture mass can be expressed in terms of a related
circularly symmetric filter function $Q(U)$ and the complex shear
field $\gamma$ in its E/B-decomposed basis:
\be 
\mathcal{M}_{{\rm ap}/\times} (\bthet_0;\thetc) \equiv \int_{\RR} \dthet_1 \gamma_{{\rm t}/\times}(\bthet_1;\bthet_0)
\wg(|\bthet_{1}-\bthet_0|;\thetc) \ , \label{eq:MapShear}
\ee
where the tangential and cross components of the shear field at
position $\bthet+\bthet_0$ with respect to the aperture center
$\bthet_0$ are defined as \citep{BartelmannSchneider2001}:
\begin{align}
\gamma_{\rm t}(\bthet;\bthet_0) + i \gamma_{\times}(\bthet;\bthet_0) & \equiv  - \gamma(\bm\theta+\bthet_0)\ex^{-2i\phi}  \label{eq:gt1} \ ,
\end{align}
in which $\phi$ denotes the polar angle associated with the vector
$\bthet$.  In the absence of systematic errors (B-modes) in the
lensing data, map-cross should vanish \citep{Schneideretal2002a}.
 
For this work we will make use of the polynomial filter function
introduced by \citet{Schneideretal1998}:
\be 
Q(\theta;\thetc) = \frac{6}{\pi \thetc^2}
\left(\frac{\theta}{\thetc}\right)^2
\left[1-\left(\frac{\theta}{\thetc}\right)^2\right]
\mathcal{H}(\thetc - \theta) \label{eq:schneiderQ} \ ,
\ee
where $\thetc$ is the characteristic scale of the filter and 
${\mathcal H}(x)$ is the Heaviside function, which guarantees
that the filter function has compact support.

%%%%%%%%%%%%%%%%%%%%%%%%%%%%%%%%%%%%%%%%%%%%%%%%%%%%%%%

\subsection{A hierarchy of aperture mass measures}

One may construct moments of the aperture mass field, and this gives
rise to the so called aperture mass statistics. At the two-point level
this gives us the variance $\MapStatEns{2}_{\rm c}(\vartheta_1)$ and
at the three-point, the skewness $\MapStatEns{3}_{\rm
  c}(\vartheta_1)$, etc., where the subscript $\rm c$ stands for the
connected cumulant obtained from the moments
\citep{ScoccimarroFrieman1996a}.  Owing to the fact that the aperture
mass is a convolution of the convergence field with a filter function,
it is possible to rewrite these moments in terms of their Fourier
space counterparts, that is the convergence spectra. For example for
the variance and skewness we have:
\begin{align}\label{eq:Map2Kappa}
  \MapStatEns{2}_{\rm c}(\vartheta) & = \int \dellnorm{1} C_{\kappa,2}(\ellvec{1})
  \ \left|\widetilde{U}_{\vartheta}(\ellvec{1})\right|^2    \ ; \\ 
  \label{eq:Map3Kappa}
  \MapStatEns{3}_{\rm c}(\vartheta)
  &=
  \int \dellnorm{1} \cdots \int \dellnorm{3} \ (2\pi)^2\delta^D\left(\sum_{i=1}^3\ellvec{i}\right) 
  \nonumber \\ &\hspace{-0.5cm} \times
  C_\kappa(\ellvec{1}, \cdots, \ellvec{3}) \
  \widetilde{U}_{\vartheta}(\ellvec{1})\widetilde{U}_{\vartheta}(\ellvec{2}) \widetilde{U}_{\vartheta}(\ellvec{3}) \  \ ,
\end{align}
where $\widetilde{U}_{\vartheta_i}$ denotes the Fourier transform of
the aperture mass filter function $U(\theta; \vartheta_i)$ and
$C_\kappa(\ellvec{1})$ denotes the convergence power spectrum, and
$C_\kappa(\ellvec{1}, \ellvec{2},\ellvec{3})$ the convergence
bispectrum. These spectra can formally be defined:
\begin{align}
    \left\langle \tilde{\kappa}(\ellvec{1}) \tilde{\kappa}(\ellvec{2}) \right\rangle_{\rm c}
    & = 
    (2\pi)^3 \delta^D\left(\ellvec{1}+\ellvec{2}\right)  \ 
    C_\kappa(\ellvec{1}) \ ; \\
    \left\langle \tilde{\kappa}(\ellvec{1}) \tilde{\kappa}(\ellvec{2})\tilde{\kappa}(\ellvec{3}) \right\rangle_{\rm c}
    & = 
    (2\pi)^3 \delta^D\left(\sum_{i=1}^3\ellvec{i}\right)  \ 
    C_\kappa(\ellvec{1},\ellvec{2},\ellvec{3})\ .
\end{align}
This of course can be generalised to $n$-point aperture mass moments:
\begin{align}\label{eq:MapnKappa}
    \MapStatEns{n}_{\rm c}(\vartheta)
    &=
    \int \dellnorm{1} \cdots \int \dellnorm{n} \ (2\pi)^2\delta^D\left(\sum_{i=1}^n\ellvec{i}\right) 
    \nonumber \\ &\hspace{-0.5cm} \times
    C_\kappa(\ellvec{1}, \cdots, \ellvec{n}) \ \widetilde{U}_{\vartheta}(\ellvec{1}) \cdots
    \widetilde{U}_{\vartheta}(\ellvec{n}) \  \ ,
\end{align}
where the $n$-point convergence spectrum is defined:
\begin{align}
  \left\langle \tilde{\kappa}(\ellvec{1}) \dots
  \tilde{\kappa}(\ellvec{n}) \right\rangle_{\rm c}
    & = 
    (2\pi)^3 \delta^D\left(\sum_{i=1}^n\ellvec{i}\right)  \ 
    C_\kappa(\ellvec{1},\dots,\ellvec{n})\ .
\end{align}

It is worth noting that due to the fact that $\widetilde{U}$ is a
sharply peaked filter function in Fourier space, the aperture mass
moment on a given scale only carries information about a specific range
of wavemodes $\vec{\ell}$ from the underlying polyspectrum. In order to
extract more of the information that is available one needs to compute
\Eqn{eq:MapnKappa} for a large set of aperture radii
\citep{Schneideretal2005a}.

%%%%%%%%%%%%%%%%%%%%%%%%%%%%%%%%%%%%%%%%%%%%%%%%%%%%%%%

\subsection{Multiscale aperture mass moments and their correlators}\label{sec:TheoryMultiscale}

Even if one considers a wide range of aperture radii there will be
certain wavemode configurations of the polyspectra that are
suppressed when compared with other configurations. This may result in
a loss of sensitivity to certain physical effects that are only
manifest in the higher-order polyspectra, such as those induced by
modifications of gravity or primordial non-Gaussianities. In order to
combat this one can further generalise the aperture mass moments in
several ways. First, if we choose different scales for the aperture
mass filter function, then we get the multiscale aperture mass moments.
For the $n$-point multiscale aperture mass moment this can be written:
\begin{align}\label{eq:MapnKappaMS}
    \MapStatEns{n}_{\rm c}(\vartheta_1, \cdots, \vartheta_n)
    &=
    \int \dellnorm{1} \cdots \int \dellnorm{n} \ (2\pi)^2\delta^D\left(\sum_{i=1}^n\ellvec{i}\right) 
    \nonumber \\ &\hspace{-0.5cm} \times
    C_\kappa(\ellvec{1}, \cdots, \ellvec{n}) \
    \widetilde{U}_{\vartheta_1}(\ellvec{1}) \cdots \widetilde{U}_{\vartheta_n}(\ellvec{n}) \  \ .
\end{align}

Second, if we correlate a set of apertures at different spatial
positions in the sky, then one can define the multiscale aperture
mass moment correlators \citep{Szapudietal1997,Munshietal2003}.  There
are two special cases where this approach can be applied, the first is
the case where the separation of the aperture is directed
perpendicular to the line of sight. The second case is where the
apertures are placed along the same line of sight, but where different
tomographic bins of source galaxies are used to estimate the aperture
mass. The former case measures the correlation of the cumulants on the
same redshift slice, but at different angular positions. The latter
case corresponds to correlating aperture measures in different surveys
with overlapping footprints, or between photometric redshift bins
within the same survey. As the aperture mass filter carries most of its weight in a compact region surrounding the aperture center one expects the signal to fall off rapidly for
aperture separations that exceed beyond a few times the aperture
radius. Generalizing the result of \citep{Schneideretal1998} we can
formally write this as follows:
\begin{align}\label{eq:MapStatCorr}
    \left\langle  
    \mathcal{M}_{\text{ap}}^n
    \mathcal{M}_{\text{ap}}^m
    \right\rangle_{\rm c}(\vartheta_1, \cdots, \vartheta_n, \vartheta'_1,
    \cdots, \vartheta'_m; \overrightarrow{\Delta})
    &= 
    \nonumber \\ &\hspace{-5cm}
    \int \dellnorm{1} \cdots \dellnorm{n+m} \ (2\pi)^2 \delta^D
    \left(\sum_{j=1}^{n+m}\vec{\ell_j}\right) \ C_\kappa(\vec{\ell}_1, \cdots, \vec{\ell}_{n+m}) \
    \nonumber \\ &\hspace{-4.5cm} \times \hspace{.2cm}
    \tilde{U}_{\vartheta_1}(\vec{\ell}_1) \cdots \tilde{U}_{\vartheta'_m}(\vec{\ell}_{n+m})
    \ e^{i\overrightarrow{\Delta} \sum_{j=1}^m\vec{\ell}_{n+j}}\ ,
\end{align}
where $\vec{\Delta}$ is a separation vector. Note that for zero
separation we recover the $(m+n)$th cumulant. In addition, we can
assess the impact of the exponential factor by evaluating the two
point cross-correlation coefficients $r_{mn}$, which are defined
in a similar way to those in \citep{Munshietal2005}:
\begin{align}\label{eq:twocrosscorrcoeff}
r_{mn}(\Delta) \equiv \frac{\left\langle X^m X^n \right\rangle_c(\Delta)}{\left\langle X^{m+n} \right\rangle_c}  \ ,
\end{align}
where for our case $X^m=\mathcal{M}_{\text{ap}}^m$.  In this work,
however, we do not consider the cosmological information contained in
\Eqn{eq:MapStatCorr}, but instead use it to assess how fast the
$r_{mn}$ converge to unity - this can be seen as a proxy for how
densely apertures need to be sampled within a survey footprint to
retrieve all available signal.

%%%%%%%%%%%%%%%%%%%%%%%%%%%%%%%%%%%%%%%%%%%%%%%%%%%%%%%

\section{Estimators for higher order aperture mass statistics}\label{sec:Estimators}

%%%%%%%%%%%%%%%%%%%%%%%%%%%%%%%%%%%%%%%%%%%%%%%%%%%%%%%
%%%%%%%%%%%%%%%%%%%%%%%%%%%%%%%%%%%%%%%%%%%%%%%%%%%%%%%

\subsection{Direct estimators for the aperture mass moments and
  their evaluation in linear order time}

In this subsection we concern ourselves with estimators for higher
order aperture mass statistics that mimic the original theoretical
expressions \Eqn{eq:MapnKappa} more closely. At first, let us
investigate the special case of all the radii being equal.

Consider an aperture of angular radius $\vartheta$, centred on the
position $\bthet_0$. The aperture contains $N$
galaxies\footnote{Strictly speaking, we select galaxies within the
support of the $Q$ filter function of that aperture. For the filter
functions we use in this work the support is always concentric around
the aperture center and linearly scaling with aperture
radius. Therefore we will continue referring to $N$ as the number of
galaxies per aperture.} with positions $\bthet_i$, complex
ellipticities $e_i$ and weights $w_i$. Then, for a single aperture,
one can write down an estimator for the $n$th order aperture mass
statistic \Eqn{eq:MapnKappa} as \citep{Schneideretal1998,
  Munshietal2003}
\begin{align}\label{eq:MapnEst}
  \MapStatEst{n} = (\pi \vartheta ^2)^n\frac{\sum_{(i_1, ..., i_n)^N} w_{i_1}Q_{i_1}e_{t,i_1}
    \cdots w_{i_n}Q_{i_n}e_{t,i_n}}{\sum_{(i_1, ..., i_n)^N}w_{i_1} \cdots w_{i_n}} \ ,
\end{align}
where we defined the shorthand notation
\be \label{eq:ShorthandSum}
\sum_{(i_1,\dots,i_n)^N} \equiv \sum_{i_1=1}^N \sum_{i_2\neq i_1}^N
\dots \sum_{i_n\neq i_{n-1}\neq\dots\neq i_{1}}^N\ .
\ee
In certain cases we might use further abbreviations, meaning that $(i_1,...,i_n)^N \equiv (i_1,...,i_n) \equiv \ \neq$. On applying the above estimator to the case of $n=2$, one can easily
show that that this estimator is unbiased after averaging over the
intrinsic ellipticity distribution, the galaxy positions within the
aperture, and finally over cosmological ensembles
\citep{Schneideretal1998, Porthetal2020}.

If we were to apply the above estimator given by \Eqn{eq:MapnEst} to
determine the hierarchy of aperture mass moments, then this naive
implementation would appear to result in an estimator that requires of
the order $N^n$ operations to compute. However, following our earlier
work \citep{Porthetal2020}, one can complete the sums to transform the
estimators into sums and products of linear order terms. In
Appendix~\ref{app:estimators1} we explicitly show, using elementary
means, how one can compute the skewness ($\MapStatEst{3}$) and
kurtosis ($\MapStatEst{4}$) using linear sums. The results for
second, third and fourth orders are:
\begin{align}
\MapEst &= \MsEst{1} \ ; \\
\MapStatEst{2} &= \frac{\MsEst{1}^{2} - \MsEst{2} }{1-\SEst{2}}  \ ;\\
\MapStatEst{3} &= \frac{\MsEst{1}^{3} - 3\MsEst{2}\MsEst{1} + 2\MsEst{3}}{1-3\SEst{2}+2\SEst{3}} \ ;\\
\MapStatEst{4} &= \frac{\MsEst{1}^{4} - 6\MsEst{2}\MsEst{1}^{2} + 3\MsEst{2}^{2} +8\MsEst{3}\MsEst{1} -6\MsEst{4}}
           {1-6\SEst{2}+3\SEstPow{2}{2}+8\SEst{3}-6\SEst{4}}\ ,
\end{align}
where we have introduced two additional quantities: $\SEst{m}$ and
$\MsEst{m}$, which are defined:
\begin{align}
  \MsEst{m} &\equiv \SEst{m} \ (\pi \vartheta^2)^m \
  \frac{\sum_{i=1}^N w_i^m Q_i^m e_{t,i}^m}{\sum_{i=1}^N w_i^m} \ ;
  \label{eq:MapnEstPowerSumsA}\\
  \SEst{m} &\equiv \frac{\sum_{i=1}^N w_i^m}{\left(\sum_{i=1}^N w_i\right)^m} \ \ ;
  \label{eq:MapnEstPowerSumsB}
\end{align}

Applying the elementary approach described in
Appendix~\ref{app:estimators1} beyond fourth order rapidly becomes
cumbersome, to say the least. We have therefore developed an analytic
method for generation of the $n$th order estimator decomposed into
linear sums. This follows from noting that the sum in \Eqn{eq:MapnEst}
runs over unequal indices and that one can express any statistic
$\MapStatEst{n}$ as a sum of the power sums
\Eqn{eq:MapnEstPowerSumsA}, where the coefficients preceding each term
are determined with the help of the complete Bell polynomials
$B_n$. Hence, for the general $n$th order estimate one has:
\begin{align}\label{eq:MapnEstBell}
\MapStatEst{n} =
\frac{B_n\left(-\MsEst{1}, -\MsEst{2}, -2\MsEst{3}, ..., -(n-1)!\MsEst{n}\right)}
     {B_n\left(-\SEst{1}, -\SEst{2}, -2\SEst{3}, ..., -(n-1)!\SEst{n}\right)} \ .
\end{align}
For full details of this derivation we refer the reader to
Appendix~\ref{app:estimators2}. Here we only note that each argument
that goes into \Eqn{eq:MapnEstBell} is a single sum over the galaxies
in the aperture and is therefore independent of the order of the
statistic. Using this formalism, we extend our decomposition to 5th
and 6th order, giving us:
\begin{align}
\MapStatEst{5} &= \frac{1}{N_5}
\left[ \MsEst{1}^5 - 10\MsEst{2}\MsEst{1}^{3} + 15\MsEst{2}^{2}\MsEst{1} +
  20\MsEst{3}\MsEst{1}^{2} \right. \nn \\
  & \left. \hspace{0.2cm} -20 \MsEst{3}\MsEst{2} -30\MsEst{4}\MsEst{1} + 24\MsEst{5} \right]\ ; \\
\MapStatEst{6} &= \frac{1}{N_6} 
\left[\MsEst{1}^{6}-15\MsEst{2}\MsEst{1}^{4}+45\MsEst{2}^{2}\MsEst{1}^{2}-15\MsEst{2}^{3}
  +40\MsEst{3}^{2}\right. \nn \\
  & \hspace{0.2cm} -90\MsEst{4}\MsEst{1}^{2}+40\MsEst{3}\MsEst{1}^{3}-120\MsEst{3}\MsEst{2}\MsEst{1} \nn \\
  & \hspace{0.2cm} \left. +90\MsEst{4}\MsEst{2}+144\MsEst{5}\MsEst{1}-120\MsEst{6}\right]
  \hspace{0.2cm} \ ,
\end{align}
where
\begin{align}
  N_5 =& 1-10\SEst{2}+15\SEstPow{2}{2}+20\SEst{3}-20\SEst{3}\SEst{2}-30\SEst{4}+24\SEst{5} \ ;\\
  N_6 =&  1-15\SEst{2}+45\SEstPow{2}{2}-15\SEstPow{2}{3}+40\SEst{3}-120\SEst{3}\SEst{2} \nn \\
  & +40\SEstPow{3}{2}-90\SEst{4}+90\SEst{4}\SEst{2}+144\SEst{5}-120\SEst{6}\ .
\end{align} 

%%%%%%%%%%%%%%%%%%%%%%%%%%%%%%%%%%%%%%%%%%%%%%%%%%%%%%%

\subsection{Direct estimators for the multiscale aperture mass moments}

In complete analogy we can write down an unbiased direct estimator for
the full multiscale aperture mass moments of \Eqn{eq:MapnKappaMS}:
\begin{align}\label{eq:MapnREst_h1}
\MapStatEst{n}(\vartheta_1, ..., \vartheta_n) &= (\pi \vartheta_1 ^2) \cdots (\pi \vartheta_n ^2)
\nonumber \\
&\hspace{-.8cm}\times \frac{\sum_{(i_1, ..., i_n)} w_{i_1}Q_{\vartheta_1,i_1}e_{t,i_1}
  \cdots w_{i_n}Q_{\vartheta_n,i_n}e_{t,i_n}}{\sum_{(i_1, ..., i_n)^N}w_{i_1} \cdots w_{i_n}} \ ,
\end{align}
where each index runs through all the galaxies within the aperture of
the largest radius. In this case the power sums of
\Eqns{eq:MapnEstPowerSumsA}{eq:MapnEstPowerSumsB} do not form a
sufficient basis to express these estimators, but we are still able to
write down the estimators from elements within the sets
\be
X_n \equiv \left\{ X_{(\mathscr{s}_1, ..., \mathscr{s}_n)}^{(m)} \  \left|  \  \mathscr{s}_i \in \{0,1\} \ ,
\  \sum_{i=1}^n\mathscr{s}_i = m \leq n \ \right. \right\} \ ,
\ee
where $X \in \{ M_{\rm s}, S \}$ and the corresponding elements
constitute of multivariate power sums being defined as
\begin{align}\label{eq:Msvar}
\MsEstPoly{m}{\mathscr{s}_1, ..., \mathscr{s}_n}
&\equiv
\left(\prod_{k=1}^n \left(\pi \vartheta^2_k\right)^{\mathscr{s}_k}\right)
\sum_{i=1}^{N(\mathscr{s})} w_i^m \prod_{j=1}^n \left[ e_{t,i}Q_{\vartheta_j,i} \right]^{\mathscr{s}_j} \ ,
\nonumber \\
\SEstPoly{m}{\mathscr{s}_1, ..., \mathscr{s}_n} &\equiv 
\sum_{i=1}^{N(\mathscr{s})} w_i^m  \ ,
\end{align}
where $N(\mathscr{s})$ denotes the number of galaxies within the aperture
of the smallest radius for which $\mathscr{s}_i$ is not zero. Despite the
more complicated looking form compared to the equal radius case these
estimators can also be computed in $\mathcal{O}(N)$ time using the
$\left|\hat{X}_n\right| = 2^n-1$ distinct multivariate power sums
\Eqn{eq:Msvar} and summing over various partitions $P$ of the set
$\{1, \cdots, n \}$:
\begin{align}\label{eq:MapnREst}
  \MapStatEst{n}&(\vartheta_1, ..., \vartheta_n) = \nonumber\\
    &\frac{\sum_{m=1}^n \sum_{\pi \in P_{n,m}} (-1)^m \prod_{i=1}^{m} (n_i-1)! \
    \MsEstPoly{n_i}{\mathscr{s}_1(\pi_i), \cdots, \mathscr{s}_n(\pi_i)}
  }{\sum_{m=1}^n \sum_{\pi \in P_{n,m}} (-1)^m \prod_{i=1}^{m} (n_i-1)! \
    \SEstPoly{n_i}{\mathscr{s}_1(\pi_i), \cdots, \mathscr{s}_n(\pi_i)}} \ .
\end{align}
In this expression the combination of the two outer sums run through
each partition $\pi$ that consists of $m$ blocks and the $\alpha(\pi_i)$
denote the value of the $\alpha$ as evaluated from the $i$th block of
the partition. For a motivation of this equation and explicit
expressions we again refer to \App{app:estimators2}.

%%%%%%%%%%%%%%%%%%%%%%%%%%%%%%%%%%%%%%%%%%%%%%%%%%%%%%%

\subsection{Estimators applied to a large survey}

In order to estimate any aperture statistics
\be \mathfrak{M}\in \left\{\left<M_{\rm ap}^{2}\right>(\vartheta_1),\ \left<M_{\rm
    ap}^{3}\right>(\vartheta_1),\dots \right\}\ee
on a contiguous survey field one can simply place an ensemble of
apertures on the field and compute their weighted means
\begin{align}\label{eq:MapStatEns}
  \hat{\mathfrak{M}} = \frac{\sum_{\rm ap} w_{\rm ap}
    \hat{\mathfrak{M}}_{\rm ap}}{\sum_{\rm ap} w_{\rm ap}} \ ,
\end{align}
where the weights $w_i$ should be chosen to minimize the variance of
the estimator. Owing to the linearity of \Eqn{eq:MapStatEns}, if the
estimator of a single aperture is unbiased, then so is
\Eqn{eq:MapStatEns}. Thus including more apertures will increase the
signal-to-noise of the ensemble estimator.

%%%%%%%%%%%%%%%%%%%%%%%%%%%%%%%%%%%%%%%%%%%%%%%%%%%%%%%

\subsection{Variance of the direct estimators}\label{sec:DirectCov}

In order to understand how to weight the apertures we need to obtain
expressions for the variance of the moment estimators. On generalizing
the prescriptions outlined in \citep{Schneideretal1998,
  Munshietal2003} to include the shear weights, as in
\citet{Porthetal2020}, one can work out expressions for the variance
of the higher-order direct estimators \Eqn{eq:MapnEst} for a given
aperture. For the explicit derivation of for the variance of the third order statistic see
Appendix D in the online supplementary material. From this analysis we see that the general
expression can be written:
\begin{align}\label{eq:MapStatCov}
\sigma^2\left[\MapStatEst{n}\right] 
&=
\sum_{\ell=0}^n  \sum_{m=\ell}^n \frac{\sum_{\neq} w_{i_1}^2\cdots w_{i_m}^2 w_{i_{m+1}}
  \cdots w_{i_n} w_{j_{m+1}}\cdots w_{j_n}}{\left( \sum_{\neq} w_{i_1}\cdots w_{i_n}\right)^2} 
\nonumber \\ 
&\times C(n,\ell,m) M_{g,2}^{\ell}\sigsq{\ell}  \MsMapStatEns{m-\ell}{2(n-m)} \ ,
\end{align}
where the sum over the galaxy weights can again be decomposed as sums
of (bivariate) power sums and the multiplicities are given by
\begin{align}
C(n, \ell, m) & = \frac{1}{\ell!(m-\ell)!} \left(\frac{n!}{(n-m)!}\right)^2 \ .
\end{align}
For a discussion of the origin for the multiplicity factor $C(n, \ell,
m)$, as well as a motivation of \Eqn{eq:MapStatCov} and some of its limits we refer the reader to \App{app:covariance}; in particular we obtain for the shape noise dominated limit
\begin{align}
    \sigma^2\left[\MapStatEst{n}\right] 
    \approx
    n!\frac{\sum_{\neq} w_{i_1}^2\cdots w_{i_n}^2}{\left( \sum_{\neq} w_{i_1}\cdots w_{i_n}\right)^2}  \ \left(\frac{\sigma^2_\epsilon}{2}\right)^n \MgStat{n} \  . \label{eq:MapStatCovShotLimit}
\end{align}
The above formula gives the variance per aperture, thus for the
estimator over the full survey field, \Eqn{eq:MapStatEns}, the
variance can be written down as:
\begin{align}\label{eq:MapStatEnsCov}
  \sigma^2\left[\hat{\mathfrak{M}}\right] &= \cov\left( \frac{\sum_i w_i \hat{\mathfrak{M}}_i}{\sum_i w_i},
  \frac{\sum_j w_j \hat{\mathfrak{M}}_j}{\sum_j w_j} \right)
    \nonumber \\ &= 
    S_2 \ \sigma^2\left[{\hat{\mathfrak{M}}_{\rm ap}}\right] +
    \frac{\sigma^2\left[{\hat{\mathfrak{M}}_{\rm ap}}\right]}{\left(\sum_i w_i\right)^2}
    \sum_{i \neq j} w_i w_j \rho (\hat{\mathfrak{M}}_i, \hat{\mathfrak{M}}_j)\ ,
\end{align}
\begin{figure}
    \centering
    \centering{
      \includegraphics[width=0.45\columnwidth]{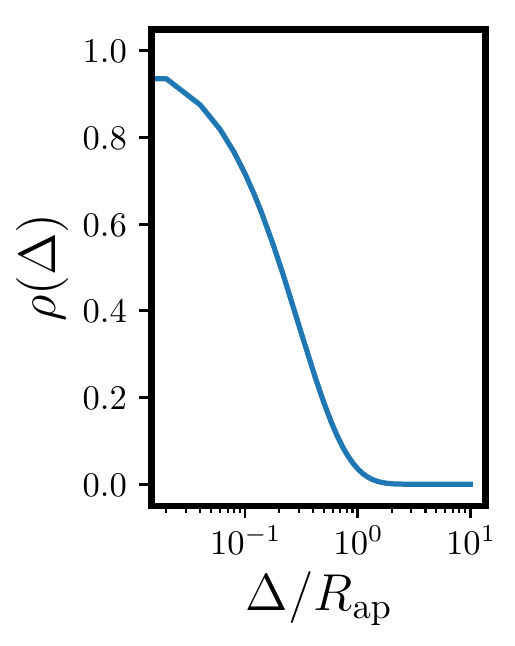}
      \includegraphics[width=0.45\columnwidth]{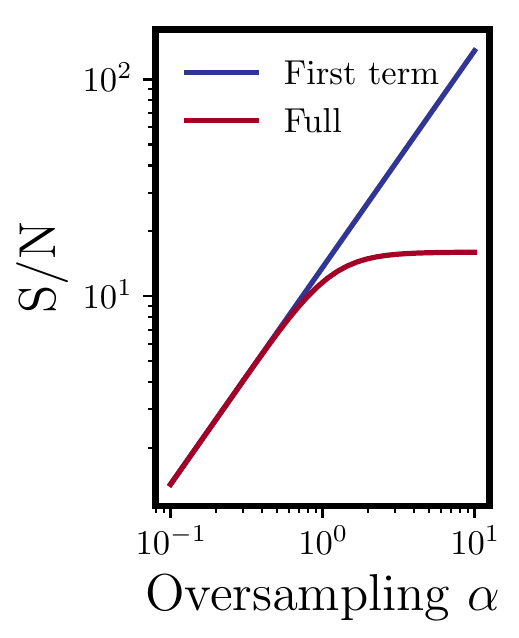}
      }
      \caption{Example configuration of the correlation coefficient $\rho$ (left) and its effect on the signal contained in a survey field as predicted from \Eqn{eq:MapStatEnsCovb} (right).}
\label{fig:SpatialCorrCoeffExample}
\end{figure}
where in the above we have defined the cross-correlation coefficient
between apertures whose centres are at position $\bm\theta_i$ and
$\bm\theta_j$ to be:
\be \label{eq:twocrosscorrcoeff2}
\rho(\mathfrak{M}_i, \mathfrak{M}_j)
\equiv \frac{\left<\mathfrak{M}_i, \mathfrak{M}_j\right>}
       {\sqrt{\left<\mathfrak{M}_i, \mathfrak{M}_i\right>\left<\mathfrak{M}_j, \mathfrak{M}_j\right>}} \ .
\ee
Note that for the case of well separated apertures, the
cross-correlation coefficient will vanish and only the first summand
needs to be taken into account, which for unity weights gives the
familiar $1/N_{\rm ap}$ scaling of the variance. If the apertures are
oversampled, this assumption is no longer valid and the term involving
$\rho$ must be included. Owing to the fact that $\rho$ should only
depend on the relative spatial distance $\Delta$ between the aperture
centres, we can rewrite \eqref{eq:MapStatEnsCov} as a weighted sum over all possible distances between aperture center pairs:
\begin{align}
    \sigma^2\left[\hat{\mathfrak{M}}\right] = \ 
    &S_2 \ \sigma^2\left[{\hat{\mathfrak{M}}_{\rm {ap}}}\right] + 
    \nonumber \\ 
    &\frac{\sigma^2\left[{\hat{\mathfrak{M}}_{\rm {ap}}}\right]}{\left(\sum_i w_i\right)^2}
    \sum_{b \in \rm bins} \left(\sum_{i,j \in \mathcal{I}_b} w_i w_j \right)
    \rho(\hat{\mathfrak{M}}, \Delta_b)
     \\ \approx
    &\frac{\sigma^2\left[{\hat{\mathfrak{M}}}_{\mathrm{ap}}\right]}{N_{\mathrm{ap}}} + 2\pi \  \frac{\sigma^2\left[{\hat{\mathfrak{M}}}_{\mathrm{ap}}\right]}{A_{\mathrm{survey}}}  \int_{R_{\mathrm{ap}}/\alpha}^\infty \d\Delta \ \Delta \  \rho (\hat{\mathfrak{M}}, \Delta)
    \ , \label{eq:MapStatEnsCovb}
\end{align}
where in the first step the bins are defined as a partition of the reals, and $\mathcal{I}_b
\equiv \{ i,j | \Delta(i,j) \in b \}$ collects all the aperture center
pairs falling into bin $b$. For the second step we make the approximation that each aperture contains roughly the same signal such that the weights can be set to unity and we furthermore rewrote the expression in a continuous version, which makes the interpretation of the cross term more concise. In particular, we parametrize the lower bound of the integral in terms of the aperture oversampling rate $\alpha \equiv R_{\mathrm{ap}}/\Delta_{\mathrm{min}}$.

In a realistic scenario we expect $\rho$ to to rapidly decrease from unity and then to approach zero for $\Delta \gg R_{\mathrm{ap}}$. An example of such a correlation coefficient is shown in \Fig{fig:SpatialCorrCoeffExample}. Here we explicitly see the importance of taking into account the cross term once there is a substantial overlap between neighbouring apertures. In this example we would infer that measuring the statistics with $\alpha \approx 2$ would be sufficient to extract most of the signal.

%%%%%%%%%%%%%%%%%%%%%%%%%%%%%%%%%%%%%%%%%%%%%%%%%%%%%%%

\subsection{Implementation and scaling of the direct estimator}

A practical implementation of \Eqn{eq:MapStatEns} consists of three
steps:
\begin{enumerate}
\item Spatially organise the shape catalog to allow for a fast
  assignment of galaxies to apertures.
\item For each aperture of the ensemble compute
  $\hat{\mathfrak{M}}_{\rm ap}$, the associated weight and optionally
  additional systematics (i.e. the coverage fraction $c_k$). Store
  each of these values in an array.
\item Based on some aperture selection and aperture weighting criteria
  $w_{\rm ap}$, update the weights and evaluate the weighted sum.
\end{enumerate}
In what follows we will explore each of these steps in more detail and
for clarity we will denote the number of galaxies in the survey and in
the aperture as $N_{\rm g}$ and $N_{\rm g, ap}$, respectively.

%%%%%%%%%%%%%%%%%%%%%%%%%%%%%%%%%%%%%%%%%%%%%%%%%%%%%%%

\subsubsection{Assigning galaxies to apertures}

For our implementation we use a spatial hashing data structure. We
start by covering the survey footprint with an equal area mesh of
$N_{\rm pix}$ pixels and create a hash table with the ID of each pixel
as the key and the galaxy IDs as values. The hash function in our case
is the ordinary pixel assignment function. For each aperture we
iterate over the associated galaxies within pixels that partially lie
within the $Q$ filter's support. The construction of the hash table
scales as $\mathcal{O}\left(\max (N_{\rm pix}, N_{\rm g}) \right)$ and the assignment is achieved in $\mathcal{O}\left(\max(N_{\rm pix,ap},
N_{\rm g,ap})\right)$ time per aperture.  We found that when
making a sensible choice of the mesh's coarseness, this data structure
is more stable than a naive KD-tree based implementation as it does not require an additional
range search operation which scales as $\mathcal{O}(\log(N_{\rm g}))$
per aperture and thus becomes a bottleneck for small apertures.
%

%%%%%%%%%%%%%%%%%%%%%%%%%%%%%%%%%%%%%%%%%%%%%%%%%%%%%%%

\begin{figure*}
\centering{
    \includegraphics[width=1.\columnwidth]{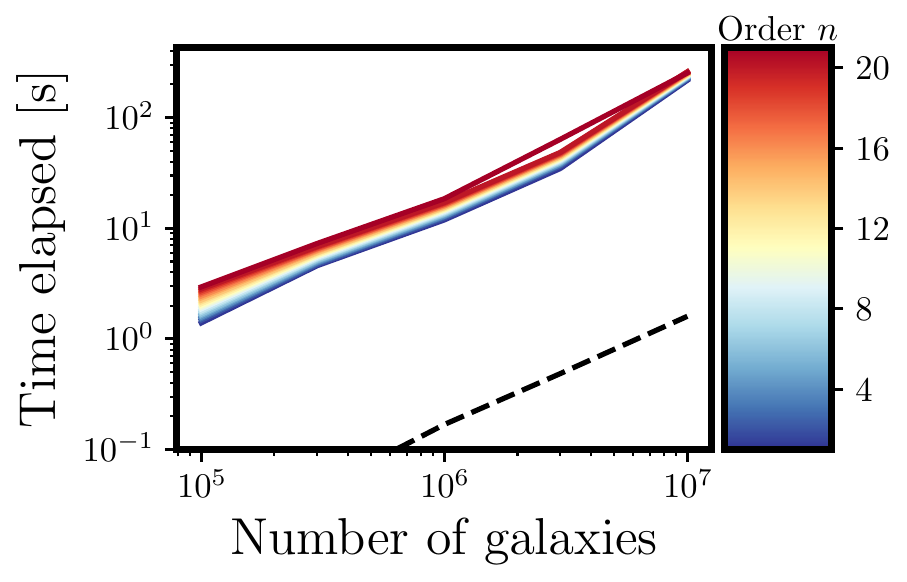}
    \includegraphics[width=1.\columnwidth]{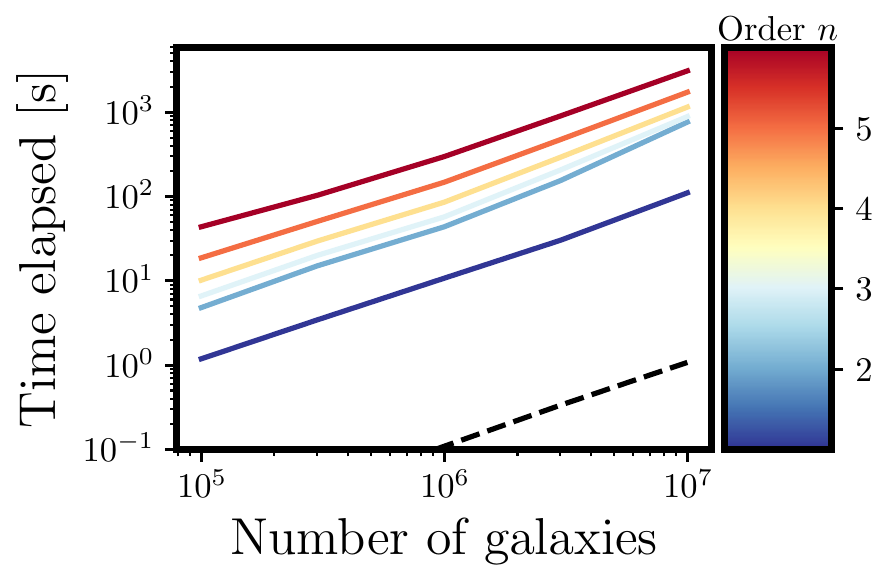}}
  \caption{\small{Computational complexity of the direct estimators
      for equal (left) and unequal (right) aperture radii as a
      function of the number of galaxies. All results are given for
      apertures of radius $10'$ which are oversampled by a factor of
      sixteen $(\alpha=4)$ on a survey field of size $(12
      \rm{deg})^2$. Different colors indicate different orders of the
      evaluated statistics.  The black dashed line indicates the time
      spent in constructing the spatial hash. We see that for equal
      aperture radii the evaluation of higher order statistics
      basically comes for free, while for unequal radii there is a
      constant multiplicative offset based on the relative size of the
      radii and on the order which traces the number of multivariate 
      power sums that need to be evaluated. All the scaling were
      obtained when running the estimator on a single CPU
      core. }}\label{fig:DirectEstimatorScaling}
\end{figure*}

%%%%%%%%%%%%%%%%%%%%%%%%%%%%%%%%%%%%%%%%%%%%%%%%%%%%%%%

\subsubsection{Computing the statistics per aperture}

For the case of all radii being equal we first compute the power sums
in \Eqns{eq:MapnEstPowerSumsA}{eq:MapnEstPowerSumsB} and then
recursively transform them to the corresponding moments via the
recurrence relation \citep{Comtet1974}
\begin{align}\label{eq:BellRecurrence}
B_{n+1}(x_1, \cdots, x_{n+1}) &=
\sum_{i=0}^n \binom{n}{i} B_n(x_1, \cdots, x_n) x_{i+1} \ ,
\end{align}
where $B_0 \equiv 1$. Evaluating each power sum is linear in $N_{\rm
  g, ap}$ and for all practical applications the time taken for
transforming to the $M_{\rm ap}^n$ basis can be neglected.

For the general case we need to compute the relevant multivariable
power sums \Eqn{eq:Msvar} and bring them to the aperture moments basis
by the transformation \Eqn{eq:MapnREst}. In order to
dynamically allocate and evaluate those expressions we use a
combinadic counting scheme to organize the power sum basis whereas the
transformation equation is generated with the help of restricted growth strings \citep{Knuth2005_D}.

%%%%%%%%%%%%%%%%%%%%%%%%%%%%%%%%%%%%%%%%%%%%%%%%%%%%%%%

\subsubsection{Choice of weights for the averaging}

Following our findings in \citet{Porthetal2020} we employ an inverse
shot noise weighting scheme with an additional hard cutoff $c_{\rm
  lim}$ for the aperture coverage $c_{\rm ap}$, which for second order
statistics was found to lower the mask induced bias while increasing
the signal-to-noise compared to equal weights. The explicit form of
the weights for the $n$th moment can be found from \Eqn{eq:MapStatCovShotLimit}
when neglecting all constant contributions:
\begin{align}\label{eq:InvShotWeight}
\mathcal{W}^{\rm (shot)}_{\rm ap}(c_{\rm lim}) \equiv 
\left[\frac{\sum_{(i_1, \cdots, i_n)} w_{i_1}^2 \cdots w_{i_n}^2 }
  {\left(\sum_{(i_1, \cdots, i_n)} w_{i_1} \cdots w_{i_n}\right)^2}\right]^{-1}
\mathcal{H}(c_{\rm ap} - c_{\rm lim}) \ .
\end{align}
Dependent on whether we are dealing with the case of equal or unequal
aperture radii the sums can be decomposed in a similar fashion as
described above and evaluated together with the corresponding
linearised direct estimator. As a further refinement one could also
include the weights and completenesses of the surrounding apertures
weighted by the spatial cross correlation coefficient $\hat{\rho}$ -
this would upweight apertures that are close to a mask as they cover
more unique area.

%%%%%%%%%%%%%%%%%%%%%%%%%%%%%%%%%%%%%%%%%%%%%%%%%%%%%%%
%%%%%%%%%%%%%%%%%%%%%%%%%%%%%%%%%%%%%%%%%%%%%%%%%%%%%%%

\section{Results: application to Gaussian mocks}
\label{sec:Measurements}

%%%%%%%%%%%%%%%%%%%%%%%%%%%%%%%%%%%%%%%%%%%%%%%%%%%%%%%

\subsection{Aperture mass statistics and Gaussian lensing fields}
%

%%%%%%%%%%%%%%%%%%%%%%%%%%%%%%%%%%%%%%%%%%%%%%%%%%%%%%%
\begin{figure*}
    \centering
    \centering{
      \includegraphics[width=1.0\columnwidth]{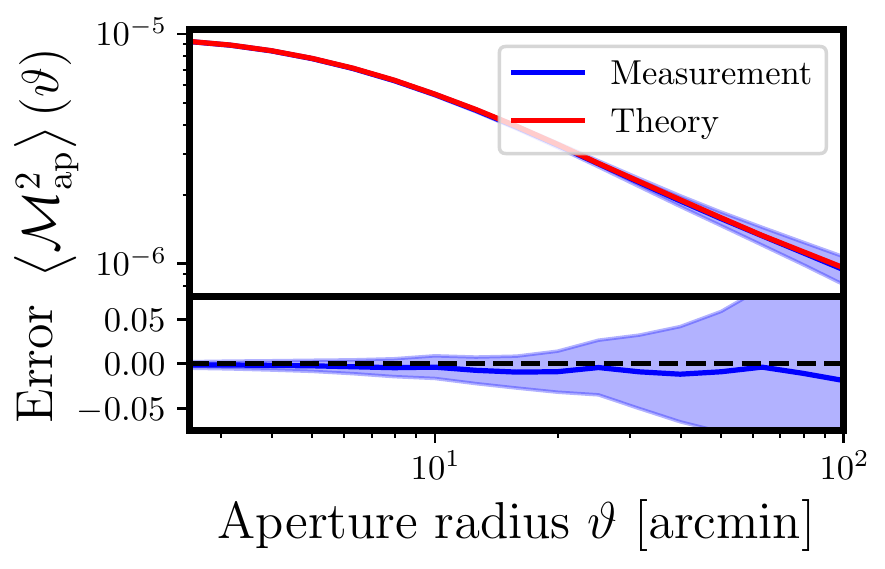}
      \includegraphics[width=1.0\columnwidth]{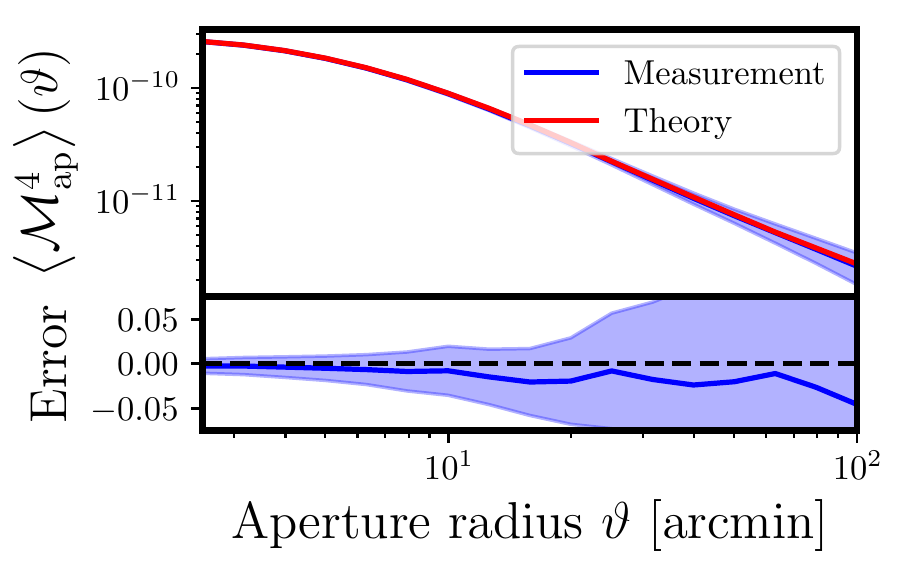}}
    \caption{\small{Comparison of the measured aperture mass moments
        with their theoretical prediction. {\bf Left panel: } The
        upper subpanel shows the aperture mass dispersion as a
        function of the aperture scale. The red line shows the
        theoretical predictions evaluated from the input power
        spectrum and the blue line shows the measurement from the
        mocks. The blue shaded regions show the standard deviation of
        the corresponding measurement across the ensemble. The lower
        subpanel shows the relative error between the measurement and
        the theory, with the line styles as before. {\bf Right panel:
        } Same as left panel, but for the kurtosis of aperture
        mass. }} \label{fig:GaussMock1}
\end{figure*}

%
%%%%%%%%%%%%%%%%%%%%%%%%%%%%%%%%%%%%%%%%%%%%%%%%%%%%%%%

In order to validate that our hierarchy of aperture mass moment
estimators are unbiased and do indeed recover correct results, we
first apply them to a set of Gaussian mock lensing simulations. In
this case, the whole moment hierarchy can be written as powers of the
variance of the aperture mass. Hence, this motivates us to define the
scaled aperture mass moments:
\begin{align} \label{eq:GaussCum}
s_n (\vartheta_1)
\equiv 
\frac{1}{(n-1)!!}\frac{\MapStatEns{n}(\vartheta_1)}{\left[\MapStatEns{2}(\vartheta_1)\right]^{n/2}}
=
\delta^K_{n,2\mathbb{N}} \ ,
\end{align}  
where the final equality is true for a Gaussian field only. 

In order to test this we generated 256 Gaussian lensing mocks. The
methodology to create each mock was as follows:
\begin{itemize}
\item We first generate a Gaussian convergence field over a $12
  \times12 \ {\rm square\ deg}$ survey area. The area is tiled by a
  rectangular mesh of $8192^2$ pixels. The variance of the convergence
  is obtained through specifying the convergence power spectrum, and
  we do this for a source distribution similar to that for the \cfhtl
  \ survey \citep{Fuetal2014}.
\item We next obtain the shear field. This is done by Fourier
  transforming the convergence field and making use of the
  \citet{Kaiseretal1993} approach\footnote{In order to suppress edge
  effects introduced by the FFT we build the pixelated convergence
  field on an a plane having $16$ times the area of the mock.}.
\item We then sample $4\times 10^6$ galaxies into the survey footprint
  and use a multilinear interpolation of the shear field onto each
  galaxy.
\end{itemize}
Note that since we are assessing the accuracy of the estimators only
we choose to set the intrinsic ellipticities of our source galaxies to
zero. On repeating the analysis below when including this term we did
not find a shift of the curves.

%%%%%%%%%%%%%%%%%%%%%%%%%%%%%%%%%%%%%%%%%%%%%%%%%%%%%%%

\subsection{Computational scaling tests}

Owing to the fact that each step of our algorithm is strictly linear,
we expect a linear relationship between the elapsed time for estimator
evaluation and the number of galaxies, for any given statistic. In
addition, for the equal radius case the order of the statistics should
not strongly impact the evaluation time. However, for the unequal
radius case, this does not necessarily hold true, since the computation
depends on the relative sizes of the apertures as well as on the order
of the statistics to be evaluated.

\Figure{fig:DirectEstimatorScaling} shows the elapsed time of the
direct estimator calculation for a Gaussian mock, where the number of
sampled galaxies in the mock is increased. Focusing on the left panel
first, this shows the case for the standard aperture mass estimators
with equal radii and here we compute all of the moments up to the 20th
order. As expected the computational time for all of the moments
scales linearly with the size of the problem and we also see that
there is no obvious drop in performance for the higher order moments.

The right panel of \Figure{fig:DirectEstimatorScaling} is the same as
the left panel, but now for the case of unequal radii aperture mass
moments, and here we only consider moments up to 6th order. There are
two differences between the equal and non-equal radius case. First, we
can see that there is a much larger multiplicative offset between
adjacent orders for the generalized statistics. This is expected as
the number of basis elements that need to be allocated in that case is
given by $2^n$ compared to the $n$ ones in the equal radius
statistics. We also observe that for the second order statistic the
unequal radius calculation does roughly need four times as long as the
equal radius one. We can explain this offset when noting that for our
example the ratio of the largest and smallest scale was set to
two. With our definition of the oversampling rate as being relative to
the smallest aperture radius this implies that we need to allocate
four times as many galaxies.

Finally, we note that the superior scaling of the direct estimator
compared to traditional estimation methods should not come as a
surprise. Looking back at the original definition \Eqn{eq:MapnKappa}
of the aperture mass one sees that it depends on the positions and
shapes of the galaxies with respect to the aperture origin. In
contrast, when switching to the description of aperture mass
in terms of the shear correlation functions
\citep[i.e.][]{Schneideretal2002a}, the main dependence shifts to the
relative distance and shapes between tuples of galaxies. This change
of reference position makes the evaluation of correlation function
based estimators intrinsically much more complex than a simple
discretization of \Eqn{eq:MapnKappa}.

%%%%%%%%%%%%%%%%%%%%%%%%%%%%%%%%%%%%%%%%%%%%%%%%%%%%%%%

\begin{figure}
  \centering{
    \includegraphics[width=1.0\columnwidth]{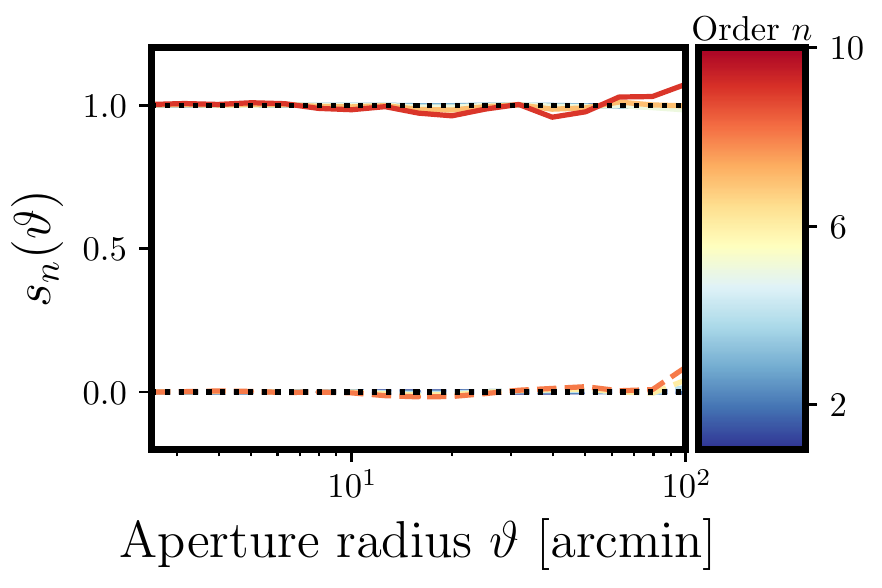}}
  \caption{\small{Scaled $n$th order aperture mass moments $s_n
      (\vartheta)$ (see Eq.\,\ref{eq:GaussCum}), measured in the
      ensemble of 256 Gaussian mocks, as a function of the aperture
      scale, for all moments up to 10th order. The solid lines of
      varying colours show the mean of the measurements. The dotted
      black lines show the Gaussian theoretical expectations. For a
      Gaussian mock, the even order $s_n (\vartheta)$ give unity, and
      the odd ones vanish. }} \label{fig:GaussMock2}
\end{figure}

%%%%%%%%%%%%%%%%%%%%%%%%%%%%%%%%%%%%%%%%%%%%%%%%%%%%%%%

\begin{figure*}
  \centering{
    \includegraphics[width=1.0\columnwidth]{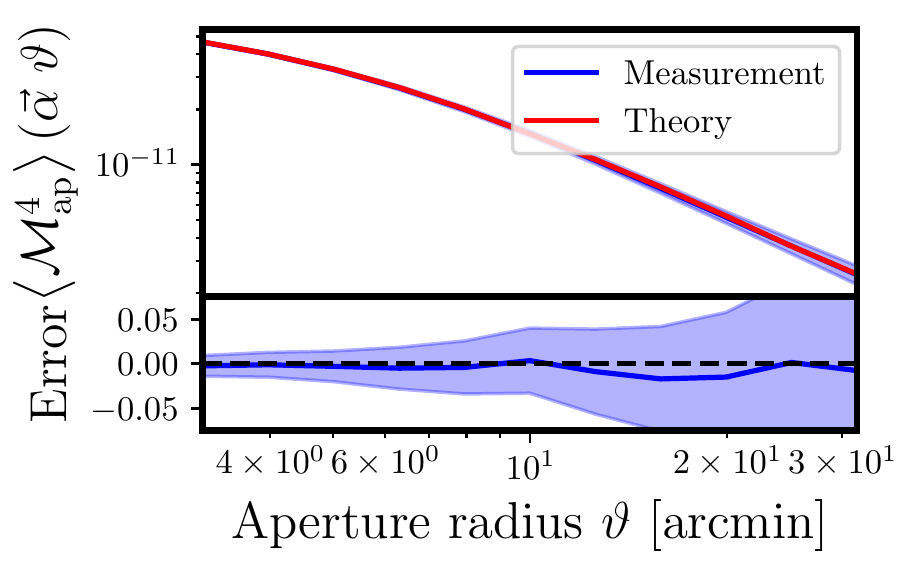}
    \includegraphics[width=1.0\columnwidth]{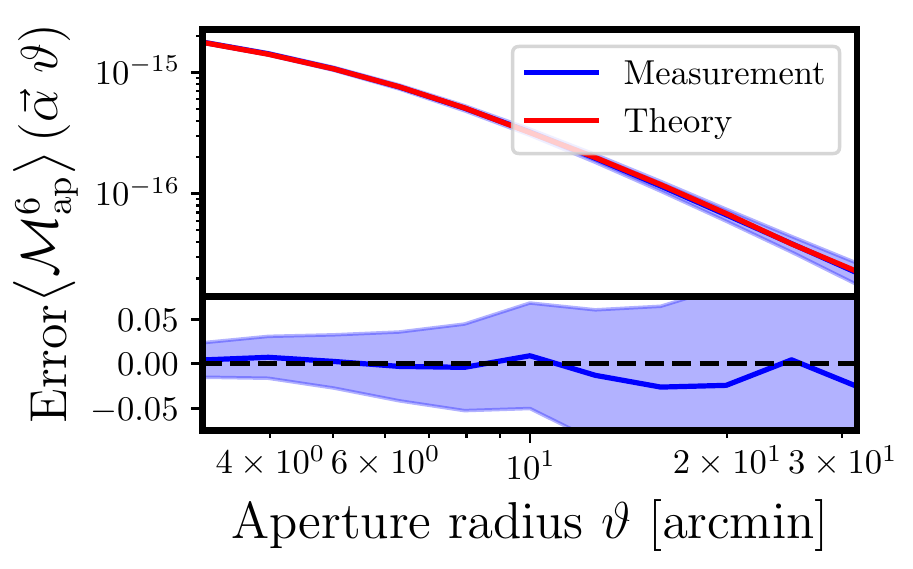}}
  \caption{\small{Multiscale aperture mass moments as a function of
      the scale parameter $\vartheta$, measured in the ensemble of 256
      Gaussian mocks. Line styles are the same as in
      \Fig{fig:GaussMock1}.  {\bf Left panel: } the fourth order
      aperture statistics. In this case, the vector of aperture scales
      was set to $\bm\alpha=(0.5,0.8,1.,2.)$. {\bf Right hand side: }
      Same as left hand side, but this time for the sixth order
      statistic with the vector of aperture scales set to $\bm{\alpha}
      = (0.5,0.7,1.,1.1,1.5,2.)$. }}
    \label{fig:GaussMock3}
\end{figure*}

%%%%%%%%%%%%%%%%%%%%%%%%%%%%%%%%%%%%%%%%%%%%%%%%%%%%%%%

\subsection{A hierarchy of aperture mass moments}

\Figure{fig:GaussMock1} shows a comparison of the direct estimators
for the second and fourth order aperture mass moments as a function of
angular scale as applied to the 256 Gaussian mocks.  Here we consider
the case where all the aperture radii are equal (recall that for a
Gaussian field all of the odd moments vanish). In both cases the
curves are in very good agreement with the Gaussian theory
predictions, indicated by the solid red lines. We also note that for
increasingly large aperture radii the measured results appear to be
slightly below the theoretical expectation. This discrepancy can be
attributed to finite field effects, as well as border effects being
introduced by the Kaiser-Squires inversion method \citep[see][for a
  discussion]{Piresetal2020}.

\Figure{fig:GaussMock2} presents the measured $s_n$ (see
\Eqn{eq:GaussCum}) for all of the aperture mass moments up to 10th
order as a function of the aperture scale. We see that they are
consistent with the Gaussian theoretical expectations. Note that in
order to obtain this good agreement and circumvent the finite field
effects described above, we used the ensemble mean of the measured
aperture mass variance as the denominator in $s_n$.

\Figure{fig:GaussMock3} displays the fourth and sixth order
multiscale aperture mass statistics as a function of the scale
parameter. Note that there are a number of options for exploring the
configuration dependence of the multiscale aperture mass moments,
here we focus on fixing the ratio of the filter lengths and varying
the overall scale of the configuration $\vec{a}$ with the parameter $\vartheta$,
e.g. for the kurtosis we would have 
\be 
\MapStatEns{4}_{\rm c}(\vec{\vartheta})
\equiv
\MapStatEns{4}_{\rm c}(a_1\vartheta, \cdots,
a_4\vartheta) \ 
\ee
where the constant $a_i \in \mathbb{R}_+$ specify the configuration. The estimates shown in the figure were obtained
using our generalized estimator \Eqn{eq:MapnREst}. As for the previous
cases, we find good agreement between the measurements and the
Gaussian predictions, which were obtained by making use of
\Eqn{eq:MapnKappaMS} and Wicks theorem for the convergence polyspectra
\citep{Bernardeauetal2002}.
%
%%%%%%%%%%%%%%%%%%%%%%%%%%%%%%%%%%%%%%%%%%%%%%%%%%%%%%%%

\begin{figure*}
\centering{
    \includegraphics[width=1.\columnwidth]{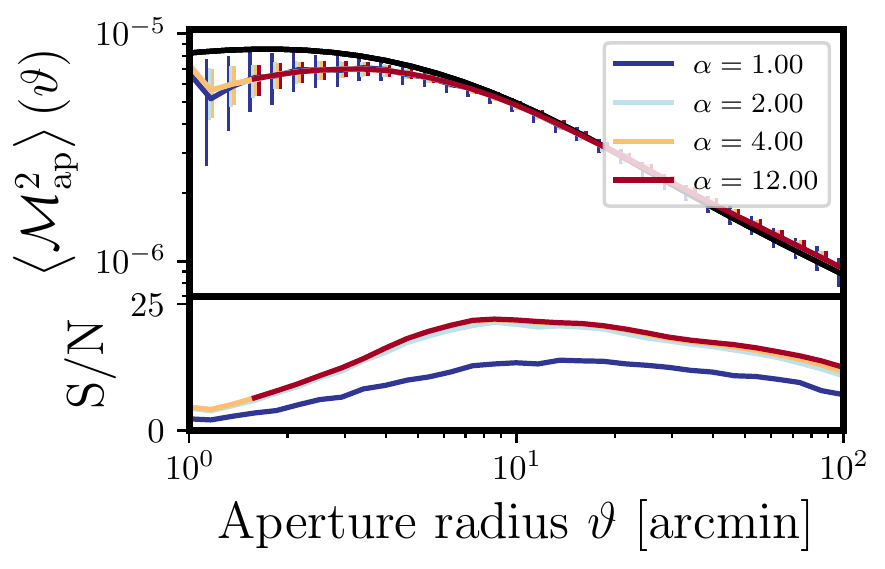}
    \includegraphics[width=1.\columnwidth]{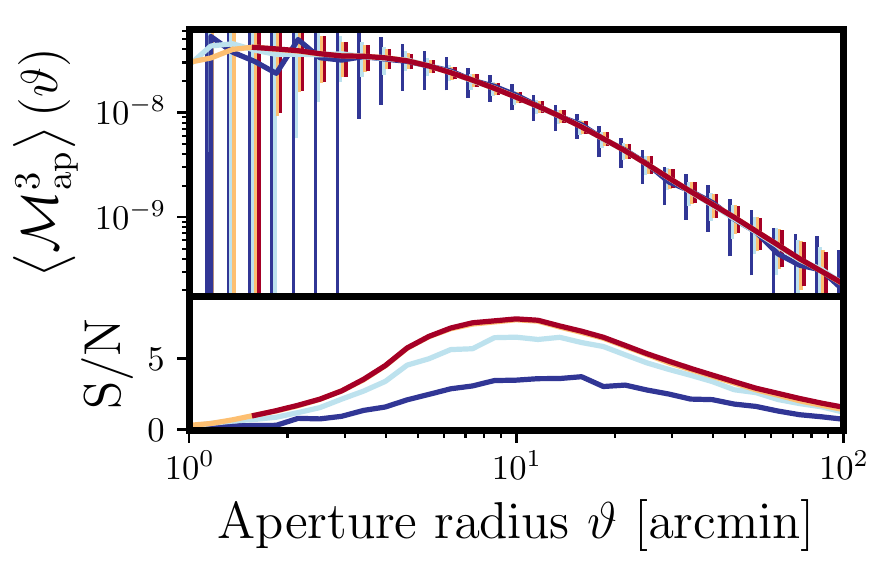}\\
    \includegraphics[width=1.\columnwidth]{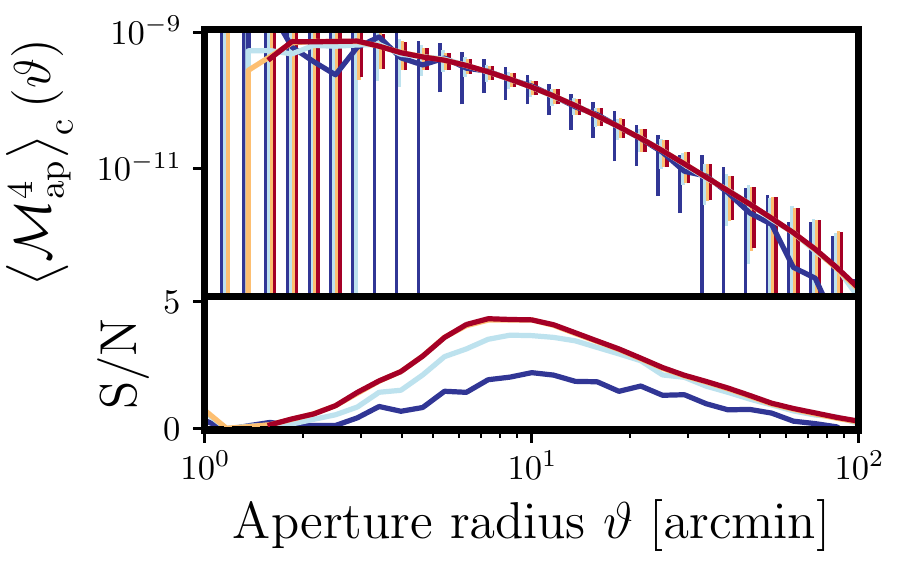}
    \includegraphics[width=1.\columnwidth]{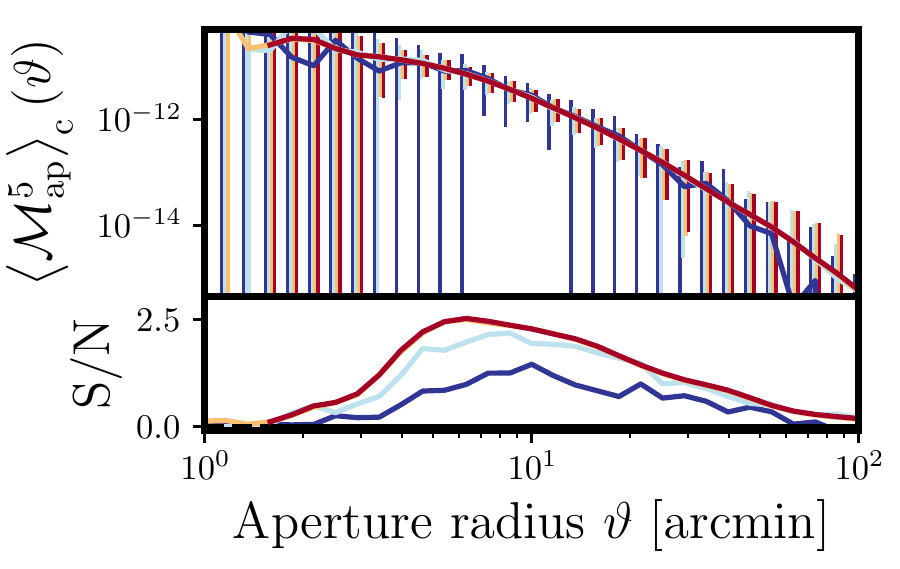}}
  \caption{\small{Measurement of the aperture mass statistics in the
      SLICS simulation suite for different aperture oversampling rates
      $\alpha$. All measurements were done on an ensemble of 819
      realizations with an angular area of 100 deg$^2$ each, where
      the $n(z)$ follows the KiDS-450 distribution.  The upper part of
      the panels correspond to the mean and rescaled standard deviation from
      the ensembles. The lower panel shows the signal-to-noise for the
      corresponding statistics when rescaled to match a 1000 deg$^2$ survey. For the aperture mass dispersion we
      additionally plot the theoretical prediction as the black
      line. For the fourth and fifth order plots we restrict ourselves
      to the contribution of the connected part of the convergence
      polyspectra. We see that choosing an oversampling parameter of
      $\alpha \gtrsim 4$ recovers most of the
      information.}}\label{fig:SLICSMockEqualSN}
\end{figure*}

%%%%%%%%%%%%%%%%%%%%%%%%%%%%%%%%%%%%%%%%%%%%%%%%%%%%%%%

\section{Results: Detection significance of higher order moments}\label{sec:DetectionSignificance}

In this section we now turn to the question of the detection
significance of higher order aperture statistics from current and
future surveys.

%%%%%%%%%%%%%%%%%%%%%%%%%%%%%%%%%%%%%%%%%%%%%%%%%%%%%%%
%
\begin{figure*}
\centering{
    \includegraphics[width=.8\columnwidth]{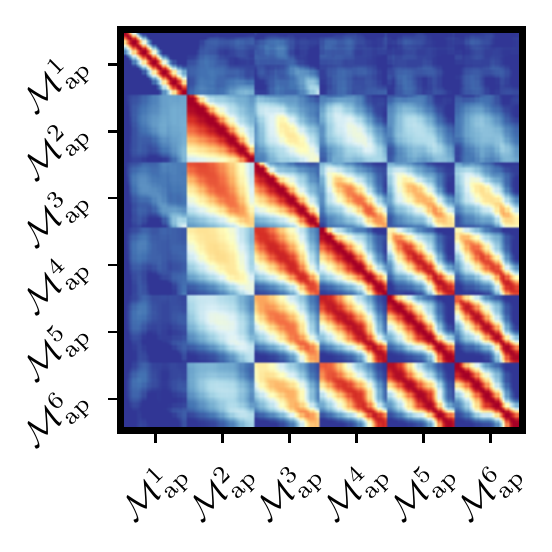}
    \includegraphics[width=1.2\columnwidth]{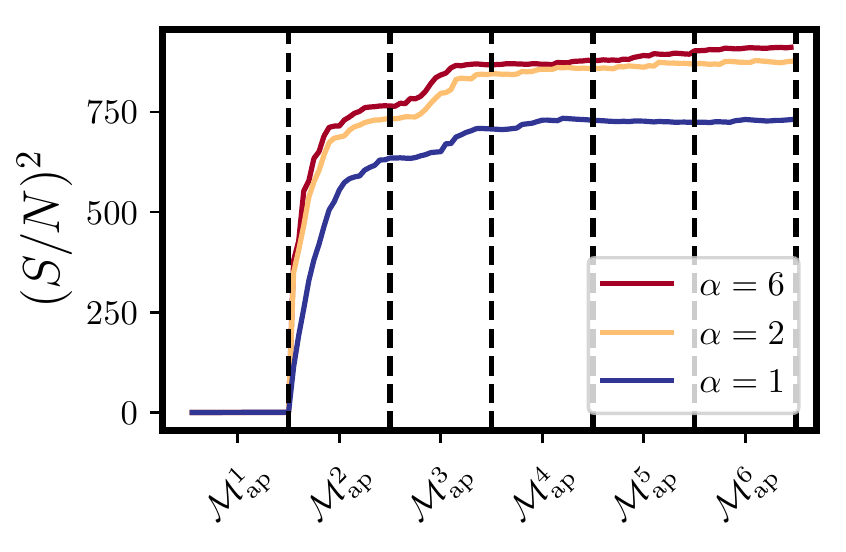}
    }
  \caption{\small{Correlation coefficient matrix (left) and cumulative detection significance (right) for connected moments of the aperture mass statistics. We take into account aperture sizes between $10'$ and $100'$. In the correlation matrix the lower triangle shows the results without shape noise while the upper part includes this term and serves as the basis for the computation of the detection significance.}}
  \label{fig:SLICSMockDetectionSignificanceEqual}
\end{figure*}
\begin{figure*}
\centering{
    \includegraphics[width=1.\columnwidth]{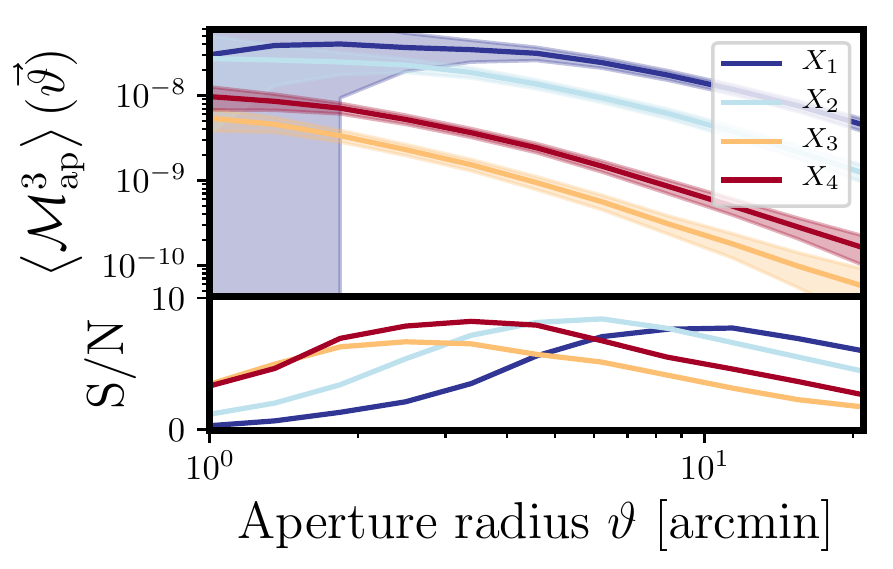}
    \includegraphics[width=1.\columnwidth]{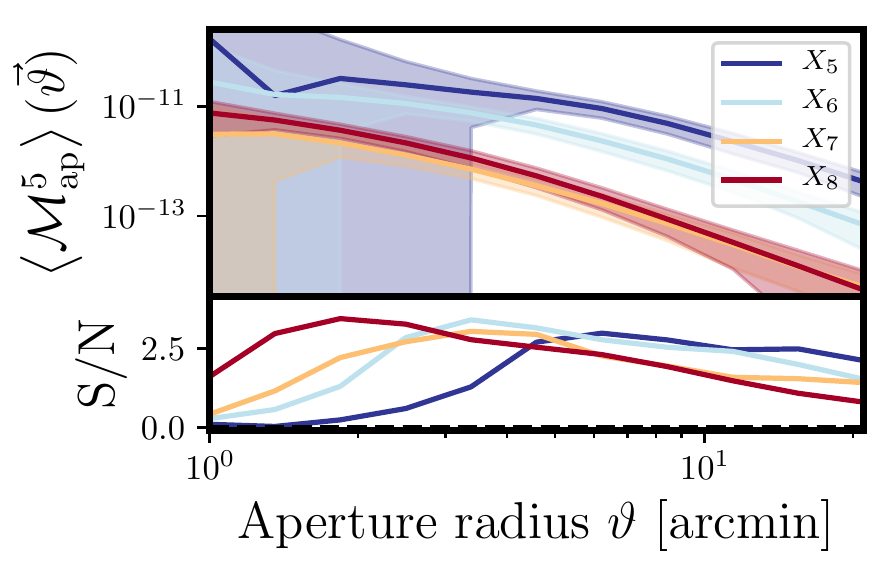}
    }
  \caption{\small{Measurements of the unequal radii aperture mass
      statistics of third (left) and fifth (right) order in the SLICS
      simulation suite. Each line corresponds to a different set of
      relative aperture sizes as given in Table
      \ref{tab:MapVaryConfig}.}}
  \label{fig:SLICSMockUnqualSN}
\end{figure*}
\begin{figure*}
\centering{
    \begin{minipage}[b]{.61\columnwidth}
    \includegraphics[width=\linewidth]{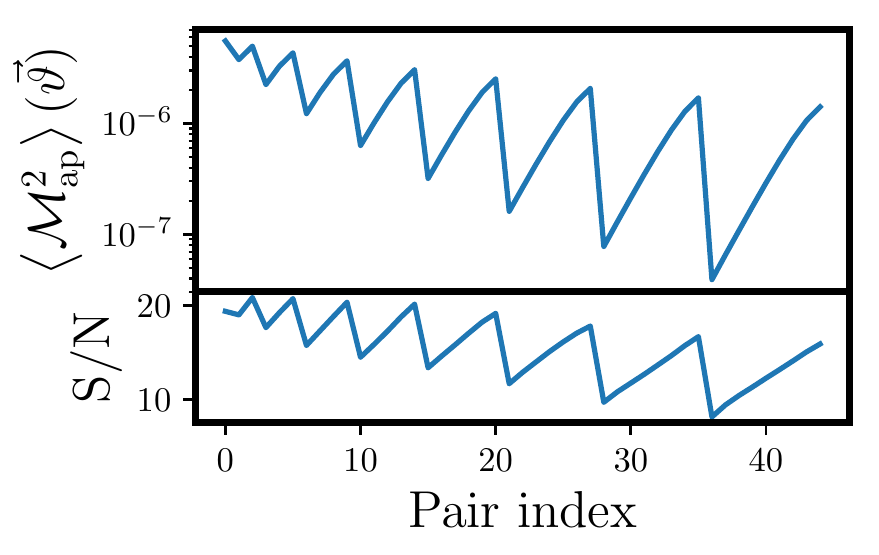}\\
    \includegraphics[width=\linewidth]{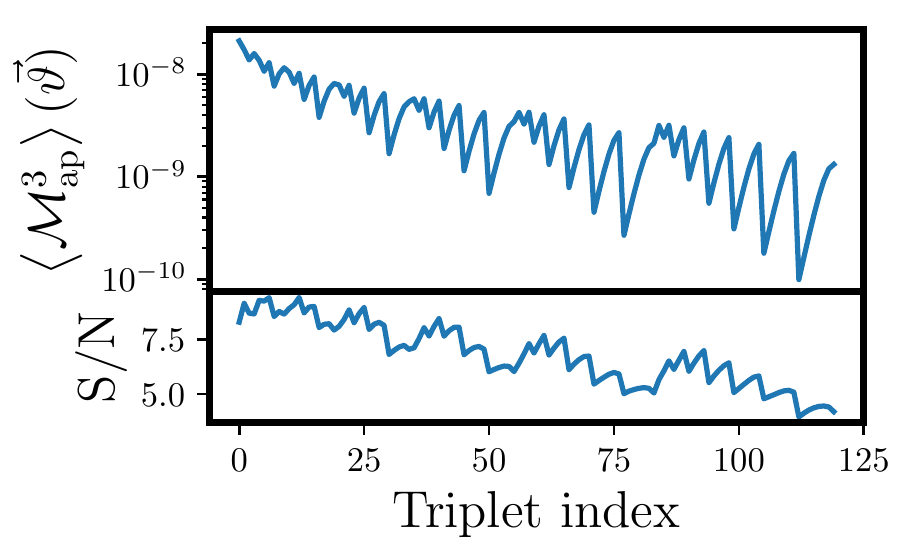}\\
    \includegraphics[width=\linewidth]{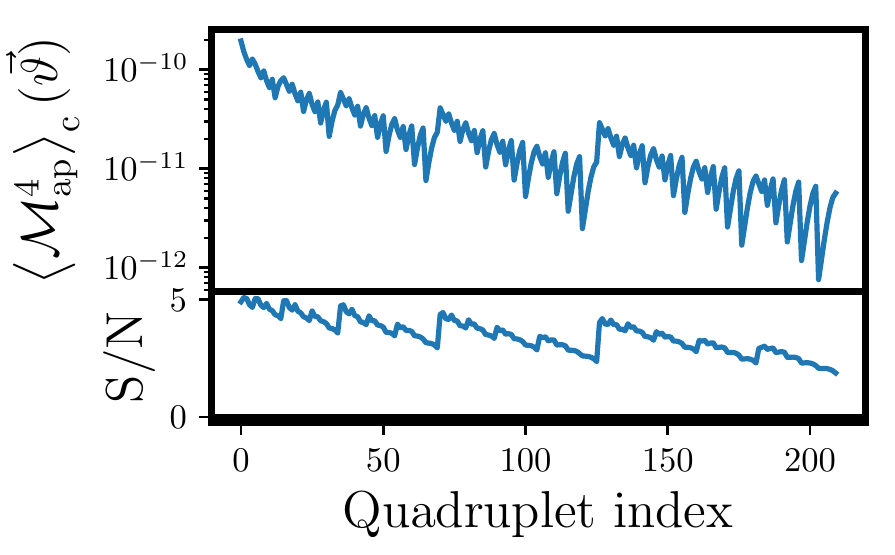}
    \end{minipage}
    \includegraphics[width=1.39\columnwidth]{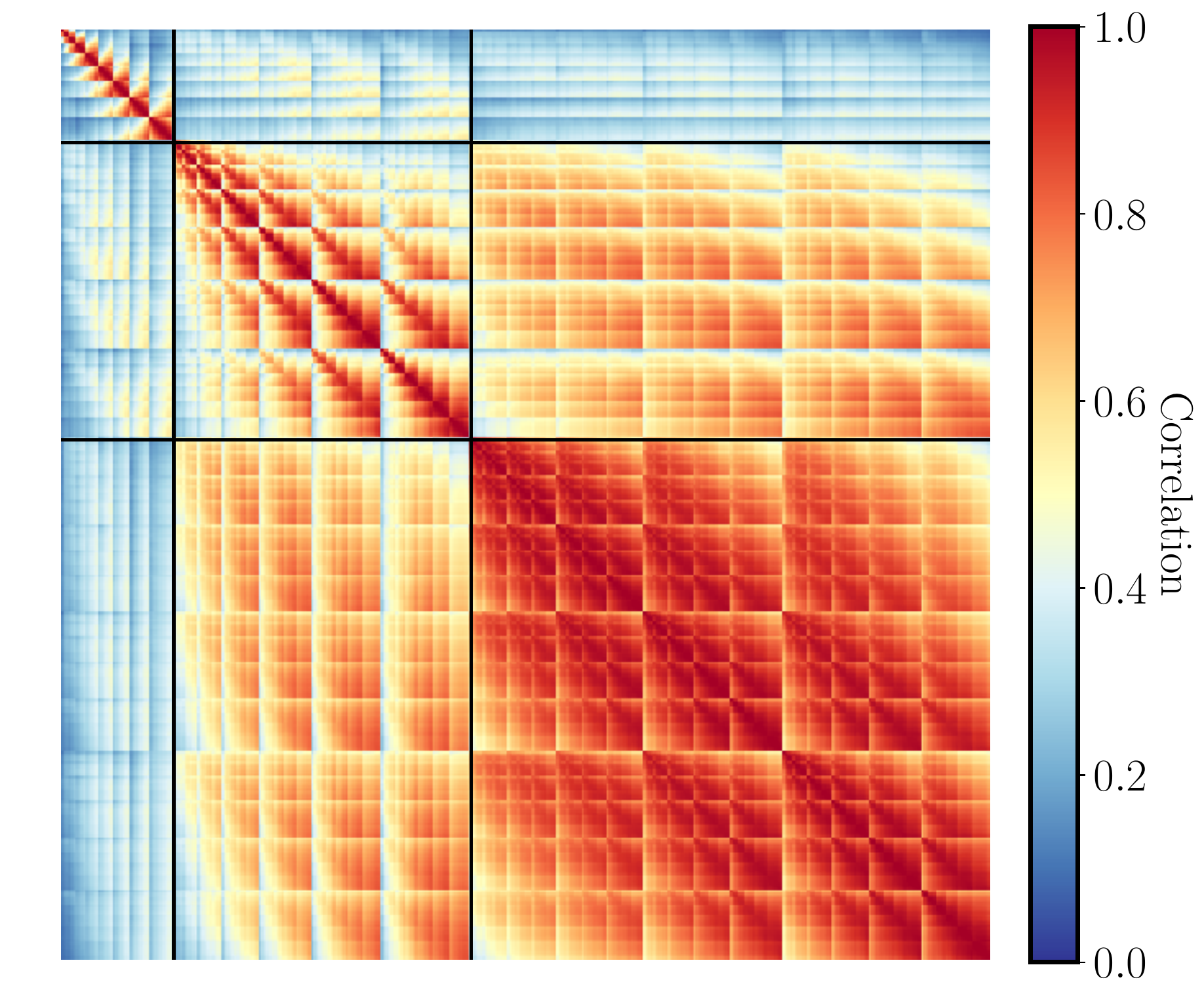}
    }
  \caption{\small{Measurements (left) and correlation matrix (right) of the multiscale aperture mass statistics of second, third and fourth order. For each of those statistics we compute all configurations for ten logarithmically spaced radii between $5'$ and $50'$ in which all the apertures have unequal radii. The black lines indicate the blocks of the (cross-) correlations of different orders.}}
  \label{fig:SLICSMockMultiscaleCorrcoef}
\end{figure*}

\begin{figure}
\centering{
    \includegraphics[width=.8\columnwidth]{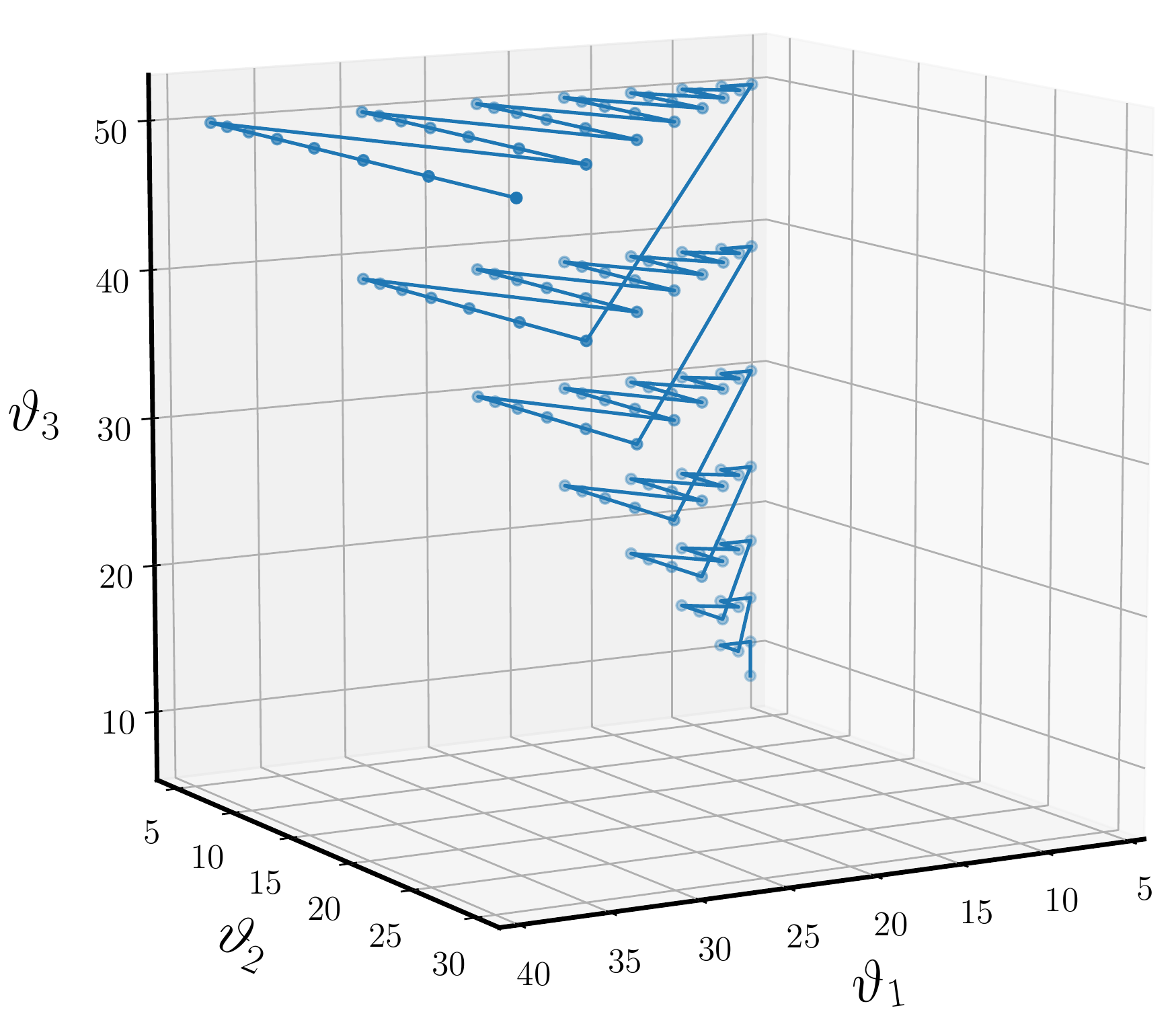}
    }
  \caption{\small{Path in which the set of non redundant aperture scales for the third order statistics is traversed. The starting point is the lower corner. The subpath in the $\vartheta_3=50'$ plane corresponds to the full path taken for the second order statistics.}}
  \label{fig:Indices3D}
\end{figure}
%%%%%%%%%%%%%%%%%%%%%%%%%%%%%%%%%%%%%%%%%%%%%%%%%%%%%%%
%%%%%%%%%%%%%%%%%%%%%%%%%%%%%%%%%%%%%%%%%%%%%%%%%%%%%%%

\subsection{The SLICS mocks}

In order to answer this question we make use of the
SLICS\footnote{\url{https://slics.roe.ac.uk/}} mocks -- this is a
large suite of lensing mock catalogues generated from a large set of
cosmological $N$-body simulations \citep[for full details
  see][]{Harnois-Derapsetal2018}. Each SLICS mock corresponds to a
survey area of $100 \deg^2$. These are generated from the past light
cone extracted from fully independent gravity-only $N$-body
simulations, which evolve $N=1532^3$ particles within a comoving box
of length $L=505\,\Mpc$. The lensing maps are constructed using the
Born approximation. We adopt the catalogues for which the galaxies are
randomly distributed within the lightcone according to the KiDS-450
source distribution \citep{Hildebrandtetal2017}. The shape noise has
been set to $\sigma_{\epsilon} = 0.29$ per shear component. In order
to mimic a constraining power that is comparable to the KiDS-1000 data while not being too noisy, we rescale the errorbars by a factor of $\sqrt{10}$. This provides us
with effectively $819$ simulated $1000({\rm deg})^2$ surveys with which to perform
our analysis.

When estimating the aperture mass statistics from the SLICS
mocks using the estimator given by \Eqn{eq:MapStatEns}, the
achievable signal-to-noise ratio will depend on the number of sampled
apertures selected. If too few are chosen then our estimate will be
inefficient, on the other hand due to the fact that there are
aperture-to-aperture correlations choosing too many will capture all
of the available information, but ultimately will be computationally
inefficient. We therefore expect that the information will saturate for
a given oversampling rate, and that to sample at a higher rate would
be of little use. To investigate this we proceed as in
\citet{Porthetal2020} and place apertures on a regular grid with
spacing $\Delta$, corresponding to an aperture oversampling rate of
$\alpha \equiv \min\left(\{\vartheta_i\}\right) /2\Delta$.

%%%%%%%%%%%%%%%%%%%%%%%%%%%%%%%%%%%%%%%%%%%%%%%%%%%%%%%

\subsection{Measurement in the SLICS mocks}

\Figure{fig:SLICSMockEqualSN} shows the detection significance of the
equal radii aperture mass statistics for the second, third, fourth and
fifth order aperture mass statistics as a function of the aperture
scale and for various choices of the oversampling rate. For the second
order statistics we also plot the theoretical prediction of the
aperture mass dispersion evaluated from \Eqn{eq:MapnKappa}, where
the convergence power spectrum was computed with
CCL\footnote{\url{https://github.com/LSSTDESC/CCL}}
\citep{Chisarietal2019} using {\tt Halofit} \citep{Smithetal2003}, but
with the modifications of \citet{Takahashietal2012}, as the matter
power spectrum. While part of the difference between the curves for
small aperture radii could be attributed to uncertainties in {\tt
  Halofit}, our suspicion is that they mostly stem from the limited
particle mass resolution in the SLICS mocks (see Fig. 6 in
\cite{Harnois-Derapsetal2018} for the resulting suppression of the
shear correlation functions for small separations).

Several important points are worth noting from these measurements.
First, we see that for a KiDS-1000 like survey there is sufficient
fidelity to detect the aperture mass statistics up to fourth order\footnote{We find the cumulative detection significance of the fifth order statistics to be at the $2.9\sigma$ level.},
with the signal-to-noise peaking at an aperture size of around
$\vartheta\approx10'$ for all statistics. This is exciting, as this
has never before been achieved with standard correlation function
based estimators, and if correct would represent the first robust
detection of these statistics using these methods. Second, while for
the case of the two-point statistics the signal-to-noise ratio (shown
in lower sub-panels for each plot) falls off slowly for larger
apertures, this ratio approaches zero faster for the connected parts
of the higher order statistics\footnote{Owing to the fact that the
aperture mass has zero mean, the full and connected moments differ
only for even order moments of four or more.}. Third, while an
aperture oversampling rate of $\alpha\approx2$ seems sufficient to
capture all the signal for second order statistics, it misses some
information for subsequent orders where it becomes necessary to use
$\alpha \gtrsim 4$.

\Figure{fig:SLICSMockDetectionSignificanceEqual} displays the correlation structure of the aperture mass cumulants as well as the cumulative detection significance. We only consider measurements with $\vartheta \geq 10'$ as this is where the SLICS mocks do agree reasonably well with the theoretical predictions and and due to the fact that the robust theoretical modelling of those statistics might reach its limits at around those scales. We see that while for shape noise free ellipticity catalogs there are strong correlations for small aperture radii, this is not the case for the realistic mocks in which those scales are still shape noise dominated. We further note large correlations around the diagonal between different orders, where the degree of correlation increases with the order of the cumulants. For the cumulative detection significance we see that that the cumulants beyond third order do not add a substantial amplitude to the cumulative signal-to-noise. This is expected, given the relatively lower signal-to-noise as well as the larger portion of cross-covariances that need to be taken into account. One should note that this type of analysis does not imply that the higher order cumulants are obsolete as they still may add complementary information by breaking cosmological parameter degeneracies.

%%%%%%%%%%%%%%%%%%%%%%%%%%%%%%%%%%%%%%%%%%%%%%%%%%%%%%%

\subsection{Multiscale aperture mass measurements}

We now shift to the measurement of the multiscale statistics for which there are a number of ways on how to select various aperture scale multiplets. In \Figure{fig:SLICSMockUnqualSN} we focus on a fixed set of aperture propositions and then simply scale them with a single parameter $\vartheta$. The different configurations $\vec{a}$ of aperture radii that we have employed are shown in Table~\ref{tab:MapVaryConfig}. We see that for both, the third and the fifth order moments there does not appear to be a strong decline in detection significance for multiscale apertures compared to the associated moments, even if the relative spread of radii is large. 

Another way to select aperture scale multiplets for a statistic of order $n$ is choose a list of $m \geq n$ aperture scales and to compute the statistics for each choice of $n$ elements within that list. For our purposes we choose the subset in which none of the aperture radii are equal, as this speeds up our calculation, see \App{app:ApplicationToMap} for the details. In the left hand side of \Fig{fig:SLICSMockMultiscaleCorrcoef} we show our measurements for the second, third and fourth order connected cumulants of the multiscale aperture mass statistic using ten logarithmically spaced scales between $5'$ and $50'$. The first index of the multiplet corresponds to the selection of the smallest possible aperture scales from which we then start choosing the next lowest radius in the subsequent dimension up until we reach the combination of the largest possible set of aperture radii - for an example of this path for the third order statistics see \Fig{fig:Indices3D}. Recalling that the second order aperture mass statistic is simply a filtered version of the power spectrum, we should not expect the multiscale extension add any information to that order\footnote{For a different form of the $Q$ filter function like the one proposed in \cite{Crittendenetal2001} one can easily work this out analytically, see .i.e. \cite{Schneideretal2005a}.}. For the three statistics we again find a detection significance that is comparable to the equal scale case, meaning that we can extract substantial signal from convergence spectra configurations which are not corresponding to regular polygons. In the right hand side of  \Fig{fig:SLICSMockMultiscaleCorrcoef} we plot the joint correlation coefficient of the multiscale cumulants. On the investigated range of scales we only find a slight to modest correlation between the higher order multiscale statistics and the second order one. It also appears that the higher order cumulants exhibit a stronger auto- and cross correlation. However, this is (at least partially) an artefact of the range and sampling density of the chosen radii.   

%%%%%%%%%%%%%%%%%%%%%%%%%%%%%%%%%%%%%%%%%%%%%%%%%%%%%%%

\begin{table}
    \centering
    \begin{tabular}[h]{|c c|c c|}
    \hline
    \multicolumn{2}{|c|}{Third order} &  \multicolumn{2}{c|}{Fifth order} \\ 
    Label  & Configuration $\vec{a}$ & Label & Configuration $\vec{a}$ \\ \hline
    $X_1$ & $(1, 1, 1)$ & $X_5$ & $(1, 1, 1, 1, 1)$ \\ 
    $X_2$ & $(1, 2, 2)$ & $X_6$  & $(1, 1, 2, 2, 2)$ \\ 
    $X_3$ & $(1, 5, 5)$ & $X_7$  & $(1, 1, 5, 5, 5)$ \\  
    $X_4$ & $(1, 3, 5)$ & $X_8$ & $(1, 2, 3, 4, 5)$ \\
    \hline
    \end{tabular}
    \caption{Cofigurations of the aperture radii displayed in \Fig{fig:SLICSMockUnqualSN}}
    \label{tab:MapVaryConfig}
\end{table}
%
%%%%%%%%%%%%%%%%%%%%%%%%%%%%%%%%%%%%%%%%%%%%%%%%%%%%%%%

\section{Conclusions and Discussion}\label{sec:Conclusions}

In this paper we have explored an alternative method for estimating
the aperture mass statistics in weak lensing cosmic shear
surveys. This study extended our previous work \citep{Porthetal2020}
in a number of ways: First, we generalized the direct estimator
approach to higher statistics, and showed how to rewrite the standard
estimator as a product of linear order time sums.  Second, we provided
the details of the computation of the variance of these estimators.
Third, we further generalised the aperture mass statistics to include
the multiscale approach. Again, we showed how one could estimate
these using linear order products of power sums. The work can be
summarised as follows:

In \Sect{sec:Theory} we reviewed the background theory of cosmological
weak lensing and showed how the connected cumulants of the aperture
mass statistics are related to the convergence polyspectra.

In \Sect{sec:Estimators} we introduced the direct estimator for
moments of the aperture mass statistics. We then gave expressions for
how the nested sums can be decomposed into a linear combination of
products of (multivariate) power sums that facilitates a linearly
scaling estimation procedure in the number of galaxies within an
aperture. We then generalized this estimator to an ensemble of
overlapping apertures and computed its variance. We argued that the
aperture cross correlation coefficient leads to a substantial
correction to the naive $1/N$ scaling if the apertures are not well
separated, and that it also can be used to assess the degree of
aperture oversampling that is necessary to capture most of the
available information. Finally, we gave a detailed explanation of the
algorithms used for our implementation.

In \Sect{sec:Measurements} we successfully validated our method on
Gaussian mock simulations and furthermore verified the linear scaling. 

In \Sect{sec:DetectionSignificance} we turned to the SLICS simulation
suite and assessed the signal-to-noise of the statistics for a $1000$
degree survey following a KiDS-450 like $n(z)$ distribution
function. We found that with these specifications significant
detections of up to fourth order can be expected for the equal and
unequal radii cumulants and that an aperture oversampling rate of at
least four extracts nearly all the signal.

In this paper we have neglected the impact of survey masks on the
measurement process and the possible bias that this could induce, the
exploration of this is sufficient to warrant its own publication and
this is the subject of our associated publication (Porth et al. in
prep.).  Throughout this paper we were mainly concerned with making
the extraction of information from higher order statistics of galaxy
shape catalogs computationally feasible and accurate. However, we
remained agnostic about further challenges that need to be addressed
before applying our methods to real data. For example, one should
investigate the required PSF modelling, shape measurement and shear
bias calibration quality to not introduce substantial biases in the
measurement. Additionally, the range of measurements that can
ultimately be used for obtaining cosmological parameter constraints
will be limited to the scales for which one can theoretically
accurately model those higher order statistics.

%%%%%%%%%%%%%%%%%%%%%%%%%%%%%%%%%%%%%%%%%%%%%%%%%%%%%%%

\section*{Acknowledgements}
We would like to thank Patrick Simon, Laura Marian, Stefan Hilbert, Peter Schneider, Cora Uhlemann and Gary Bernstein for useful discussions, as well as the anonymous referee for helpful comments. We would like to thank Joachim
Harnois-Deraps for making public the SLICS mock data, which can be
found at http://slics.roe.ac.uk/. 
LP acknowledges support from a STFC Research Training Grant (grant
number ST/R505146/1). RES acknowledges support from the STFC (grant
number ST/P000525/1, ST/T000473/1). This work used the DiRAC@Durham
facility managed by the Institute for Computational Cosmology on
behalf of the STFC DiRAC HPC Facility (www.dirac.ac.uk). The equipment
was funded by BEIS capital funding via STFC capital grants
ST/K00042X/1, ST/P002293/1, ST/R002371/1 and ST/S002502/1, Durham
University and STFC operations grant ST/R000832/1. DiRAC is part of
the National eInfrastructure. This research made use of numpy, a
library used for scientific computing and technical computing and
matplotlib, a Python library for publication quality graphics
\citep{Harris2020, Hunter2007}.

%%%%%%%%%%%%%%%%%%%%%%%%%%%%%%%%%%%%%%%%%%%%%%%%%%

\section*{Data Availability}
The SLICS mock catalogs are available at \url{https://slics.roe.ac.uk/}. Additional data underlying this article will be shared on reasonable request to the corresponding author.



\bibliographystyle{mnras}
\bibliography{refs}

\begin{thebibliography}{}
\makeatletter
\relax
\def\mn@urlcharsother{\let\do\@makeother \do\$\do\&\do\#\do\^\do\_\do\%\do\~}
\def\mn@doi{\begingroup\mn@urlcharsother \@ifnextchar [ {\mn@doi@}
  {\mn@doi@[]}}
\def\mn@doi@[#1]#2{\def\@tempa{#1}\ifx\@tempa\@empty \href
  {http://dx.doi.org/#2} {doi:#2}\else \href {http://dx.doi.org/#2} {#1}\fi
  \endgroup}
\def\mn@eprint#1#2{\mn@eprint@#1:#2::\@nil}
\def\mn@eprint@arXiv#1{\href {http://arxiv.org/abs/#1} {{\tt arXiv:#1}}}
\def\mn@eprint@dblp#1{\href {http://dblp.uni-trier.de/rec/bibtex/#1.xml}
  {dblp:#1}}
\def\mn@eprint@#1:#2:#3:#4\@nil{\def\@tempa {#1}\def\@tempb {#2}\def\@tempc
  {#3}\ifx \@tempc \@empty \let \@tempc \@tempb \let \@tempb \@tempa \fi \ifx
  \@tempb \@empty \def\@tempb {arXiv}\fi \@ifundefined
  {mn@eprint@\@tempb}{\@tempb:\@tempc}{\expandafter \expandafter \csname
  mn@eprint@\@tempb\endcsname \expandafter{\@tempc}}}

\bibitem[\protect\citeauthoryear{{Aihara} et~al.,}{{Aihara}
  et~al.}{2018}]{HSC2018}
{Aihara} H.,  et~al., 2018, \mn@doi [\pasj] {10.1093/pasj/psx066}, \href
  {https://ui.adsabs.harvard.edu/abs/2018PASJ...70S...4A} {70, S4}

\bibitem[\protect\citeauthoryear{{Asgari} et~al.,}{{Asgari}
  et~al.}{2021}]{Asgarietal2021}
{Asgari} M.,  et~al., 2021, \mn@doi [Astronomy & Astrophysics]
  {10.1051/0004-6361/202039070}, 645, A104

\bibitem[\protect\citeauthoryear{{Bacon}, {Refregier}  \& {Ellis}}{{Bacon}
  et~al.}{2000}]{Baconetal2000}
{Bacon} D.~J.,  {Refregier} A.~R.,   {Ellis} R.~S.,  2000, \mn@doi [\mnras]
  {10.1046/j.1365-8711.2000.03851.x}, \href
  {https://ui.adsabs.harvard.edu/abs/2000MNRAS.318..625B} {318, 625}

\bibitem[\protect\citeauthoryear{{Bartelmann} \& {Schneider}}{{Bartelmann} \&
  {Schneider}}{2001}]{BartelmannSchneider2001}
{Bartelmann} M.,  {Schneider} P.,  2001, \mn@doi [\physrep]
  {10.1016/S0370-1573(00)00082-X}, \href
  {http://esoads.eso.org/abs/2001PhR...340..291B} {340, 291}

\bibitem[\protect\citeauthoryear{{Barthelemy}, {Codis}  \&
  {Bernardeau}}{{Barthelemy} et~al.}{2020}]{Barthelemyetal2020}
{Barthelemy} A.,  {Codis} S.,   {Bernardeau} F.,  2020, Probability
  distribution function of the aperture mass field with large deviation theory
  (\mn@eprint {arXiv} {2012.03831})

\bibitem[\protect\citeauthoryear{{Bernardeau} \& {Valageas}}{{Bernardeau} \&
  {Valageas}}{2000}]{Bernardeauetal2000}
{Bernardeau} F.,  {Valageas} P.,  2000, \aap, \href
  {https://ui.adsabs.harvard.edu/abs/2000A&A...364....1B} {364, 1}

\bibitem[\protect\citeauthoryear{{Bernardeau}, {Colombi}, {Gazta{\~n}aga}  \&
  {Scoccimarro}}{{Bernardeau} et~al.}{2002}]{Bernardeauetal2002}
{Bernardeau} F.,  {Colombi} S.,  {Gazta{\~n}aga} E.,   {Scoccimarro} R.,  2002,
  \physrep, \href {http://esoads.eso.org/abs/2002PhR...367....1B} {367, 1}

\bibitem[\protect\citeauthoryear{{Blandford}, {Saust}, {Brainerd}  \&
  {Villumsen}}{{Blandford} et~al.}{1991}]{Blandfordetal1991}
{Blandford} R.~D.,  {Saust} A.~B.,  {Brainerd} T.~G.,   {Villumsen} J.~V.,
  1991, \mn@doi [\mnras] {10.1093/mnras/251.4.600}, \href
  {https://ui.adsabs.harvard.edu/abs/1991MNRAS.251..600B} {251, 600}

\bibitem[\protect\citeauthoryear{{Byun}, {Eggemeier}, {Regan}, {Seery}  \&
  {Smith}}{{Byun} et~al.}{2017}]{Byunetal2017}
{Byun} J.,  {Eggemeier} A.,  {Regan} D.,  {Seery} D.,   {Smith} R.~E.,  2017,
  \mn@doi [\mnras] {10.1093/mnras/stx1681}, \href
  {https://ui.adsabs.harvard.edu/abs/2017MNRAS.471.1581B} {471, 1581}

\bibitem[\protect\citeauthoryear{{Chisari} et~al.,}{{Chisari}
  et~al.}{2019}]{Chisarietal2019}
{Chisari} N.~E.,  et~al., 2019, \mn@doi [\apjs] {10.3847/1538-4365/ab1658},
  \href {https://ui.adsabs.harvard.edu/abs/2019ApJS..242....2C} {242, 2}

\bibitem[\protect\citeauthoryear{{Comtet}}{{Comtet}}{1974}]{Comtet1974}
{Comtet} L.,  1974, {Advanced Combinatorics: The Art of Finite and Infinite
  Expansions}.
{Springer Netherlands}

\bibitem[\protect\citeauthoryear{{Crittenden}, {Natarajan}, {Pen}  \&
  {Theuns}}{{Crittenden} et~al.}{2001}]{Crittendenetal2001}
{Crittenden} R.~G.,  {Natarajan} P.,  {Pen} U.-L.,   {Theuns} T.,  2001,
  \mn@doi [\apj] {10.1086/322370}, \href
  {https://ui.adsabs.harvard.edu/abs/2001ApJ...559..552C} {559, 552}

\bibitem[\protect\citeauthoryear{{Dodelson}}{{Dodelson}}{2003}]{Dodelson2003}
{Dodelson} S.,  2003, {Modern cosmology}.
Academic Press, San Diego, CA, \url {https://cds.cern.ch/record/1282338}

\bibitem[\protect\citeauthoryear{{Dodelson}}{{Dodelson}}{2017}]{Dodelson2017}
{Dodelson} S.,  2017, Gravitational Lensing.
Cambridge University Press, \mn@doi{10.1017/9781316424254}

\bibitem[\protect\citeauthoryear{{Fabbian}, {Calabrese}  \&
  {Carbone}}{{Fabbian} et~al.}{2018}]{Fabbianetal2018}
{Fabbian} G.,  {Calabrese} M.,   {Carbone} C.,  2018, \mn@doi [\jcap]
  {10.1088/1475-7516/2018/02/050}, \href
  {https://ui.adsabs.harvard.edu/abs/2018JCAP...02..050F} {2018, 050}

\bibitem[\protect\citeauthoryear{{Friedrich}, {Seitz}, {Eifler}  \&
  {Gruen}}{{Friedrich} et~al.}{2016}]{Friedrichetal2016}
{Friedrich} O.,  {Seitz} S.,  {Eifler} T.~F.,   {Gruen} D.,  2016, \mn@doi
  [\mnras] {10.1093/mnras/stv2833}, \href
  {https://ui.adsabs.harvard.edu/abs/2016MNRAS.456.2662F} {456, 2662}

\bibitem[\protect\citeauthoryear{{Fu} et~al.,}{{Fu} et~al.}{2014}]{Fuetal2014}
{Fu} L.,  et~al., 2014, \mn@doi [\mnras] {10.1093/mnras/stu754}, \href
  {https://ui.adsabs.harvard.edu/abs/2014MNRAS.441.2725F} {441, 2725}

\bibitem[\protect\citeauthoryear{{Harnois-D{\'e}raps}
  et~al.,}{{Harnois-D{\'e}raps} et~al.}{2018}]{Harnois-Derapsetal2018}
{Harnois-D{\'e}raps} J.,  et~al., 2018, \mn@doi [\mnras]
  {10.1093/mnras/sty2319}, \href
  {https://ui.adsabs.harvard.edu/abs/2018MNRAS.481.1337H} {481, 1337}

\bibitem[\protect\citeauthoryear{{Harris} et~al.,}{{Harris}
  et~al.}{2020}]{Harris2020}
{Harris} C.~R.,  et~al., 2020, \mn@doi [Nature] {10.1038/s41586-020-2649-2},
  585, 357

\bibitem[\protect\citeauthoryear{{Heydenreich}, {Brück}  \&
  {Harnois-Déraps}}{{Heydenreich} et~al.}{2020}]{Heydenreichetal2020}
{Heydenreich} S.,  {Brück} B.,   {Harnois-Déraps} J.,  2020, Persistent
  homology in cosmic shear: constraining parameters with topological data
  analysis (\mn@eprint {arXiv} {2007.13724})

\bibitem[\protect\citeauthoryear{{Hikage} et~al.,}{{Hikage}
  et~al.}{2019}]{Hikageetal2019}
{Hikage} C.,  et~al., 2019, \mn@doi [\pasj] {10.1093/pasj/psz010}, \href
  {https://ui.adsabs.harvard.edu/abs/2019PASJ...71...43H} {71, 43}

\bibitem[\protect\citeauthoryear{{Hilbert}, {Hartlap}, {White}  \&
  {Schneider}}{{Hilbert} et~al.}{2009}]{Hilbertetal2009}
{Hilbert} S.,  {Hartlap} J.,  {White} S.~D.~M.,   {Schneider} P.,  2009,
  \mn@doi [\aap] {10.1051/0004-6361/200811054}, \href
  {http://adsabs.harvard.edu/abs/2009A%26A...499...31H} {499, 31}

\bibitem[\protect\citeauthoryear{{Hilbert}, {Marian}, {Smith}  \&
  {Desjacques}}{{Hilbert} et~al.}{2012}]{Hilbertetal2012}
{Hilbert} S.,  {Marian} L.,  {Smith} R.~E.,   {Desjacques} V.,  2012, \mn@doi
  [\mnras] {10.1111/j.1365-2966.2012.21841.x}, \href
  {https://ui.adsabs.harvard.edu/abs/2012MNRAS.426.2870H} {426, 2870}

\bibitem[\protect\citeauthoryear{{Hildebrandt} et~al.,}{{Hildebrandt}
  et~al.}{2017}]{Hildebrandtetal2017}
{Hildebrandt} H.,  et~al., 2017, \mn@doi [\mnras] {10.1093/mnras/stw2805},
  \href {https://ui.adsabs.harvard.edu/abs/2017MNRAS.465.1454H} {465, 1454}

\bibitem[\protect\citeauthoryear{{Hunter}}{{Hunter}}{2007}]{Hunter2007}
{Hunter} J.~D.,  2007, \mn@doi [Computing in Science \& Engineering]
  {10.1109/MCSE.2007.55}, 9, 90

\bibitem[\protect\citeauthoryear{{Jain} \& {Seljak}}{{Jain} \&
  {Seljak}}{1997}]{JainSeljak1997}
{Jain} B.,  {Seljak} U.,  1997, \mn@doi [\apj] {10.1086/304372}, \href
  {https://ui.adsabs.harvard.edu/abs/1997ApJ...484..560J} {484, 560}

\bibitem[\protect\citeauthoryear{{Jarvis}, {Bernstein}, {Fischer}, {Smith},
  {Jain}, {Tyson}  \& {Wittman}}{{Jarvis} et~al.}{2003}]{Jarvisetal2003}
{Jarvis} M.,  {Bernstein} G.~M.,  {Fischer} P.,  {Smith} D.,  {Jain} B.,
  {Tyson} J.~A.,   {Wittman} D.,  2003, \mn@doi [\aj] {10.1086/367799}, \href
  {http://adsabs.harvard.edu/abs/2003AJ....125.1014J} {125, 1014}

\bibitem[\protect\citeauthoryear{{Jarvis}, {Bernstein}  \& {Jain}}{{Jarvis}
  et~al.}{2004}]{Jarvisetal2004}
{Jarvis} M.,  {Bernstein} G.,   {Jain} B.,  2004, \mn@doi [\mnras]
  {10.1111/j.1365-2966.2004.07926.x}, \href
  {https://ui.adsabs.harvard.edu/abs/2004MNRAS.352..338J} {352, 338}

\bibitem[\protect\citeauthoryear{{Kacprzak} et~al.,}{{Kacprzak}
  et~al.}{2016}]{Kacprzaketal2016}
{Kacprzak} T.,  et~al., 2016, \mn@doi [\mnras] {10.1093/mnras/stw2070}, \href
  {https://ui.adsabs.harvard.edu/abs/2016MNRAS.463.3653K} {463, 3653}

\bibitem[\protect\citeauthoryear{{Kaiser}}{{Kaiser}}{1995}]{Kaiser1995}
{Kaiser} N.,  1995, \mn@doi [\apjl] {10.1086/187730}, \href
  {https://ui.adsabs.harvard.edu/abs/1995ApJ...439L...1K} {439, L1}

\bibitem[\protect\citeauthoryear{{Kaiser}}{{Kaiser}}{1998}]{Kaiser1998}
{Kaiser} N.,  1998, \mn@doi [\apj] {10.1086/305515}, \href
  {https://ui.adsabs.harvard.edu/abs/1998ApJ...498...26K} {498, 26}

\bibitem[\protect\citeauthoryear{{Kaiser} \& {Squires}}{{Kaiser} \&
  {Squires}}{1993}]{Kaiseretal1993}
{Kaiser} N.,  {Squires} G.,  1993, \mn@doi [\apj] {10.1086/172297}, \href
  {https://ui.adsabs.harvard.edu/abs/1993ApJ...404..441K} {404, 441}

\bibitem[\protect\citeauthoryear{{Kaiser}, {Wilson}  \& {Luppino}}{{Kaiser}
  et~al.}{2000}]{Kaiseretal2000}
{Kaiser} N.,  {Wilson} G.,   {Luppino} G.~A.,  2000, arXiv e-prints, \href
  {https://ui.adsabs.harvard.edu/abs/2000astro.ph..3338K} {pp
  astro--ph/0003338}

\bibitem[\protect\citeauthoryear{{Kayo}, {Takada}  \& {Jain}}{{Kayo}
  et~al.}{2013}]{Kayoetal2013}
{Kayo} I.,  {Takada} M.,   {Jain} B.,  2013, \mn@doi [\mnras]
  {10.1093/mnras/sts340}, \href
  {https://ui.adsabs.harvard.edu/abs/2013MNRAS.429..344K} {429, 344}

\bibitem[\protect\citeauthoryear{{Kilbinger}}{{Kilbinger}}{2015}]{Kilbinger2015}
{Kilbinger} M.,  2015, \mn@doi [Reports on Progress in Physics]
  {10.1088/0034-4885/78/8/086901}, 78, 086901

\bibitem[\protect\citeauthoryear{{Kilbinger} \& {Schneider}}{{Kilbinger} \&
  {Schneider}}{2005}]{KilbingerSchneider2005}
{Kilbinger} M.,  {Schneider} P.,  2005, \mn@doi [\aap]
  {10.1051/0004-6361:20053531}, \href
  {https://ui.adsabs.harvard.edu/abs/2005A&A...442...69K} {442, 69}

\bibitem[\protect\citeauthoryear{{Knuth}}{{Knuth}}{2005}]{Knuth2005_D}
{Knuth} D.~E.,  2005, The Art of Computer Programming, Volume 4, Fascicle 3:
  Generating All Combinations and Partitions.
Addison-Wesley Professional

\bibitem[\protect\citeauthoryear{{LSST}}{{LSST}}{2009}]{LSST2009}
{LSST} 2009, preprint, \href {http://esoads.eso.org/abs/2009arXiv0912.0201L} {}
  (\mn@eprint {arXiv} {0912.0201})

\bibitem[\protect\citeauthoryear{{Laureijs} et~al.,}{{Laureijs}
  et~al.}{2011}]{Euclid2011}
{Laureijs} R.,  et~al., 2011, preprint, \href
  {http://adsabs.harvard.edu/abs/2011arXiv1110.3193L} {} (\mn@eprint {arXiv}
  {1110.3193})

\bibitem[\protect\citeauthoryear{{Mandelbaum}}{{Mandelbaum}}{2018}]{Mandelbaum2018}
{Mandelbaum} R.,  2018, \mn@doi [Annual Review of Astronomy and Astrophysics]
  {10.1146/annurev-astro-081817-051928}, 56, 393

\bibitem[\protect\citeauthoryear{{Marian}, {Smith}, {Hilbert}  \&
  {Schneider}}{{Marian} et~al.}{2012}]{Marianetal2012}
{Marian} L.,  {Smith} R.~E.,  {Hilbert} S.,   {Schneider} P.,  2012, \mn@doi
  [\mnras] {10.1111/j.1365-2966.2012.20992.x}, 423, 1711

\bibitem[\protect\citeauthoryear{{Marian}, {Smith}, {Hilbert}  \&
  {Schneider}}{{Marian} et~al.}{2013}]{Marianetal2013}
{Marian} L.,  {Smith} R.~E.,  {Hilbert} S.,   {Schneider} P.,  2013, \mn@doi
  [\mnras] {10.1093/mnras/stt552}, \href
  {https://ui.adsabs.harvard.edu/abs/2013MNRAS.432.1338M} {432, 1338}

\bibitem[\protect\citeauthoryear{{Martinet}, {Harnois-Déraps}, {Jullo}  \&
  {Schneider}}{{Martinet} et~al.}{2021}]{Martinetetal2021}
{Martinet} N.,  {Harnois-Déraps} J.,  {Jullo} E.,   {Schneider} P.,  2021,
  Probing dark energy with tomographic weak-lensing aperture mass statistics
  (\mn@eprint {arXiv} {2010.07376})

\bibitem[\protect\citeauthoryear{{Massey} et~al.,}{{Massey}
  et~al.}{2013}]{Masseyetal2013}
{Massey} R.,  et~al., 2013, \mn@doi [\mnras] {10.1093/mnras/sts371}, \href
  {https://ui.adsabs.harvard.edu/abs/2013MNRAS.429..661M} {429, 661}

\bibitem[\protect\citeauthoryear{{Miralda-Escude}}{{Miralda-Escude}}{1991}]{Miralda-Escude1991}
{Miralda-Escude} J.,  1991, \mn@doi [\apj] {10.1086/170555}, \href
  {http://adsabs.harvard.edu/abs/1991ApJ...380....1M} {380, 1}

\bibitem[\protect\citeauthoryear{{Munshi} \& {Coles}}{{Munshi} \&
  {Coles}}{2003}]{Munshietal2003}
{Munshi} D.,  {Coles} P.,  2003, \mn@doi [Monthly Notices of the Royal
  Astronomical Society] {10.1046/j.1365-8711.2003.06136.x}, 338, 846

\bibitem[\protect\citeauthoryear{{Munshi} \& {Valageas}}{{Munshi} \&
  {Valageas}}{2005}]{Munshietal2005}
{Munshi} D.,  {Valageas} P.,  2005, \mn@doi [Monthly Notices of the Royal
  Astronomical Society] {10.1111/j.1365-2966.2004.08462.x}, 356, 439

\bibitem[\protect\citeauthoryear{{Munshi}, {Valageas}  \& {Barber}}{{Munshi}
  et~al.}{2004}]{Munshietal2004}
{Munshi} D.,  {Valageas} P.,   {Barber} A.~J.,  2004, \mn@doi [Monthly Notices
  of the Royal Astronomical Society] {10.1111/j.1365-2966.2004.07553.x}, 350,
  77

\bibitem[\protect\citeauthoryear{{Pires} et~al.,}{{Pires}
  et~al.}{2020}]{Piresetal2020}
{Pires} S.,  et~al., 2020, \mn@doi [\aap] {10.1051/0004-6361/201936865}, \href
  {https://ui.adsabs.harvard.edu/abs/2020A&A...638A.141P} {638, A141}

\bibitem[\protect\citeauthoryear{{Porth}, {Smith}, {Simon}, {Marian}  \&
  {Hilbert}}{{Porth} et~al.}{2020}]{Porthetal2020}
{Porth} L.,  {Smith} R.~E.,  {Simon} P.,  {Marian} L.,   {Hilbert} S.,  2020,
  \mn@doi [Monthly Notices of the Royal Astronomical Society]
  {10.1093/mnras/staa2900}, 499, 2474

\bibitem[\protect\citeauthoryear{{Pratten} \& {Lewis}}{{Pratten} \&
  {Lewis}}{2016}]{PrattenLewis2016}
{Pratten} G.,  {Lewis} A.,  2016, \mn@doi [\jcap]
  {10.1088/1475-7516/2016/08/047}, \href
  {https://ui.adsabs.harvard.edu/abs/2016JCAP...08..047P} {2016, 047}

\bibitem[\protect\citeauthoryear{{Sato}, {Takada}, {Hamana}  \&
  {Matsubara}}{{Sato} et~al.}{2011}]{Satoetal2011}
{Sato} M.,  {Takada} M.,  {Hamana} T.,   {Matsubara} T.,  2011, \mn@doi [\apj]
  {10.1088/0004-637X/734/2/76}, \href
  {https://ui.adsabs.harvard.edu/abs/2011ApJ...734...76S} {734, 76}

\bibitem[\protect\citeauthoryear{{Schneider}}{{Schneider}}{1996}]{Schneider1996}
{Schneider} P.,  1996, \mn@doi [\mnras] {10.1093/mnras/283.3.837}, \href
  {https://ui.adsabs.harvard.edu/abs/1996MNRAS.283..837S} {283, 837}

\bibitem[\protect\citeauthoryear{{Schneider}}{{Schneider}}{1998}]{Schneider1998}
{Schneider} P.,  1998, \mn@doi [\apj] {10.1086/305559}, \href
  {https://ui.adsabs.harvard.edu/abs/1998ApJ...498...43S} {498, 43}

\bibitem[\protect\citeauthoryear{{Schneider}}{{Schneider}}{2006a}]{Schneider2006p1}
{Schneider} P.,  2006a, in {Meylan} G.,  {Jetzer} P.,  {North} P.,  {Schneider}
  P.,  {Kochanek} C.~S.,   {Wambsganss} J.,  eds, Saas-Fee Advanced Course 33:
  Gravitational Lensing: Strong, Weak and Micro. pp 1--89

\bibitem[\protect\citeauthoryear{{Schneider}}{{Schneider}}{2006b}]{Schneider2006p3}
{Schneider} P.,  2006b, in {Meylan} G.,  {Jetzer} P.,  {North} P.,  {Schneider}
  P.,  {Kochanek} C.~S.,   {Wambsganss} J.,  eds, Saas-Fee Advanced Course 33:
  Gravitational Lensing: Strong, Weak and Micro. pp 269--451

\bibitem[\protect\citeauthoryear{{Schneider} \& {Kilbinger}}{{Schneider} \&
  {Kilbinger}}{2007}]{Schneideretal2007}
{Schneider} P.,  {Kilbinger} M.,  2007, \mn@doi [\aap]
  {10.1051/0004-6361:20065532}, \href
  {https://ui.adsabs.harvard.edu/abs/2007A&A...462..841S} {462, 841}

\bibitem[\protect\citeauthoryear{{Schneider} \& {Lombardi}}{{Schneider} \&
  {Lombardi}}{2003}]{SchneiderLombardi2003}
{Schneider} P.,  {Lombardi} M.,  2003, \mn@doi [\aap]
  {10.1051/0004-6361:20021541}, \href
  {https://ui.adsabs.harvard.edu/abs/2003A&A...397..809S} {397, 809}

\bibitem[\protect\citeauthoryear{{Schneider}, {van Waerbeke}, {Jain}  \&
  {Kruse}}{{Schneider} et~al.}{1998}]{Schneideretal1998}
{Schneider} P.,  {van Waerbeke} L.,  {Jain} B.,   {Kruse} G.,  1998, \mn@doi
  [\mnras] {10.1046/j.1365-8711.1998.01422.x}, \href
  {https://ui.adsabs.harvard.edu/abs/1998MNRAS.296..873S} {296, 873}

\bibitem[\protect\citeauthoryear{{Schneider}, {van Waerbeke}  \&
  {Mellier}}{{Schneider} et~al.}{2002a}]{Schneideretal2002a}
{Schneider} P.,  {van Waerbeke} L.,   {Mellier} Y.,  2002a, \mn@doi [\aap]
  {10.1051/0004-6361:20020626}, \href
  {https://ui.adsabs.harvard.edu/abs/2002A&A...389..729S} {389, 729}

\bibitem[\protect\citeauthoryear{{Schneider}, {van Waerbeke}, {Kilbinger}  \&
  {Mellier}}{{Schneider} et~al.}{2002b}]{Schneideretal2002b}
{Schneider} P.,  {van Waerbeke} L.,  {Kilbinger} M.,   {Mellier} Y.,  2002b,
  \mn@doi [\aap] {10.1051/0004-6361:20021341}, \href
  {https://ui.adsabs.harvard.edu/abs/2002A&A...396....1S} {396, 1}

\bibitem[\protect\citeauthoryear{{Schneider}, {Kilbinger}  \&
  {Lombardi}}{{Schneider} et~al.}{2005}]{Schneideretal2005a}
{Schneider} P.,  {Kilbinger} M.,   {Lombardi} M.,  2005, \mn@doi [\aap]
  {10.1051/0004-6361:20034217}, \href
  {https://ui.adsabs.harvard.edu/abs/2005A&A...431....9S} {431, 9}

\bibitem[\protect\citeauthoryear{{Schneider}, {Eifler}  \&
  {Krause}}{{Schneider} et~al.}{2010}]{Schneideretal2010}
{Schneider} P.,  {Eifler} T.,   {Krause} E.,  2010, \mn@doi [A\&A]
  {10.1051/0004-6361/201014235}, 520, A116

\bibitem[\protect\citeauthoryear{{Scoccimarro} \& {Frieman}}{{Scoccimarro} \&
  {Frieman}}{1996}]{ScoccimarroFrieman1996a}
{Scoccimarro} R.,  {Frieman} J.,  1996, \mn@doi [\apjs] {10.1086/192306}, \href
  {http://esoads.eso.org/abs/1996ApJS..105...37S} {105, 37}

\bibitem[\protect\citeauthoryear{{Seitz}, {Schneider}  \& {Ehlers}}{{Seitz}
  et~al.}{1994}]{Seitzetal1994}
{Seitz} S.,  {Schneider} P.,   {Ehlers} J.,  1994, \mn@doi [Classical and
  Quantum Gravity] {10.1088/0264-9381/11/9/016}, \href
  {https://ui.adsabs.harvard.edu/abs/1994CQGra..11.2345S} {11, 2345}

\bibitem[\protect\citeauthoryear{{Semboloni}, {Hoekstra}, {Schaye}, {van
  Daalen}  \& {McCarthy}}{{Semboloni} et~al.}{2011}]{Sembolonietal2011}
{Semboloni} E.,  {Hoekstra} H.,  {Schaye} J.,  {van Daalen} M.~P.,   {McCarthy}
  I.~G.,  2011, \mn@doi [\mnras] {10.1111/j.1365-2966.2011.19385.x}, \href
  {http://adsabs.harvard.edu/abs/2011MNRAS.417.2020S} {417, 2020}

\bibitem[\protect\citeauthoryear{{Smith} et~al.,}{{Smith}
  et~al.}{2003}]{Smithetal2003}
{Smith} R.~E.,  et~al., 2003, \mn@doi [\mnras]
  {10.1046/j.1365-8711.2003.06503.x}, \href
  {http://esoads.eso.org/abs/2003MNRAS.341.1311S} {341, 1311}

\bibitem[\protect\citeauthoryear{{Szapudi} \& {Szalay}}{{Szapudi} \&
  {Szalay}}{1997}]{Szapudietal1997}
{Szapudi} I.,  {Szalay} A.~S.,  1997, \mn@doi [\apj] {10.1086/310641}, 481, L1

\bibitem[\protect\citeauthoryear{{Takahashi}, {Sato}, {Nishimichi}, {Taruya}
  \& {Oguri}}{{Takahashi} et~al.}{2012}]{Takahashietal2012}
{Takahashi} R.,  {Sato} M.,  {Nishimichi} T.,  {Taruya} A.,   {Oguri} M.,
  2012, \mn@doi [\apj] {10.1088/0004-637X/761/2/152}, \href
  {http://adsabs.harvard.edu/abs/2012ApJ...761..152T} {761, 152}

\bibitem[\protect\citeauthoryear{{Troxel} \& {Ishak}}{{Troxel} \&
  {Ishak}}{2015}]{TroxelIshak2015}
{Troxel} M.~A.,  {Ishak} M.,  2015, \mn@doi [\physrep]
  {10.1016/j.physrep.2014.11.001}, \href
  {https://ui.adsabs.harvard.edu/abs/2015PhR...558....1T} {558, 1}

\bibitem[\protect\citeauthoryear{{Troxel} et~al.,}{{Troxel}
  et~al.}{2018}]{Troxeletal2018}
{Troxel} M.~A.,  et~al., 2018, \mn@doi [\prd] {10.1103/PhysRevD.98.043528},
  \href {https://ui.adsabs.harvard.edu/abs/2018PhRvD..98d3528T} {98, 043528}

\bibitem[\protect\citeauthoryear{{Van Waerbeke} et~al.,}{{Van Waerbeke}
  et~al.}{2000}]{VanWaerbekeetal2000}
{Van Waerbeke} L.,  et~al., 2000, \aap, \href
  {https://ui.adsabs.harvard.edu/abs/2000A&A...358...30V} {358, 30}

\bibitem[\protect\citeauthoryear{{Wittman}, {Tyson}, {Kirkman}, {Dell'Antonio}
  \& {Bernstein}}{{Wittman} et~al.}{2000}]{Wittmanetal2000}
{Wittman} D.~M.,  {Tyson} J.~A.,  {Kirkman} D.,  {Dell'Antonio} I.,
  {Bernstein} G.,  2000, \mn@doi [\nat] {10.1038/35012001}, \href
  {https://ui.adsabs.harvard.edu/abs/2000Natur.405..143W} {405, 143}

\bibitem[\protect\citeauthoryear{{Zhang}, {Liguori}, {Bean}  \&
  {Dodelson}}{{Zhang} et~al.}{2007}]{ZhangPetal2007}
{Zhang} P.,  {Liguori} M.,  {Bean} R.,   {Dodelson} S.,  2007, \mn@doi [\prl]
  {10.1103/PhysRevLett.99.141302}, \href
  {https://ui.adsabs.harvard.edu/abs/2007PhRvL..99n1302Z} {99, 141302}

\makeatother
\end{thebibliography}

%%%%%%%%%%%%%%%%%%%%%%%%%%%%%%%%%%%%%%%%%%%%%%%%%%
%%%%%%%%%%%%%%%%% APPENDICES %%%%%%%%%%%%%%%%%%%%%

\onecolumn

\appendix

%%%%%%%%%%%%%%%%%%%%%%%%%%%%%%%%%%%%%%%%%%%%%%%%%%%%%%%

\section{Derivations of aperture mass skewness and kurtosis estimators}\label{app:estimators1}

\newcommand{\sumss}[2]{\sum_{\substack{{#1}\\{#2}}}}
\newcommand{\nsumss}[2]{\sum_{\substack{({#1})\\{#2}}}}

In the following we will derive the accelerated direct estimators for the third and fourth order aperture mass moments. For properly treating summation indices we add to our notation \eqref{eq:ShorthandSum} the following generalizations that deal with individual indices being set equal with each other:

\begin{align}
    \sumss{i_1,\cdots,i_l\cdots i_{m-1}, i_{m+1},\cdots, i_n}{i_l=i_m}
    &\equiv \sum_{i_1} \cdots \sum_{i_{m-1}} \sum_{i_{m+1}} \cdots \sum_{i_n}
    \\
    \nsumss{i_1,\cdots,i_l\cdots i_{m-1}, i_{m+1} \cdots i_n}{i_l=i_m}
    &\equiv \sum_{i_1} \sum_{i_2 \neq i_1} \cdots \sum_{i_{m-1}\neq\cdots\neq i_1} \sum_{i_{m+1}\neq\cdots\neq i_1} \cdots \sum_{i_n\neq\cdots\neq i_1}
\end{align}
%%%%%%%%%%%%%%%%%%%%%%%%%%%%%%%%%%%%%%%%%%%%%%%%%%%%%%%

\subsection{Derivation of the estimator for \texorpdfstring{$\MapStatEst{3}$}{} }

Let us compute the derivation of the skewness $\MapStatEst{3}$ of the
aperture mass. The standard direct estimator is given by:
\be \MapStatEst{3} = (\pi \vartheta ^2)^3
\frac{\sum_{(i,j,k)}^N w_{i}w_{j}w_{k}Q_{i}Q_{j}Q_{k}e_{t,i}e_{t,j}e_{t,k} }
     {\sum_{(i,j,k)}^N w_{i}w_{j}w_{k} } \ . \label{eq:Map3E1}
\ee
It can be shown using the methods described in
\citet{Schneideretal1998} and \citet{Porthetal2020} that this leads to
an unbiased estimator of the skewness. We can rewrite the above
estimator by noting that an unconstrained triple sum can be
decomposed into the following partial sums:
\begin{align} \sum_{i,j,k}^{N} &= 
\sum_{(i,j,k)}^N + \nsumss{i,j}{i=k}^{N} + \nsumss{i,j}{j=k}^{N} + \nsumss{i,k}{i=j}^{N} + \sum_{i=j=k}^N
\end{align}
This can be rearranged to give:
\begin{align} \sum_{(i,j,k)}^{N} & =
  \sum_{i,j,k}^N - \nsumss{i,j}{i=k}^{N} - \nsumss{i,j}{j=k}^{N} - \nsumss{i,k}{i=j}^{N} - \sum_{i=j=k}^N \label{eq:Map3E2}
\end{align}
Similarly, the unconstrained double sum can be decomposed and
rearranged in the following manner:
\be
\sum_{i,j} = \sum_{(i,j)}+\sum_{i=j} \ \ \Rightarrow \ \  \sum_{(i,j)} = \sum_{i,j} - \sum_{i=j}\ .
\label{eq:Map3E3}\ee     
Using this result repeatedly in \Eqn{eq:Map3E2} allows us to rewrite the constrained sums as unconstrained sums:
\begin{align}
  \sum_{(i,j,k)}^{N}  & =
  \sum_{i,j,k}^{N}
  - \left(\sumss{i,j}{i=k}^{N}-\sum_{i=j=k}^{N}\right)
  - \left(\sumss{i,j}{j=k}^{N}-\sum_{i=j=k}^{N}\right)
  - \left(\sumss{i,k}{i=j}^{N}-\sum_{i=j=k}^{N}\right) 
  -\sum_{i=j=k}^{N} \nn \\
  &  =
  \sum_{i,j,k}^{N}
  - \sumss{i,j}{i=k}^{N}
  - \sumss{i,j}{j=k}^{N}
  - \sumss{i,k}{i=j}^{N}
  +2\sum_{i=j=k}^{N}\ = \sum_{i,j,k}^{N}\left[1-\delta^{K}_{j,k}-\delta^{K}_{k,i}-\delta^{K}_{i,j}
    +2\delta^{K}_{i,j}\delta^{K}_{i,k}\right]\ .
    \label{eq:Map3E4}
\end{align}
Hence, on repeatedly using this result we can rewrite the sum in the
numerator and denominator of \Eqn{eq:Map3E1} to give us an alternate
form for the skewness as:
\begin{align} \MapStatEst{3} = (\pi \vartheta ^2)^3
    \frac{\left[
    \sum_{i,j,k}^{N} w_{i}w_{j}w_{k}Q_{i}Q_{j}Q_{k}e_{t,i}e_{t,j}e_{t,k}
    -3\sum_{i,j}^{N} w_{i}w_{j}^2Q_{i}Q_{j}^2e_{t,i}e_{t,j}^2
    +2\sum_{i}^{N} w_{i}^3Q^3_{i}e^3_{t,i} \right]}
  {\left[\sum_{i,j,k}^{N} w_{i}w_{j}w_{k}-3\sum_{i,j}^{N} w_{i}w_{j}^2+2\sum_{i}^{N} w_{i}^3\right]}\ .
\end{align}
If we now divide through each term by $(\sum_i^N w_i)^3$ and recall
expressions \Eqns{eq:MapnEstPowerSumsA}{eq:MapnEstPowerSumsB} we see that
our estimator becomes:
\begin{align}
  \MapStatEst{3} &= \frac{\MsEst{1}^{3} - 3\MsEst{2}\MsEst{1} + 2\MsEst{3}}{1-3\SEst{2}+2\SEst{3}} \ .
\end{align}
%

%%%%%%%%%%%%%%%%%%%%%%%%%%%%%%%%%%%%%%%%%%%%%%%%%%%%%%%

\subsection{Derivation of the estimator for \texorpdfstring{$\MapStatEst{4}$}{}}

The standard direct estimator for the kurtosis of aperture mass is given by:
\be \MapStatEst{4} = (\pi \vartheta ^2)^4
\frac{\sum_{(i,j,k,l)}^N w_{i}w_{j}w_{k}w_{l}Q_{i}Q_{j}Q_{k}Q_{l}e_{t,i}e_{t,j}e_{t,k} e_{t,l} }
     {\sum_{(i,j,k,l)}^Nw_{i}w_{j}w_{k}w_{l} } \ . \label{eq:Map4E1}
\ee
We follow similar steps to the derivation of the skewness and note that
the unconstrained quadruple sum can be written:
\begin{align} \sum_{i,j,k,l}^{N} = & \sum_{(i,j,k,l)}^{N}
  + \left[\nsumss{i,j,k}{i=l}^{N}+ \ 5 \ {\rm perms}\right]
  + \left[\nsumss{i,j}{i=k,j=l}^{N}+\nsumss{i,j}{i=l,j=k}^{N}+\nsumss{i,k}{i=j,k=l}^{N}\right]
  + \left[\nsumss{i,j}{j=k=l}^N +  \nsumss{i,l}{i=j=k}^N + \nsumss{i,k}{i=j=l}^N + \nsumss{i,j}{i=k=l}^N\right]
  + \sum_{i=j=k=l}^{N} \label{eq:Map4E2}\ ,
\end{align}
which on rearranging leads us to:
\begin{align} \sum_{(i,j,k,l)}^{N}  = &
\sum_{i,j,k,l}^{N}
  - \left[\nsumss{i,j,k}{i=l}^{N}+ \ 5 \ {\rm perms}\right]
  - \left[\nsumss{i,j}{i=k,j=l}^{N}+\nsumss{i,j}{i=l,j=k}^{N}+\nsumss{i,k}{i=j,k=l}^{N}\right]
  - \left[\nsumss{i,j}{j=k=l}^N +  \nsumss{i,l}{i=j=k}^N + \nsumss{i,k}{i=j=l}^N + \nsumss{i,j}{i=k=l}^N\right]
  - \sum_{i=j=k=l}^{N} \label{eq:Map4E3} \ .
\end{align}
We now make use of our previous results to rewrite the constrained
sums on the right-hand side of the expression as unconstrained sums:
\begin{align} \sum_{(i,j,k,l)}^{N}
  & = \sum_{i,j,k,l}^{N} 
  - \left\{\left[\sumss{i,j,k}{i=l}^{N}
  - \sumss{i,j}{i=k=l}^{N}
  - \sumss{i,j}{j=k,i=l}^{N}
  - \sumss{i,k}{i=j=l}^{N}
  + \ 2\sum_{i=j=k=l}^{N} \right] + \ 5 \ {\rm perms}\right\}
    - \left\{
    \left[\sumss{i,j}{i=k,j=l}^{N}-\sum_{i=j= k=l}^{N}\right]
  + \left[\sumss{i,j}{i=l,j=k}^{N}-\sum_{i=j=k=l}^{N}\right]\right.\nn \\
  &+ \left.\left[\sumss{i,k}{i=j,k=l}^{N}-\sum_{i=j=k=l}^{N}\right]\right\}
  - \left\{
    \left[\sumss{i,j}{j=k=l}^{N}-\sum_{i=j=k=l}^{N}\right]
   +\left[\sumss{i,l}{i=j=k}^{N}-\sum_{i=j=k=l}^{N}\right]
   +\left[\sumss{i,k}{i=j=l}^{N}-\sum_{i=j=k=l}^{N}\right]
   +\left[\sumss{i,j}{i=k=l}^{N}-\sum_{i=j=k=l}^{N}\right]\right\}
  - \sum_{i=j=k=l}^{N} \label{eq:Map4E4}\ .
\end{align}
On making repeated use of the Kroneker delta symbol this can now be
compactly written as:
\begin{align} \sum_{(i,j,k,l) }^{N}
  & = \sum_{i,j,k,l}^{N} \Bigg[1
  - \left\{\left[\delta^K_{k,l}
    -\delta^K_{j,k}\delta^K_{k,l} 
    -\delta^K_{i,k}\delta^K_{k,l}
    -\delta^K_{i,j}\delta^K_{k,l}
    +2\delta^K_{i,j}\delta^K_{i,k}\delta^K_{i,l} \right] +5 \ {\rm perms}\right\}
  - \Big\{
  \left[\delta^K_{i,j}\delta^K_{k,l}-\delta^K_{i,j}\delta^K_{i,k}\delta^K_{i,l} \right]
  \nn \\
  &
  + \left[\delta^K_{i,k}\delta^K_{j,l}-\delta^K_{i,j}\delta^K_{i,k}\delta^K_{i,l}\right]
  + \left[\delta^K_{i,l}\delta^K_{j,k}-\delta^K_{i,j}\delta^K_{i,k}\delta^K_{i,l}\right]\Big\}
    - \Big\{
    \left[\delta^K_{i,j}\delta^K_{i,k}-\delta^K_{i,j}\delta^K_{i,k}\delta^K_{i,l}\right]
    \nn \\
  &
   +\left[\delta^K_{i,j}\delta^K_{i,l}-\delta^K_{i,j}\delta^K_{i,k}\delta^K_{i,l}\right]
   +\left[\delta^K_{i,k}\delta^K_{i,l}-\delta^K_{i,j}\delta^K_{i,k}\delta^K_{i,l}\right]
   +\left[\delta^K_{j,k}\delta^K_{j,l}-\delta^K_{i,j}\delta^K_{i,k}\delta^K_{i,l}\right]\Big\}
   - \delta^K_{i,j}\delta^K_{i,k}\delta^K_{i,l}\Bigg]
  \label{eq:Map4E5}\ .
\end{align}
On collecting, cancelling and grouping like terms we see that this can be written:
\begin{align} \sum_{(i,j,k,l)}^{N}
  & = \sum_{i,j,k,l}^{N} \Bigg[1
    - \left(
    \delta^K_{i,j}+\delta^K_{i,k}+\delta^K_{i,l}+\delta^K_{j,k}+\delta^K_{j,l}+\delta^K_{k,l}\right)
    +\left\{\delta^K_{i,j}\delta^K_{k,l}+\delta^K_{i,k}\delta^K_{j,l}+\delta^K_{i,l}\delta^K_{j,k}\right\}\nn \\
    &
    +\left\{
    \delta^K_{i,j}\delta^K_{i,k}+\delta^K_{i,j}\delta^K_{i,l}+\delta^K_{i,k}\delta^K_{i,l}
    +\delta^K_{j,i}\delta^K_{j,k}+\delta^K_{j,i}\delta^K_{j,l}+\delta^K_{j,k}\delta^K_{j,l}
    +\delta^K_{k,i}\delta^K_{k,l}+\delta^K_{k,j}\delta^K_{k,l}\right\}
    -6\delta^K_{i,j}\delta^K_{i,k}\delta^K_{i,l} \Bigg]
  \label{eq:Map4E7}\ .
\end{align}
Hence, on making repeated use of \Eqn{eq:Map4E7} in \Eqn{eq:Map4E1}
and along with \Eqns{eq:MapnEstPowerSumsA}{eq:MapnEstPowerSumsB},
the estimator for kurtosis of aperture mass becomes:
\begin{align}
  \MapStatEst{4} &= \frac{\MsEst{1}^{4} - 6\MsEst{2}\MsEst{1}^2 + 3\MsEst{2}^2+8\MsEst{3}\MsEst{1}-6\MsEst{4}}
             {1-6\SEst{2}+3\SEst{2}^2+8\SEst{3}-6\SEst{4}} \ .
\end{align}
     
%%%%%%%%%%%%%%%%%%%%%%%%%%%%%%%%%%%%%%%%%%%%%%%%%%%%%%%

\section{A proof of the general theorem for arbitrary order aperture mass statistics}
\label{app:estimators2}

%%%%%%%%%%%%%%%%%%%%%%%%%%%%%%%%%%%%%%%%%%%%%%%%%%%%%%%

In this section we provide a derivation of the the general form of the
$n$-point aperture mass statistic estimator given by
\Eqn{eq:MapnEstBell}. At the time of writing, we are not aware that
the combinatoric methods that we have used in the derivation of the
general expression have been used before in the cosmological context,
and therefore provide a brief overview of them -- in particular the
Bell polynomials. In what follows we will try to not rely on advanced mathematical
methods, but instead use a basic framework to explain
how the Bell polynomials are linked to set partitions, and finally
how they are connected to the aperture mass estimators.

%%%%%%%%%%%%%%%%%%%%%%%%%%%%%%%%%%%%%%%%%%%%%%%%%%%%%%%

\subsection{Set partitions and Bell polynomials}

We begin by defining a partition $\pi$ of a set ${\bf n} = \{1, 2,
\cdots, n \}$ as a collection of mutually exclusive subsets (blocks)
of ${\bf n}$ whose union equals ${\bf n}$. In our case all these
partitions can be mapped onto an associated partition $\lambda$ being
defined as the number of elements of each block in $\pi$. Each element
$\lambda$ can be represented as $(n_1, n_2, \cdots, n_m)$ or as
$\left( 1^{m_1}, \ 2^{m_2}, \ \cdots , n^{m_n} \right)$ where for the
former expression the $n_i$ denote the length of the $i$th block while
for the latter case the $m_i$ represent the number of occurrences of a
block of length $i$ in $\pi$. If $\pi$ is a partition of ${\bf n}$
having $m$ blocks this implies that $\sum_i m_i = m$ and $\sum_i i
\ m_i = n$. We will now show that the following proposition holds:

\vspace{0.2cm}
\noindent
\textbf{Proposition}: \\For the set ${\bf n}$ and a partition $\lambda$ of
length $m$ given as $\left( 1^{m_1}, \ 2^{m_2}, \ \cdots ,
\ell^{m_\ell} \right)$ there are $\frac{n!}{\prod_{i=1}^\ell m_i!
  (i!)^{m_i}}$ partitions $\pi$ of ${\bf n}$ having the same
$\lambda(\pi)$.  \\
\textbf{Proof}:\\ As a first step we just look at the number of ways
the $m$ subsets can be chosen from $\bm n$. This can easily be worked
out when noting that for the first subset there are $\binom{n}{n_1}$
choices, for the following $\binom{n-n_1}{n_2}$ etc. Following through
all of the subsets we then have
\be
\binom{n}{n_1}\binom{n-n_1}{n_2}\cdots\binom{n-n_1 - \cdots n_{m-2}}{n_{m-1}} \binom{n_m}{n_m}
=
\frac{n!}{n_1!(n-n_1)!}\frac{(n-n_1)!}{n_2!(n-n_1-n_2)!} \cdots
\frac{(n-n_1 - \cdots n_{m-2})!}{n_{m-1}!n_m!} \frac{n_m!}{n_m!}
=
\frac{n!}{n_1!n_2!\cdots n_m!}
\ee
possibilities. Shifting this expression to the representation of
$\lambda$ given above we see that many of them give the identical
partition $\pi$; to get rid of those ones we need to divide by the
number of ways all the equal size blocks themselves can be permuted
with each other. Applying those conditions we have
\be
\frac{1}{\text{Norm.}} \times \frac{n!}{n_1!n_2!\cdots n_m!}
=
\frac{1}{m_1! m_2! \cdots m_\ell !} \times \frac{n!}{(1!)^{m_1} (2!)^{m_2} \cdots (\ell!)^{m_\ell}}
=
\frac{n!}{\prod_{i=1}^\ell m_i! (i!)^{m_i}}
\ee
possibilities remaining, which is exactly the proposed expression.

With this result in hand we are now in position to understand the form
of the partial Bell polynomial $B_{n,m}$ being defined as
\begin{align}\label{eq:PartialBell}
    B_{n,m}(x_1 \cdots, x_{n-m+1}) 
    &=
    \sum_{(m_1, \cdots, m_{n-m+1}) \in P_{n,m}} \frac{n!}{m_1! \cdots m_{n-m+1}!}
    \left(\frac{x_1}{1!}\right)^{m_1} \cdots \left(\frac{x_{n-m+1}}{(n-m+1)!}\right)^{m_{n-m+1}}
    \ \ , 
\end{align}
where
\begin{align}
  P_{n,m} 
    &\equiv
    \left\{ (m_1, \cdots m_{n-m+1}) \ \in \ \mathbb{N}_0^{n-m+1} \ | \
    \sum_{i=1}^{n-m+1} m_i = m \ \ , \ \  \sum_{i=1}^{n-m+1} i \ m_i = n \right\} \ . \nonumber
\end{align}
Comparing the prefactors and the index set\footnote{The upper limit is
given by the partition having the largest possible block size, namely
$(1^{m-1}, \ 2^0, \cdots, \ (n-(m-1))^1)$} with our discussion above
we see that the partial Bell polynomials simply sum over all the
partitions $\lambda$ of ${\bf n}$ having a fixed $m$, i.e. they list
the number of ways a set consisting the $n$ objects can be partitioned
into $m$ blocks. For example, looking at $B_{4,2}$ the allowed index
combinations are $\{ (0,2,0), (1,0,1) \}$ such that
\Eqn{eq:PartialBell} evaluates to $B_{4,2} = 4x_1x_3 + 3x_2^2$. We
note in the passing that these expressions generate the same
prefactors that arise in the halo model, i.e. we can relate the
structure of $B_{4,2}$ to the two-halo term of the halo model
trispectrum.

Finally, we define the complete Bell polynomial $B_n$ which list all
possible partitions of $n$ objects:
\begin{align}
  B_n(x_1 \cdots, x_{n-m+1}) &
  = \sum_{m=1}^n  B_{n,m}(x_1 \cdots, x_{n-m+1})
  = \sum_{m=1}^n \sum_{\pi \in P_{n,m}} \prod_{i=1}^{n-m(\lambda(\pi))+1} x_i^{m_i(\lambda (\pi))} \ , \label{eq:CompleteBell}
\end{align}
where the first equality states the formal definition and the second
one rewrites it into an explicit sum over all the partitions of the
set $\bm n$.
%
%%%%%%%%%%%%%%%%%%%%%%%%%%%%%%%%%%%%%%%%%%%%%%%%%%%%%%%

\subsection{Sums over unequal indices and Bell polynomials}

Let us look at the simple expression $\sum_{i=1}^N \sum_{j\neq i}^N
x_i x_j$. A naive implementation of this double sum would imply a
quadratic complexity of the corresponding program. A much faster way
resulting in linear complexity can be achieved when noting that
$\left( \sum_{i_1=1}^N x_{i_1} \right) \left( \sum_{i_2=1}^N x_{i_2}
\right) = \sum_{i_1=1}^N \sum_{i_2\neq i_1}^N x_{i_1} x_{i_2} +
\sum_{i_1=1}^N x_1^2 \ $.
We can easily generalize this pattern by treating the number of
indices as the set ${\bf n}$ from the previous subsection. Then all
the different partitions $\lambda$ of this set correspond to different
ways these indices can be set equal with one another; the
corresponding prefactors can be obtained via the Bell polynomial. To
clarify this statement we write down as an example the expression for
$n=4$:
\begin{align}\label{eq:helperAppA1_1}
    \left( \sum_{i} x_{i} \right)^4 
    &=
    \sum_{i_1\neq i_2 \neq i_3 \neq i_4 } x_{i_1}x_{i_2}x_{i_3}x_{i_4}
    \ + \
    \left( \sum_{i_1 \neq i_3 \neq i_4 } x_{i_1}^2x_{i_3}x_{i_4} + 5 \text{perm.} \right)
    \ + \ 
    \left( \sum_{i_1 \neq i_2 } x_{i_1}^3x_{i_2} + 3 \text{perm.} \right)
    \ + \ 
    \left( \sum_{i_1 \neq i_3 } x_{i_1}^2x_{i_3}^2 + 2 \text{perm.} \right)
    \ + \ 
    \sum_{i_1} x_{i_1}^4
    \\ &\sim
    (1^4 , \ 2^0 \ , \ 3^0 \ , \ 4^0 ) 
    \ \ + \ \ 6 \times (1^2 \ , \ 2^1 \ , \ 3^0 \ , \ 4^0 ) 
    \ \ + \ \  4 \times (1^1 \ , \ 2^0 \ , \ 3^1 \ , \ 4^0) 
    \ \ + \ \ 3 \times (1^0 \ , \ 2^2 \ , \ 3^0 \ , \ 4^0)
    \ \ + \ \  (1^0\ , \ 2^0 \ , \ 3^0 \ , \ 4^1) \nonumber
\end{align}
From here we see that we can express a sum over $n$ unequal indices in
terms of two power sums and a set of related sums over at most $n-1$
unequal indices. Repeating the same argument on the latter sums one
eventually arrives at an expression only involving power
sums. Carrying out aforementioned calculations along the lines of \App{app:estimators1} for our example this
yields
\begin{align}\label{eq:helperAppA1_2}
    \sum_{i_1\neq i_2 \neq i_3 \neq i_4 } x_{i_1}x_{i_2}x_{i_3}x_{i_4}
    =
    \left(\sum_{i} x_{i}\right)^4
    - 6 \left(\sum_{i} x_i\right)^2 \left( \sum_i x_i^2\right)
    + \ 8 \left(\sum_{i} x_i\right) \left( \sum_i x_i^3\right)
    + \ 3 \left(\sum_{i} x_i^2\right)^2
    - \ 6  \left(\sum_{i} x_i^4\right) \ \ .
\end{align}
Comparing the latter two expressions we note that their index
partitions are the same, but that they differ in some signs and
prefactors; namely there is a negative sign for an odd partition
length $m$ and an additional multiplicative factor of $(i-1)!$ for
each block of length $i$. Looking at the structure of
\Eqn{eq:helperAppA1_2}, i.e. the fact that all of its summands
correspond to a partition of an integer set and that furthermore it
constitutes of $n$ different building blocks we might be tempted to
cast it in terms of Bell polynomials with the identifying the $x_\ell$
from \Eqn{eq:CompleteBell} with the power sums $ c_\ell \sum_i
x_i^\ell \ \ : \ \ c_\ell \in \mathbb{R}$.  In the next paragraphs we
formalize these observations and from there determine the $c_\ell$.

The first difference can be motivated most easily by choosing a
graphical representation in which we draw each index as a single
point. Then the prefactors in \Eqn{eq:helperAppA1_1} are given by
the number of ways one can group together different points such that
they constitute the corresponding partition whereas for
\Eqn{eq:helperAppA1_2} it additionally matters in which order these
points have been set equal with each other, which in mathematical
terms is described by how many closed cycles one can draw between
them. The induced correction of $(\ell-1)!$ for a block of length
$\ell$ can be absorbed in the Bell polynomial by setting $c_\ell =
(\ell-1)!$.

The second observation can be generalized inductively. Looking at our
example of $n=4$ we see that the sign for each partition $\lambda$ is
given by $\text{sgn}(\lambda)$ = $\prod_{i=1}^n (-1)^{m_i(\lambda)
  ((i+1) \text{mod} 2)} $, that is each block of even length
contributes a negative sign. Performing the induction step we have
\be
\sum_{i_1 \neq \cdots i_n \neq i_{n+1} } x_{i_1}\cdots x_{i_{n+1}}
=
\left(\sum_{i_{n+1}} x_{i_{n+1}}\right) \left(\sum_{i_1 \neq \cdots i_3 \neq i_{n} } x_{i_1}\cdots x_{i_{n}}\right)
-
\left[ \left(\sum_{i_1 \neq \cdots i_3 \neq i_{n}} x_{i_1}^2 x_{i_2}\cdots x_{i_{n}}\right) + (n-1) \ \text{perm.} \right] \ .
\ee
Looking at the modification of the partitions, for the first term we
have $m_1 \rightarrow m_1+1$ for all $\lambda$ such that we would not
have expected any sign flips. For the second term, we need to update
the block in which the identical index sits, assuming it had length
$k$ we have $m_k \rightarrow m_k-1$ and $m_{k+1} \rightarrow
m_{k+1}+1$. In case of an even $k$ reducing its occurrence by one
induces an additional sign flip whereas for odd $k$s we get a sign
flip for the increase of $m_{k+1}$. Putting things together we
conclude that we could predict the correct signs by examining the
partition structures. Therefore, setting $c_\ell = (-1)^{(\ell+1)
  \ \text{mod} \ 2} (\ell-1)!$ in \Eqn{eq:CompleteBell} will
reproduce generalizations of $\Eqn{eq:helperAppA1_2}$. We can brush
this in a nicer shape by setting $c_\ell = -(\ell-1)!$ and furthermore
multiplying $B_n$ by $(-1)^n$; this modification effectively just
multiplies each term of the previous result by an even power of
negative one.
% Check that this works:
% 1) n even --> Overall prefactor is +
%  * In case we have an even number of blocks of even length in a partition we also have an even number of blocks of odd length in that partition --> Signs agree (+,+)
%  * In case we have an odd number of blocks of even length in a partition we have an even number of blocks of odd length in that partition --> Signs agree (-,-)
% 2) n odd --> Overall prefactor is -
%  * In case we have an even number of blocks of even length in a partition we have an odd number of blocks of odd length in that partition --> Signs agree (+,+)
%  * In case we have an odd number of blocks of even length in a partition we have an odd number of blocks of odd length in that partition --> Signs agree (-,-)
% Checking those four cases should be sufficient...

With these two modifications in hand we can finally write down the
main result of this subsection, namely the way on how to transform a
sum over unequal indices into a sum over products of power sums:
\begin{align}\label{eq:UnequalSumBell}
    \sum_{i_1 \neq \cdots \neq i_n } x_{i_1}\cdots x_{i_n}
    = (-1)^n \ B_n \left(-0! \ \sum_{i} x_{i}, \ -1! \ \sum_{i} x_{i}^2, \cdots , \ -(n-1)! \ \sum_{i} x_{i}^n \right)
\end{align}

%%%%%%%%%%%%%%%%%%%%%%%%%%%%%%%%%%%%%%%%%%%%%%%%%%%%%%%

\subsection{Application to the aperture mass estimator}
\label{app:ApplicationToMap}

Looking at the form of \Eqn{eq:UnequalSumBell}, the expression for
the direct estimator of the aperture statistics with equal aperture
radii \Eqn{eq:MapnEstBell} immediately follows when identifying the
arguments in the nominator and denominator with the power sums
$\MsEst{m}$ and $\SEst{m}$ and cancelling the overall sign.  \\

For the case of unequal aperture radii we still need to do a bit more
work. Looking back to our previous example \Eqn{eq:helperAppA1_1},
having unequal aperture radii induces different values of the $Q$
filters such that the $x_i$ cannot be taken to be the same variable
anymore. Hence we have to replace the prefactors in
\Eqn{eq:helperAppA1_1} by a sum over all the possible ways the
different radii can be partitioned. The second set of prefactors that
arises when going to \Eqn{eq:helperAppA1_2} still applies in the
case of unequal radii as it effectively corresponds to swapping two
aperture radii in the corresponding multivariate power sum
\Eqn{eq:Msvar}. Thus it seems appropriate to formulate the solution
via summing over partitions, such that we can rewrite
\Eqn{eq:MapnREst_h1} as
\begin{align}\label{eq:_MapnREst}
  \MapStatEst{n}(\vartheta_1, ..., \vartheta_n) = 
    \frac{\sum_{m=1}^n \sum_{\pi \in P_{n,m}} (-1)^m \prod_{i=1}^{m} (n_i-1)! \
    \MsEstPoly{n_i}{\mathscr{s}_1(\pi_i), \cdots, \mathscr{s}_n(\pi_i)}
  }{\sum_{m=1}^n \sum_{\pi \in P_{n,m}} (-1)^m \prod_{i=1}^{m} (n_i-1)! \
    \SEstPoly{n_i}{\mathscr{s}_1(\pi_i), \cdots, \mathscr{s}_n(\pi_i)}} \ .
\end{align}
We note that from this formulation one can build an efficient way of computing \Eqn{eq:MapnREst} within the subset $\mathcal{U}$ of the datacube $[R_1, \cdots R_m]^n \ \ (m \geq n)$ in which neither of the indices are equal:
This is due to the fact that the number of power sums in which $1 \leq i \leq n$ radii are selected is simply given by $\binom{m}{i}$ and therefore the n-dimensional hypercube of aperture radii can be constructed from a set consisting of just $\sum_{i=1}^n \binom{m}{i}$ power sums.
After allocating those power sums for all the galaxies within an aperture we can then enumerate through the relevant aperture radii multiplets, select the relevant subsets of the power sums, and then again apply the transformation equation \eqref{eq:MapnREst} to transform to the multiscale aperture mass moments, or equivalently to their corresponding connected parts. With the help of this procedure we were able to conduct the full analysis displayed in \Fig{fig:SLICSMockMultiscaleCorrcoef} on the SLICS ensemble (a total of around 2.5 billion galaxies) within just $6000$ CPU hours. 

%%%%%%%%%%%%%%%%%%%%%%%%%%%%%%%%%%%%%%%%%%%%%%%%%%%%%%%

\subsection{Expressions of the accelerated estimator for low orders (unequal radii)}  

In order to save space we only write down the expressions for the
nominator or \Eqn{eq:MapnREst}, the denominator will have an
identical structure. As expected, the number of sums in the $n$th
order estimator equals the $n$th Bell number.
\begin{align}
\MapStatEst{1}(\vartheta_1) = \frac{1}{\text{norm}} &\times
\MsEstPoly{1}{1} \ ; \\
%%%%%
\MapStatEst{2}(\vartheta_1,\vartheta_2) = \frac{1}{\text{norm}} &\times \left\{
\MsEstPoly{1}{1,0}\MsEstPoly{1}{0,1} - \MsEstPoly{2}{1,1}\right\}\ ; \\
%%%%%
\MapStatEst{3}(\vartheta_1,\vartheta_2,\vartheta_3) = \frac{1}{\text{norm}} &\times \left\{
\MsEstPoly{1}{1,0,0}\MsEstPoly{1}{0,1,0}\MsEstPoly{1}{0,0,1} 
- \left[\MsEstPoly{2}{1,1,0} \ \MsEstPoly{1}{0,0,1} \ + \ 2\text{ perm.}\right] 
+ 2 \  \MsEstPoly{3}{1,1,1}\right\}
\ ; \\
%%%%%
\MapStatEst{4}(\vartheta_1,\vartheta_2,\vartheta_3,\vartheta_4) = \frac{1}{\text{norm}} &\times \left\{ 
\MsEstPoly{1}{1,0,0,0}\MsEstPoly{1}{0,1,0,0}\MsEstPoly{1}{0,0,1,0}\MsEstPoly{1}{0,0,0,1}
- \left[\MsEstPoly{2}{1,1,0,0} \MsEstPoly{1}{0,0,1,0}  \MsEstPoly{1}{0,0,0,1} \
  + \ 5\text{ perm.}\right] \right.
\nn \\ &+ \left.
\left[\MsEstPoly{2}{1,1,0,0} \MsEstPoly{2}{0,0,1,1}  \ + \ 2 \text{ perm.}\right]
+ 2\left[ \MsEstPoly{3}{1,1,1,0} \MsEstPoly{1}{0,0,0,1}  \ + \ 3 \text{ perm.} \right]
-6 \ \MsEstPoly{4}{1,1,1,1}  \right\}
\ ; \\
%%%%%
\MapStatEst{5}(\vartheta_1,\vartheta_2,\vartheta_3,\vartheta_4,\vartheta_5) =
\frac{1}{\text{norm}} &\times \left\{ 
\MsEstPoly{1}{1,0,0,0,0}\MsEstPoly{1}{0,1,0,0,0}\MsEstPoly{1}{0,0,1,0,0}
\MsEstPoly{1}{0,0,0,1,0}\MsEstPoly{1}{0,0,0,0,1} \right.
\nn \\
&-  \left[\MsEstPoly{2}{1,1,0,0,0} \MsEstPoly{1}{0,0,1,0,0}
  \MsEstPoly{1}{0,0,0,1,0}\MsEstPoly{1}{0,0,0,0,1} \ + \ 9\text{ perm.}\right]
\nn \\ 
&+ \left[\MsEstPoly{2}{1,1,0,0,0} \MsEstPoly{2}{0,0,1,1,0}\MsEstPoly{1}{0,0,0,0,1}
  \ + \ 14 \text{ perm.}\right]
\nn \\ 
&+ 2\left[ \MsEstPoly{3}{1,1,1,0,0} \MsEstPoly{1}{0,0,0,1,0} \MsEstPoly{1}{0,0,0,0,1}
  + 9 \text{ perm.} \right] - 2\left[\MsEstPoly{3}{1,1,1,0,0} \MsEstPoly{2}{0,0,0,1,1} \ + \ 9 \text{ perm.}\right]
\nn \\ 
&+ \left. 6 \left[ \MsEstPoly{4}{1,1,1,1,0} \MsEstPoly{1}{0,0,0,0,1}  \ + \ 4 \text{ perm.} \right]
+24 \ \MsEstPoly{5}{1,1,1,1,1}  \right\}
\ ; \\
%%%%%
\MapStatEst{6}(\vartheta_1,\vartheta_2,\vartheta_3,\vartheta_4,\vartheta_5,\vartheta_6)
= \frac{1}{\text{norm}} &\times \left\{ 
\MsEstPoly{1}{1,0,0,0,0,0}\MsEstPoly{1}{0,1,0,0,0,0}\MsEstPoly{1}{0,0,1,0,0,0}
\MsEstPoly{1}{0,0,0,1,0,0}\MsEstPoly{1}{0,0,0,0,1,0}\MsEstPoly{1}{0,0,0,0,0,1} \right.
\nn \\
&-  \left[\MsEstPoly{2}{1,1,0,0,0,0} \MsEstPoly{1}{0,0,1,0,0,0}
  \MsEstPoly{1}{0,0,0,1,0,0}\MsEstPoly{1}{0,0,0,0,1,0}\MsEstPoly{1}{0,0,0,0,1,0} \ + \ 14\text{ perm.}\right]
\nn \\ 
&+ \left[\MsEstPoly{2}{1,1,0,0,0,0} \MsEstPoly{2}{0,0,1,1,0,0}\MsEstPoly{1}{0,0,0,0,1,0}
  \MsEstPoly{1}{0,0,0,0,0,1}  \ + \ 44 \text{ perm.}\right]
\nn \\ 
&- \left[\MsEstPoly{2}{1,1,0,0,0,0} \MsEstPoly{2}{0,0,1,1,0,0}\MsEstPoly{2}{0,0,0,0,1,1}
  \ + \ 14 \text{ perm.}\right]
\nn \\ 
&+ 2\left[ \MsEstPoly{3}{1,1,1,0,0,0} \MsEstPoly{1}{0,0,0,1,0,0} \MsEstPoly{1}{0,0,0,0,1,0}
  \MsEstPoly{1}{0,0,0,0,1,0}  + 19 \text{ perm.} \right] 
\nn \\ 
&- 2\left[\MsEstPoly{3}{1,1,1,0,0,0} \MsEstPoly{2}{0,0,0,1,1,0}\MsEstPoly{1}{0,0,0,0,1,0} \ + \ 59 \text{ perm.}\right]
\nn \\ 
&+ 4\left[\MsEstPoly{3}{1,1,1,0,0,0} \MsEstPoly{3}{0,0,0,1,1,1} \ + \ 9 \text{ perm.}\right]
\nn \\ 
&- 6\left[\MsEstPoly{4}{1,1,1,1,0,0} \MsEstPoly{1}{0,0,0,0,1,0} \MsEstPoly{1}{0,0,0,0,0,1} \ + \ 14 \text{ perm.}\right]
\nn \\ 
&+ 6\left[\MsEstPoly{4}{1,1,1,1,0,0} \MsEstPoly{2}{0,0,0,0,1,1}  \ + \ 14 \text{ perm.}\right]
+ 24\left[\MsEstPoly{5}{1,1,1,1,1,0} \MsEstPoly{1}{0,0,0,0,0,1}  \ + \ 5 \text{ perm.}\right]
\nn \\ 
&- \left. 
120 \ \MsEstPoly{6}{1,1,1,1,1,1}  \right\}
\end{align}
%
%
%%%%%%%%%%%%%%%%%%%%%%%%%%%%%%%%%%%%%%%%%%%%%%%%%%%%%%%

\section{Variance of the direct estimator}\label{app:covariance}

\subsection{Motivation of the shape and  multiplicity factor}
We recall the definition of the $\MapStatEst{n}$ variance: 
\begin{align}\label{eq:MapnCovDef}
\sigma^2\left[\MapStatEst{n}\right] 
= 
\mathbb{E}\left[\left(\MapStatEst{n}\right)^2\right] - \MapStatEns{n}^2
=
\frac{\left(\pi \vartheta^2\right)^{2n}}{\left(\sum_{\neq} w_{j_1}\cdots w_{j_n}\right)^2}
\ \cdot \ \mathbb{E}\left [
\sum_{\neq} w_{i_1}\cdots w_{i_n}x_{i_1}\cdots x_{i_n}
\cdot
 \sum_{\neq} w_{j_1}\cdots w_{j_n}x_{j_1}\cdots x_{j_n}
\right]
- \MapStatEns{n}^2 \ ,
\end{align}
where we defined $x_i \equiv Q_i e_{t,i}$ for notational simplicity. We proceed along the standard lines by decomposing the expectation value in an averaging step $A$ over the intrinsic ellipticity distribution, another one $P$ over the galaxy positions, and finally one over the cosmological ensemble. Let us start by applying the ellipticity averaging procedure for which $A(e_i, e_j) \equiv \frac{\sigma^2_\epsilon}{2} \delta^K_{i,j} + \gamma_i\gamma_j\left(1-\delta^K_{i,j}\right)$. Noting that each summation sign in \eqref{eq:MapnCovDef} runs over an index set where all the indices are unequal, we see that only indices between the two sums can be contracted to yield the shape noise expression. We can represent the index structure graphically as 
$ \left| 
\ i_1 \  \cdots \  i_n \  | \ j_1 \ \cdots  \ j_n 
\right| $
and define a contraction as a line between two indices of the $i$ and $j$ set. The prefactor of the term in the $A$-averaging is then given by the number of possible contractions. \\
As an example, let us compute the prefactor when applying two contractions in the variance of the third order statistics. For the first contraction there are $9$ possibilities, while for each second one there are only for indices remaining, giving $4$ further possibilities. As the contractions are interchangeable we need to divide the result by two to yield a prefactor of $18$. A graphical representation of this explanation would look as follows: 
$$
\left|\contraction{}{i_1}{\ i_2\  i_3\  | \ }{j_1}
 \contraction[2ex]{i_1 \ }{i_2}{\ i_3\  j_1\  | \ }{j_2}
i_1 \ i_2 \ i_3 \ | \ j_1 \ j_2 \ j_3 \right| 
= \frac{9}{2} \times 
\left| \contraction{}{i_2}{ \  i_3 \  | \ }{j_2}
i_2 \ i_3 \ | \ j_2 \ j_3 \right|
=\frac{9\cdot4}{2!} = 18 \ .
$$
This scheme allows us to easily generalize our example to performing $\ell$ contractions on the $n$th order statistics, giving a prefactor of $C_2(n,\ell) \equiv \frac{n^2(n-1)^2\cdots(n-\ell-1)^2}{\ell!}$ \ .
\\
For the position averaging we can repeat the same argument, as $P(Q_i\gamma_iQ_j\gamma_j) \sim \MsEst{2}\delta^K_{i,j} + \MapStat{2} \left(1-\delta^K_{i,j}\right)$. If we already have performed $\ell$ contractions for the $A$-averaging, there are only $(n-\ell)$ free indices left in each block - hence there will be $C_2(n-\ell,p)$ possibilities to perform $p$ additional contractions in the $P$-averaging.
\\
Next we compute the expectation value for a given index set in which we have performed $\ell$ contractions in the $A$-averaging and $p$ contractions in the $P$-averaging:
\begin{align*}
    \left\langle P \left(
    \sum_{\neq}\right.\right.
    w^2_{i_1} Q_{i_1}^2 &\cdots w^2_{i_\ell}  Q_{i_\ell}^2 \ 
    w^2_{i_{\ell+1}} Q^2_{i_{\ell+1}}\gamma^2_{t,i_{\ell+1}} \cdots w^2_{i_{\ell+p}} Q^2_{i_{\ell+p}}\gamma^2_{t,i_{\ell+p}} 
    \\
    &\left.\left. \vphantom{\sum_\neq}
    w_{i_{\ell+p+1}}Q_{i_{\ell+p+1}} \gamma_{t,i_{\ell+p+1}} \cdots  w_{i_{n}}Q_{i_{n}} \gamma_{t,i_{n}}
     \ 
    w_{j_{\ell+p+1}}Q_{j_{\ell+p+1}} \gamma_{t,j_{\ell+p+1}} \cdots  w_{j_{n}}Q_{i_{n}} \gamma_{t,j_{n}}
    \right)
    \right\rangle
    \\&\hspace{-2cm} \equiv 
    \left\langle \prod_{i=1}^N \int_{\rm{Ap.}} \frac{\d ^2 \theta_i}{\pi \vartheta^2} 
    \sum_{\neq}
    w^2_{i_1} Q_{i_1}^2 \cdots w^2_{i_\ell}  Q_{i_\ell}^2 \ 
    w^2_{i_{\ell+1}} Q^2_{i_{\ell+1}}\gamma^2_{t,i_{\ell+1}} \cdots w^2_{i_{\ell+p}} Q^2_{i_{\ell+p}}\gamma^2_{t,i_{\ell+p}} 
    \right.
    \\
    &\left. \vphantom{\sum_\neq}
    w_{i_{\ell+p+1}}Q_{i_{\ell+p+1}} \gamma_{t,i_{\ell+p+1}} \cdots  w_{i_{n}}Q_{i_{n}} \gamma_{t,i_{n}}
     \ 
    w_{j_{\ell+p+1}}Q_{j_{\ell+p+1}} \gamma_{t,j_{\ell+p+1}} \cdots  w_{j_{n}}Q_{i_{n}} \gamma_{t,j_{n}}
    \right\rangle
    \\&\hspace{-2cm} =
    \sum_\neq w^2_{i_1} \cdots w^2_{i_{\ell+p}} w_{i_{\ell+p+1}} \cdots w_{i_n} w_{j_{\ell+p+1}} \cdots w_{j_n} 
    \\  
    &\left(\prod_{i \in \{i_{1},\cdots, i_{\ell}\}} \int_{\rm{Ap.}} \frac{\d ^2 \theta_i}{\pi \vartheta^2} Q_i^2 \right)
     \left\langle  \left(\prod_{i \in \{i_{\ell+1},\cdots, i_{\ell+p}\}} \int_{\rm{Ap.}} \frac{\d ^2 \theta_i}{\pi \vartheta^2} Q_i^2 \gamma^2_{t,i}\right) \left(\prod_{i \in \{i_{\ell+p+1},\cdots, j_{n}\}} \int_{\rm{Ap.}} \frac{\d ^2 \theta_i}{\pi \vartheta^2} Q_i \gamma_{t,i} \right)  \right\rangle \left(\int_{\rm{Ap.}} \frac{\d ^2 \theta_i}{\pi \vartheta^2}\right)^{N-2(\ell+p)}
     \\&\hspace{-2cm} =
     \sum_\neq w^2_{i_1} \cdots w^2_{i_{\ell+p}} w_{i_{\ell+p+1}} \cdots w_{i_n} w_{j_{\ell+p+1}} \cdots w_{j_n} \ \times \ 
     \prod_{i=1}^\ell \left(  \int_{\rm{Ap.}} \frac{\d ^2 \theta_i}{\pi \vartheta^2} Q_i^2  \right)
     \left\langle \prod_{j=1}^p \left(  \int_{\rm{Ap.}} \frac{\d ^2 \theta_j}{\pi \vartheta^2} Q_j^2 \gamma_j^2 \right) \prod_{k=1}^{2(n-\ell-p)} \left(  \int_{\rm{Ap.}} \frac{\d ^2 \theta_k}{\pi \vartheta^2} Q_k \gamma_k \right) \right\rangle
     \\&\hspace{-2cm} \equiv
     \frac{\sum_\neq w^2_{i_1} \cdots w^2_{i_{\ell+p}} w_{i_{\ell+p+1}} \cdots w_{i_n} w_{j_{\ell+p+1}} \cdots w_{j_n}}{(\pi \vartheta)^{2n}} \ \times \
     \MgStat{\ell} \MsMapStatEns{p}{2(n-\ell-p)} \ .
\end{align*}
Note that in this derivation the order of the contracted indices does not matter as they all end up to be integration variables. If we now combine this result together with the multiplicity factors we can write a closed form expresson for \eqref{eq:MapnCovDef}:
\begin{align}
    \sigma^2\left[\MapStatEst{n}\right] 
    &=
    \sum_{\ell=0}^n C_2(n,\ell) \left(\frac{\sigma^2_\epsilon}{2}\right)^\ell \MgStat{\ell}  \sum_{p=0}^{n-\ell} \frac{\sum_{\neq} w_{i_1}^2\cdots w_{i_{\ell+p}}^2 w_{i_{\ell+p+1}} \cdots w_{i_n} w_{j_{\ell+p+1}} \cdots w_{j_n}}{\left( \sum_{\neq} w_{i_1}\cdots w_{i_n}\right)^2} C_2(n-\ell,p) \MsMapStatEns{p}{2(n-\ell-p)} \ \ - \ \  \MapStatEns{n}^2 
    \nonumber \\ &\approx
    \sum_{\ell=0}^n  \frac{\sum_{\neq} w_{i_1}^2\cdots w_{i_{\ell}}^2 w_{i_{\ell+1}} \cdots w_{i_n} w_{j_{\ell+1}} \cdots w_{j_n}}{\left( \sum_{\neq} w_{i_1}\cdots w_{i_n}\right)^2}  \ \ell!\binom{n}{\ell}\binom{n}{\ell} \ \left(\frac{\sigma^2_\epsilon}{2}\right)^\ell \MgStat{\ell} \ \MapStatEns{2(n-\ell)} \ \ - \ \  \MapStatEns{n}^2 
    \nonumber \\ &\approx
    n!\frac{\sum_{\neq} w_{i_1}^2\cdots w_{i_n}^2}{\left( \sum_{\neq} w_{i_1}\cdots w_{i_n}\right)^2}  \ \left(\frac{\sigma^2_\epsilon}{2}\right)^n \MgStat{n} \  . 
\end{align}
The first line is equivalent to \eqref{eq:MapStatCov} when combining the multiplicity factors and adjusting the indices. The second line makes the approximation that each of the $\mathcal{M}_{\rm{s,2}}$ are negligible (which is true for large $N$); for the final line we only keep the shot noise contribution.
\subsection{Modifications for unequal aperture radii}
In case of multiple apertures the structure of the variance is basically unchanged, the only thing we need to adjust is to use the multivariate version of the power sums and to replace the multiplicity factor with a sum over the actual multivariate expressions such that their radii correspond to the structure of the contracted indices. If we then take the shot noise dominated case we end up with:
\begin{align}
    \sigma^2_{\text{shot}}\left[\MapStatEst{n}(R_1, \cdots, R_n) \right] = 
    \frac{\sum^{'}_\neq w_{i_1}^2 \cdots w_{i_n}^2}{\left(\sum^{'}_\neq w_{i_1} \cdots w_{i_n}\right)^2} \left(\frac{\sigma^2_\epsilon}{2}\right)^n \sum_{\beta_1 \neq \cdots \neq \beta_n} \prod_{i=1}^{n} G_2\left(\frac{\max\left(\{R_i, R_{\beta_i}\}\right)}{\min\left(\{R_i, R_{\beta_i}\}\right)}\right) \ ,
\end{align}
where we define $G_2$ as the multiple radii generalization of $\mathcal{M}_{\rm{g,2}}$:
\begin{align*}
    G_2(\beta) 
    \equiv
    \pi R^2 \int \dthet \ Q_R(\theta) Q_{\beta R}(\theta)
    = 
    \frac{72}{\beta^2} \left[ \frac{1}{24} - \frac{1}{8\beta^2} + \frac{1}{10\beta^4}\right
    ] \ \ \ \ \ (\beta \geq 1)
\end{align*}
where the second equality denotes the corresponding equation for the polynomial filter. Note that for the corresponding inverse shot noise weighting scheme only the sum over the weights matters, as the remainder of the above expression is constant and can be factored out.
\subsection{Explicit expressions for low orders}
Here we collect the lowest order explicit expressions for \eqref{eq:MapStatCov}. The second order expression was first derived in \citep{Schneider1998}. Note that our prefactors differ from the ones defined in \citep{Munshietal2003}. 
\begin{align*}
%Map1
\sigma^2\left[\widehat{\mathcal{M}_{\mathrm {ap}}}\right] 
&= 
\frac{1}{\left(\sum_\neq w_{i_1}\right)^2} \left\{ \sum_\neq w_{i_1}w_{j_1}\left\langle\mathcal{M}_{\mathrm {ap}}^{2}\right\rangle +  \sum_\neq w_{i_1}^{2}\left\langle\mathcal{M}_{s,2}\right\rangle + 1\mathcal{M}_{g,2}\left(\frac{\sigma^2_\epsilon}{2}\right) \sum_\neq w_{i_1}^{2} \right\} - \left\langle\mathcal{M}_{\mathrm {ap}}\right\rangle^2\\
%Map2
\sigma^2\left[\widehat{\mathcal{M}_{\mathrm {ap}}^2}\right] 
&= 
\frac{1}{\left(\sum_\neq  w_{i_1}w_{i_2}\right)^2} \left\{  \vphantom{\left(\frac{\sigma^2_\epsilon}{2}\right)} \right. \sum_\neq w_{i_1}w_{j_1}w_{i_2}w_{j_2}\left\langle\mathcal{M}_{\mathrm {ap}}^{4}\right\rangle + 4 \sum_\neq w_{i_1}^{2}w_{i_2}w_{j_2}\left\langle\mathcal{M}_{s,2}\mathcal{M}_{\mathrm {ap}}^{2}\right\rangle + 2 \sum_\neq w_{i_1}^{2}w_{i_2}^{2}\left\langle\mathcal{M}_{s,2}^{2}\right\rangle 
\\ & \left. + 4\mathcal{M}_{g,2}\left(\frac{\sigma^2_\epsilon}{2}\right)\left[ \sum_\neq w_{i_1}^{2}w_{i_2}w_{j_2}\left\langle\mathcal{M}_{\mathrm {ap}}^{2}\right\rangle +  \sum_\neq w_{i_1}^{2}w_{i_2}^{2}\left\langle\mathcal{M}_{s,2}\right\rangle \right] + 2\mathcal{M}_{g,2}^{2}\left(\frac{\sigma^2_\epsilon}{2}\right)^{2} \sum_\neq w_{i_1}^{2}w_{i_2}^{2} \right\} - \left\langle\mathcal{M}_{\mathrm {ap}}^{2}\right\rangle^2
\\
%Map3
\sigma^2\left[\widehat{\mathcal{M}_{\mathrm {ap}}^3}\right] 
&= 
\frac{1}{\left(\sum_\neq w_{i_1}w_{i_2}w_{i_3}\right)^2} \left\{ \sum_\neq w_{i_1}w_{j_1}w_{i_2}w_{j_2}w_{i_3}w_{j_3}\left\langle\mathcal{M}_{\mathrm {ap}}^{6}\right\rangle + 9 \sum_\neq w_{i_1}^{2}w_{i_2}w_{j_2}w_{i_3}w_{j_3}\left\langle\mathcal{M}_{s,2}\mathcal{M}_{\mathrm {ap}}^{4}\right\rangle 
\vphantom{\left(\frac{\sigma^2_\epsilon}{2}\right)} \right. \\
&+ 18 \sum_\neq w_{i_1}^{2}w_{i_2}^{2}w_{i_3}w_{j_3}\left\langle\mathcal{M}_{s,2}^{2}\mathcal{M}_{\mathrm {ap}}^{2}\right\rangle + 6 \sum_\neq w_{i_1}^{2}w_{i_2}^{2}w_{i_3}^{2}\left\langle\mathcal{M}_{s,2}^{3}\right\rangle + 9\mathcal{M}_{g,2}\left(\frac{\sigma^2_\epsilon}{2}\right)\left[ \sum_\neq w_{i_1}^{2}w_{i_2}w_{j_2}w_{i_3}w_{j_3}\left\langle\mathcal{M}_{\mathrm {ap}}^{4}\right\rangle 
\right. \\
&+ 4 \sum_\neq w_{i_1}^{2}w_{i_2}^{2}w_{i_3}w_{j_3}\left\langle\mathcal{M}_{s,2}\mathcal{M}_{\mathrm {ap}}^{2}\right\rangle + 2 \sum_\neq w_{i_1}^{2}w_{i_2}^{2}w_{i_3}^{2}\left\langle\mathcal{M}_{s,2}^{2}\right\rangle \left.\vphantom{\left(\frac{\sigma^2_\epsilon}{2}\right)} \right] +
18\mathcal{M}_{g,2}^{2}\left(\frac{\sigma^2_\epsilon}{2}\right)^{2}\left[ \sum_\neq w_{i_1}^{2}w_{i_2}^{2}w_{i_3}w_{j_3}\left\langle\mathcal{M}_{\mathrm {ap}}^{2}\right\rangle 
\right. \\ 
&+  \sum_\neq w_{i_1}^{2}w_{i_2}^{2}w_{i_3}^{2}\left\langle\mathcal{M}_{s,2}\right\rangle \left.\vphantom{\left(\frac{\sigma^2_\epsilon}{2}\right)} \right] \left.+ 6\mathcal{M}_{g,2}^{3}\left(\frac{\sigma^2_\epsilon}{2}\right)^{3} \sum_\neq w_{i_1}^{2}w_{i_2}^{2}w_{i_3}^{2} \right\} - \left\langle\mathcal{M}_{\mathrm {ap}}^{3}\right\rangle^2
\\
%Map4
\sigma^2\left[\widehat{\mathcal{M}_{\mathrm {ap}}^4}\right] 
&= 
\frac{1}{\left(\sum_\neq w_{i_1}w_{i_2}w_{i_3}w_{i_4}\right)^2} \left\{ \sum_\neq w_{i_1}w_{j_1}w_{i_2}w_{j_2}w_{i_3}w_{j_3}w_{i_4}w_{j_4}\left\langle\mathcal{M}_{\mathrm {ap}}^{8}\right\rangle + 16 \sum_\neq w_{i_1}^{2}w_{i_2}w_{j_2}w_{i_3}w_{j_3}w_{i_4}w_{j_4}\left\langle\mathcal{M}_{s,2}\mathcal{M}_{\mathrm {ap}}^{6}\right\rangle \right.
\\
&+ 72 \sum_\neq w_{i_1}^{2}w_{i_2}^{2}w_{i_3}w_{j_3}w_{i_4}w_{j_4}\left\langle\mathcal{M}_{s,2}^{2}\mathcal{M}_{\mathrm {ap}}^{4}\right\rangle + 96 \sum_\neq w_{i_1}^{2}w_{i_2}^{2}w_{i_3}^{2}w_{i_4}w_{j_4}\left\langle\mathcal{M}_{s,2}^{3}\mathcal{M}_{\mathrm {ap}}^{2}\right\rangle + 24 \sum_\neq w_{i_1}^{2}w_{i_2}^{2}w_{i_3}^{2}w_{i_4}^{2}\left\langle\mathcal{M}_{s,2}^{4}\right\rangle
\\
&+ 16\mathcal{M}_{g,2}\left(\frac{\sigma^2_\epsilon}{2}\right)\left[ \sum_\neq w_{i_1}^{2}w_{i_2}w_{j_2}w_{i_3}w_{j_3}w_{i_4}w_{j_4}\left\langle\mathcal{M}_{\mathrm {ap}}^{6}\right\rangle + 9 \sum_\neq w_{i_1}^{2}w_{i_2}^{2}w_{i_3}w_{j_3}w_{i_4}w_{j_4}\left\langle\mathcal{M}_{s,2}\mathcal{M}_{\mathrm {ap}}^{4}\right\rangle 
\right. \\
&+ 18 \sum_\neq w_{i_1}^{2}w_{i_2}^{2}w_{i_3}^{2}w_{i_4}w_{j_4}\left\langle\mathcal{M}_{s,2}^{2}\mathcal{M}_{\mathrm {ap}}^{2}\right\rangle + 6 \sum_\neq w_{i_1}^{2}w_{i_2}^{2}w_{i_3}^{2}w_{i_4}^{2}\left\langle\mathcal{M}_{s,2}^{3}\right\rangle \left.\vphantom{\left(\frac{\sigma^2_\epsilon}{2}\right)}\right] + 72\mathcal{M}_{g,2}^{2}\left(\frac{\sigma^2_\epsilon}{2}\right)^{2}
\\ &\hspace{.5cm}\left[ \sum_\neq w_{i_1}^{2}w_{i_2}^{2}w_{i_3}w_{j_3}w_{i_4}w_{j_4}\left\langle\mathcal{M}_{\mathrm {ap}}^{4}\right\rangle + 4 \sum_\neq w_{i_1}^{2}w_{i_2}^{2}w_{i_3}^{2}w_{i_4}w_{j_4}\left\langle\mathcal{M}_{s,2}\mathcal{M}_{\mathrm {ap}}^{2}\right\rangle + 2 \sum_\neq w_{i_1}^{2}w_{i_2}^{2}w_{i_3}^{2}w_{i_4}^{2}\left\langle\mathcal{M}_{s,2}^{2}\right\rangle \right] 
\\
&+ 96\mathcal{M}_{g,2}^{3}\left(\frac{\sigma^2_\epsilon}{2}\right)^{3}\left[ \sum_\neq w_{i_1}^{2}w_{i_2}^{2}w_{i_3}^{2}w_{i_4}w_{j_4}\left\langle\mathcal{M}_{\mathrm {ap}}^{2}\right\rangle +  \sum_\neq w_{i_1}^{2}w_{i_2}^{2}w_{i_3}^{2}w_{i_4}^{2}\left\langle\mathcal{M}_{s,2}\right\rangle \right] + 24\mathcal{M}_{g,2}^{4}\left(\frac{\sigma^2_\epsilon}{2}\right)^{4} \sum_\neq w_{i_1}^{2}w_{i_2}^{2}w_{i_3}^{2}w_{i_4}^{2} \left.\vphantom{\left(\frac{\sigma^2_\epsilon}{2}\right)}\right\} - \left\langle\mathcal{M}_{\mathrm {ap}}^{4}\right\rangle^2
\\
%Map5
\sigma^2\left[\widehat{\mathcal{M}_{\mathrm {ap}}^5}\right] 
&= 
\frac{1}{\left(\sum_\neq w_{i_1}w_{i_2}w_{i_3}w_{i_4}w_{i_5}\right)^2} \left\{ \sum_\neq w_{i_1}w_{j_1}w_{i_2}w_{j_2}w_{i_3}w_{j_3}w_{i_4}w_{j_4}w_{i_5}w_{j_5}\left\langle\mathcal{M}_{\mathrm {ap}}^{10}\right\rangle  \vphantom{\left(\frac{\sigma^2_\epsilon}{2}\right)}\right.
\\
&+ 25 \sum_\neq w_{i_1}^{2}w_{i_2}w_{j_2}w_{i_3}w_{j_3}w_{i_4}w_{j_4}w_{i_5}w_{j_5}\left\langle\mathcal{M}_{s,2}\mathcal{M}_{\mathrm {ap}}^{8}\right\rangle + 200 \sum_\neq w_{i_1}^{2}w_{i_2}^{2}w_{i_3}w_{j_3}w_{i_4}w_{j_4}w_{i_5}w_{j_5}\left\langle\mathcal{M}_{s,2}^{2}\mathcal{M}_{\mathrm {ap}}^{6}\right\rangle 
\\
&+ 600 \sum_\neq w_{i_1}^{2}w_{i_2}^{2}w_{i_3}^{2}w_{i_4}w_{j_4}w_{i_5}w_{j_5}\left\langle\mathcal{M}_{s,2}^{3}\mathcal{M}_{\mathrm {ap}}^{4}\right\rangle + 600 \sum_\neq w_{i_1}^{2}w_{i_2}^{2}w_{i_3}^{2}w_{i_4}^{2}w_{i_5}w_{j_5}\left\langle\mathcal{M}_{s,2}^{4}\mathcal{M}_{\mathrm {ap}}^{2}\right\rangle 
\\
&+ 120 \sum_\neq w_{i_1}^{2}w_{i_2}^{2}w_{i_3}^{2}w_{i_4}^{2}w_{i_5}^{2}\left\langle\mathcal{M}_{s,2}^{5}\right\rangle + 25\mathcal{M}_{g,2}\left(\frac{\sigma^2_\epsilon}{2}\right)\left[ \sum_\neq w_{i_1}^{2}w_{i_2}w_{j_2}w_{i_3}w_{j_3}w_{i_4}w_{j_4}w_{i_5}w_{j_5}\left\langle\mathcal{M}_{\mathrm {ap}}^{8}\right\rangle \right.
\\ 
&+ 16 \sum_\neq w_{i_1}^{2}w_{i_2}^{2}w_{i_3}w_{j_3}w_{i_4}w_{j_4}w_{i_5}w_{j_5}\left\langle\mathcal{M}_{s,2}\mathcal{M}_{\mathrm {ap}}^{6}\right\rangle + 72 \sum_\neq w_{i_1}^{2}w_{i_2}^{2}w_{i_3}^{2}w_{i_4}w_{j_4}w_{i_5}w_{j_5}\left\langle\mathcal{M}_{s,2}^{2}\mathcal{M}_{\mathrm {ap}}^{4}\right\rangle 
\\
&+ 96 \sum_\neq w_{i_1}^{2}w_{i_2}^{2}w_{i_3}^{2}w_{i_4}^{2}w_{i_5}w_{j_5}\left\langle\mathcal{M}_{s,2}^{3}\mathcal{M}_{\mathrm {ap}}^{2}\right\rangle + 24 \sum_\neq w_{i_1}^{2}w_{i_2}^{2}w_{i_3}^{2}w_{i_4}^{2}w_{i_5}^{2}\left\langle\mathcal{M}_{s,2}^{4}\right\rangle \left.\vphantom{\left(\frac{\sigma^2_\epsilon}{2}\right)}\right] + 200\mathcal{M}_{g,2}^{2}\left(\frac{\sigma^2_\epsilon}{2}\right)^{2}
\\
&\hspace{.5cm}\left[ \sum_\neq w_{i_1}^{2}w_{i_2}^{2}w_{i_3}w_{j_3}w_{i_4}w_{j_4}w_{i_5}w_{j_5}\left\langle\mathcal{M}_{\mathrm {ap}}^{6}\right\rangle + 9 \sum_\neq w_{i_1}^{2}w_{i_2}^{2}w_{i_3}^{2}w_{i_4}w_{j_4}w_{i_5}w_{j_5}\left\langle\mathcal{M}_{s,2}\mathcal{M}_{\mathrm {ap}}^{4}\right\rangle \right.
\\
&\hspace{.5cm}+ 18 \sum_\neq w_{i_1}^{2}w_{i_2}^{2}w_{i_3}^{2}w_{i_4}^{2}w_{i_5}w_{j_5}\left\langle\mathcal{M}_{s,2}^{2}\mathcal{M}_{\mathrm {ap}}^{2}\right\rangle + 6 \sum_\neq w_{i_1}^{2}w_{i_2}^{2}w_{i_3}^{2}w_{i_4}^{2}w_{i_5}^{2}\left\langle\mathcal{M}_{s,2}^{3}\right\rangle \left.\vphantom{\left(\frac{\sigma^2_\epsilon}{2}\right)}\right] + 600\mathcal{M}_{g,2}^{3}\left(\frac{\sigma^2_\epsilon}{2}\right)^{3}
\\
&\hspace{.5cm}\left[ \sum_\neq w_{i_1}^{2}w_{i_2}^{2}w_{i_3}^{2}w_{i_4}w_{j_4}w_{i_5}w_{j_5}\left\langle\mathcal{M}_{\mathrm {ap}}^{4}\right\rangle + 4 \sum_\neq w_{i_1}^{2}w_{i_2}^{2}w_{i_3}^{2}w_{i_4}^{2}w_{i_5}w_{j_5}\left\langle\mathcal{M}_{s,2}\mathcal{M}_{\mathrm {ap}}^{2}\right\rangle + 2 \sum_\neq w_{i_1}^{2}w_{i_2}^{2}w_{i_3}^{2}w_{i_4}^{2}w_{i_5}^{2}\left\langle\mathcal{M}_{s,2}^{2}\right\rangle \right] 
\\
&+ 600\mathcal{M}_{g,2}^{4}\left(\frac{\sigma^2_\epsilon}{2}\right)^{4}\left[ \sum_\neq w_{i_1}^{2}w_{i_2}^{2}w_{i_3}^{2}w_{i_4}^{2}w_{i_5}w_{j_5}\left\langle\mathcal{M}_{\mathrm {ap}}^{2}\right\rangle +  \sum_\neq w_{i_1}^{2}w_{i_2}^{2}w_{i_3}^{2}w_{i_4}^{2}w_{i_5}^{2}\left\langle\mathcal{M}_{s,2}\right\rangle \right] 
\\
&+ 120\mathcal{M}_{g,2}^{5}\left(\frac{\sigma^2_\epsilon}{2}\right)^{5} \sum_\neq w_{i_1}^{2}w_{i_2}^{2}w_{i_3}^{2}w_{i_4}^{2}w_{i_5}^{2} \left.\vphantom{\left(\frac{\sigma^2_\epsilon}{2}\right)}\right\} - \left\langle\mathcal{M}_{\mathrm {ap}}^{5}\right\rangle^2
\\
%Map6
\sigma^2\left[\widehat{\mathcal{M}_{\mathrm {ap}}^6}\right] 
&= 
\frac{1}{\left(\sum_\neq w_{i_1}w_{i_2}w_{i_3}w_{i_4}w_{i_5}w_{i_6}\right)^2} \left\{ \sum_\neq w_{i_1}w_{j_1}w_{i_2}w_{j_2}w_{i_3}w_{j_3}w_{i_4}w_{j_4}w_{i_5}w_{j_5}w_{i_6}w_{j_6}\left\langle\mathcal{M}_{\mathrm {ap}}^{12}\right\rangle \right.
\\
&+ 36 \sum_\neq w_{i_1}^{2}w_{i_2}w_{j_2}w_{i_3}w_{j_3}w_{i_4}w_{j_4}w_{i_5}w_{j_5}w_{i_6}w_{j_6}\left\langle\mathcal{M}_{s,2}\mathcal{M}_{\mathrm {ap}}^{10}\right\rangle + 450 \sum_\neq w_{i_1}^{2}w_{i_2}^{2}w_{i_3}w_{j_3}w_{i_4}w_{j_4}w_{i_5}w_{j_5}w_{i_6}w_{j_6}\left\langle\mathcal{M}_{s,2}^{2}\mathcal{M}_{\mathrm {ap}}^{8}\right\rangle 
\\
&+ 2400 \sum_\neq w_{i_1}^{2}w_{i_2}^{2}w_{i_3}^{2}w_{i_4}w_{j_4}w_{i_5}w_{j_5}w_{i_6}w_{j_6}\left\langle\mathcal{M}_{s,2}^{3}\mathcal{M}_{\mathrm {ap}}^{6}\right\rangle + 5400 \sum_\neq w_{i_1}^{2}w_{i_2}^{2}w_{i_3}^{2}w_{i_4}^{2}w_{i_5}w_{j_5}w_{i_6}w_{j_6}\left\langle\mathcal{M}_{s,2}^{4}\mathcal{M}_{\mathrm {ap}}^{4}\right\rangle 
\\
&+ 4320 \sum_\neq w_{i_1}^{2}w_{i_2}^{2}w_{i_3}^{2}w_{i_4}^{2}w_{i_5}^{2}w_{i_6}w_{j_6}\left\langle\mathcal{M}_{s,2}^{5}\mathcal{M}_{\mathrm {ap}}^{2}\right\rangle + 720 \sum_\neq w_{i_1}^{2}w_{i_2}^{2}w_{i_3}^{2}w_{i_4}^{2}w_{i_5}^{2}w_{i_6}^{2}\left\langle\mathcal{M}_{s,2}^{6}\right\rangle + 36\mathcal{M}_{g,2}\left(\frac{\sigma^2_\epsilon}{2}\right)
\\
&\hspace{.5cm}\left[ \sum_\neq w_{i_1}^{2}w_{i_2}w_{j_2}w_{i_3}w_{j_3}w_{i_4}w_{j_4}w_{i_5}w_{j_5}w_{i_6}w_{j_6}\left\langle\mathcal{M}_{\mathrm {ap}}^{10}\right\rangle + 25 \sum_\neq w_{i_1}^{2}w_{i_2}^{2}w_{i_3}w_{j_3}w_{i_4}w_{j_4}w_{i_5}w_{j_5}w_{i_6}w_{j_6}\left\langle\mathcal{M}_{s,2}\mathcal{M}_{\mathrm {ap}}^{8}\right\rangle 
\right.
\\
&\hspace{.5cm}+ 200 \sum_\neq w_{i_1}^{2}w_{i_2}^{2}w_{i_3}^{2}w_{i_4}w_{j_4}w_{i_5}w_{j_5}w_{i_6}w_{j_6}\left\langle\mathcal{M}_{s,2}^{2}\mathcal{M}_{\mathrm {ap}}^{6}\right\rangle + 600 \sum_\neq w_{i_1}^{2}w_{i_2}^{2}w_{i_3}^{2}w_{i_4}^{2}w_{i_5}w_{j_5}w_{i_6}w_{j_6}\left\langle\mathcal{M}_{s,2}^{3}\mathcal{M}_{\mathrm {ap}}^{4}\right\rangle 
\\
&\hspace{.5cm}+ 600 \sum_\neq w_{i_1}^{2}w_{i_2}^{2}w_{i_3}^{2}w_{i_4}^{2}w_{i_5}^{2}w_{i_6}w_{j_6}\left\langle\mathcal{M}_{s,2}^{4}\mathcal{M}_{\mathrm {ap}}^{2}\right\rangle + 120 \sum_\neq w_{i_1}^{2}w_{i_2}^{2}w_{i_3}^{2}w_{i_4}^{2}w_{i_5}^{2}w_{i_6}^{2}\left\langle\mathcal{M}_{s,2}^{5}\right\rangle \left.\vphantom{\left(\frac{\sigma^2_\epsilon}{2}\right)}\right] + 450\mathcal{M}_{g,2}^{2}\left(\frac{\sigma^2_\epsilon}{2}\right)^{2}
\\
&\hspace{.5cm}\left[ \sum_\neq w_{i_1}^{2}w_{i_2}^{2}w_{i_3}w_{j_3}w_{i_4}w_{j_4}w_{i_5}w_{j_5}w_{i_6}w_{j_6}\left\langle\mathcal{M}_{\mathrm {ap}}^{8}\right\rangle + 16 \sum_\neq w_{i_1}^{2}w_{i_2}^{2}w_{i_3}^{2}w_{i_4}w_{j_4}w_{i_5}w_{j_5}w_{i_6}w_{j_6}\left\langle\mathcal{M}_{s,2}\mathcal{M}_{\mathrm {ap}}^{6}\right\rangle 
\right.\\
&\hspace{.5cm}+ 72 \sum_\neq w_{i_1}^{2}w_{i_2}^{2}w_{i_3}^{2}w_{i_4}^{2}w_{i_5}w_{j_5}w_{i_6}w_{j_6}\left\langle\mathcal{M}_{s,2}^{2}\mathcal{M}_{\mathrm {ap}}^{4}\right\rangle + 96 \sum_\neq w_{i_1}^{2}w_{i_2}^{2}w_{i_3}^{2}w_{i_4}^{2}w_{i_5}^{2}w_{i_6}w_{j_6}\left\langle\mathcal{M}_{s,2}^{3}\mathcal{M}_{\mathrm {ap}}^{2}\right\rangle 
\\
&\hspace{.5cm}+ 24 \sum_\neq w_{i_1}^{2}w_{i_2}^{2}w_{i_3}^{2}w_{i_4}^{2}w_{i_5}^{2}w_{i_6}^{2}\left\langle\mathcal{M}_{s,2}^{4}\right\rangle \left.\vphantom{\left(\frac{\sigma^2_\epsilon}{2}\right)}\right] + 2400\mathcal{M}_{g,2}^{3}\left(\frac{\sigma^2_\epsilon}{2}\right)^{3}\left[ \sum_\neq w_{i_1}^{2}w_{i_2}^{2}w_{i_3}^{2}w_{i_4}w_{j_4}w_{i_5}w_{j_5}w_{i_6}w_{j_6}\left\langle\mathcal{M}_{\mathrm {ap}}^{6}\right\rangle \right.
\\
&\hspace{.5cm}+ 9 \sum_\neq w_{i_1}^{2}w_{i_2}^{2}w_{i_3}^{2}w_{i_4}^{2}w_{i_5}w_{j_5}w_{i_6}w_{j_6}\left\langle\mathcal{M}_{s,2}\mathcal{M}_{\mathrm {ap}}^{4}\right\rangle + 18 \sum_\neq w_{i_1}^{2}w_{i_2}^{2}w_{i_3}^{2}w_{i_4}^{2}w_{i_5}^{2}w_{i_6}w_{j_6}\left\langle\mathcal{M}_{s,2}^{2}\mathcal{M}_{\mathrm {ap}}^{2}\right\rangle 
\\
&\hspace{.5cm}+ 6 \sum_\neq w_{i_1}^{2}w_{i_2}^{2}w_{i_3}^{2}w_{i_4}^{2}w_{i_5}^{2}w_{i_6}^{2}\left\langle\mathcal{M}_{s,2}^{3}\right\rangle \left.\vphantom{\left(\frac{\sigma^2_\epsilon}{2}\right)}\right] + 5400\mathcal{M}_{g,2}^{4}\left(\frac{\sigma^2_\epsilon}{2}\right)^{4}\left[ \sum_\neq w_{i_1}^{2}w_{i_2}^{2}w_{i_3}^{2}w_{i_4}^{2}w_{i_5}w_{j_5}w_{i_6}w_{j_6}\left\langle\mathcal{M}_{\mathrm {ap}}^{4}\right\rangle \right.
\\
&\hspace{.5cm}+ 4 \sum_\neq w_{i_1}^{2}w_{i_2}^{2}w_{i_3}^{2}w_{i_4}^{2}w_{i_5}^{2}w_{i_6}w_{j_6}\left\langle\mathcal{M}_{s,2}\mathcal{M}_{\mathrm {ap}}^{2}\right\rangle + 2 \sum_\neq w_{i_1}^{2}w_{i_2}^{2}w_{i_3}^{2}w_{i_4}^{2}w_{i_5}^{2}w_{i_6}^{2}\left\langle\mathcal{M}_{s,2}^{2}\right\rangle \left.\vphantom{\left(\frac{\sigma^2_\epsilon}{2}\right)}\right] + 4320\mathcal{M}_{g,2}^{5}\left(\frac{\sigma^2_\epsilon}{2}\right)^{5}
\\
&\hspace{.5cm}\left[ \sum_\neq w_{i_1}^{2}w_{i_2}^{2}w_{i_3}^{2}w_{i_4}^{2}w_{i_5}^{2}w_{i_6}w_{j_6}\left\langle\mathcal{M}_{\mathrm {ap}}^{2}\right\rangle +  \sum_\neq w_{i_1}^{2}w_{i_2}^{2}w_{i_3}^{2}w_{i_4}^{2}w_{i_5}^{2}w_{i_6}^{2}\left\langle\mathcal{M}_{s,2}\right\rangle \right] + 720\mathcal{M}_{g,2}^{6}\left(\frac{\sigma^2_\epsilon}{2}\right)^{6} \sum_\neq w_{i_1}^{2}w_{i_2}^{2}w_{i_3}^{2}w_{i_4}^{2}w_{i_5}^{2}w_{i_6}^{2} \left.\vphantom{\left(\frac{\sigma^2_\epsilon}{2}\right)}\right\} 
\\
&- \left\langle\mathcal{M}_{\mathrm {ap}}^{6}\right\rangle^2
\end{align*}

\newpage
%%%%%%%%%%%%%%%%%%%%%%%%%%%%%%%%%%%%%%%%%%%%%%%%%%%%%%%

\section{Variance of the direct estimator for the aperture mass skewness}\label{app:covariance3}

%%%%%%%%%%%%%%%%%%%%%%%%%%%%%%%%%%%%%%%%%%%%%%%%%%%%%%%

\def\ncn{\neq \cdots \neq}
\def\mapthree{\langle \mathcal{M}_{\text{ap}}^3 \rangle}
\def\epsti{\epsilon_{t,i}}
\def\sigsqone{\left(\frac{\sigma_\epsilon^2}{2}\right)}
\def\msone{\langle \mathcal{M}_{s,2} \rangle}
\newcommand{\epsts}[2]{\epsilon_{t,{#1}_{#2}}}
\newcommand{\gamts}[2]{\gamma_{t,{#1}_{#2}}}
\newcommand{\mapn}[1]{\langle \mathcal{M}_{\text{ap}}^{#1} \rangle}
\newcommand{\msn}[1]{\langle \mathcal{M}_{s,2}^{#1} \rangle}
\newcommand{\msmap}[2]{\langle \mathcal{M}_{s,2}^{#1} \mathcal{M}_{\text{ap}}^{#2} \rangle}
\newcommand{\mmsmap}[2]{\ifnum\value{#1}=1 {\langle
    \mathcal{M}_{s,2}^{\vphantom{#1}} \mathcal{M}_{\text{ap}}^{#2} \rangle}
  \else {\langle \mathcal{M}_{s,2}^{#1} \mathcal{M}_{\text{ap}}^{#2} \rangle} \fi }
\newcommand{\deltaij}[2]{\delta^K_{i_{#1},j_{#2}}}

%%%%%%%%%%%%%%%%%%%%%%%%%%%%%%%%%%%%%%%%%%%%%%%%%%%%%%%

\subsection{Notation}

Let us begin this section by defining some useful notation. Unless
otherwise specified, for an $n$th order computation we assume
apertures with $N_g>n$ galaxies within them.
\begin{align}
  M_{s,n} & \equiv \frac{\sum_{i=1}^{N_g} (w_i Q_i \epsti)^n}{\left(\sum_{i=1}^{N_g}
    w_i\right)^n} \equiv \frac{\sum_{i=1}^{N_g}  (w_i x_i)^n}{\left(\sum_{i=1}^{N_g} w_i\right)^n}
  \ \ \ ;\ \ \ \ x_i\equiv Q_i \epsti\ \ ; \\
  M_{g,n} & \equiv \frac{\sum_{i=1}^{N_g} (w_i Q_i)^n}{\left(\sum_{i=1}^{N_g} w_i\right)^n}\ .
\end{align}
From now on we assume all sums with no explicit upper limit to run up
to $N_g$. As the summation indices do become rather messy, we shall
also define the following simplifying shorthands:
\begin{align}
\sum_{\neq} &\equiv \sum_{i_1} \sum_{i_2 \neq i_1} \cdots
\sum_{i_n\neq i_{n-1} \ncn i_1} \ ; \\
\sum_{\substack{\neq \\i_a=i_b}} &\equiv \sum_{i_1} \sum_{i_2 \neq i_1} \cdots \sum_{i_a \ncn i_1}
\cdots \sum_{i_{b-1}\ncn i_1} \sum_{i_{b+1}\ncn i_1} \cdots \sum_{i_n
  \ncn i_1} \ .
\end{align} 

%%%%%%%%%%%%%%%%%%%%%%%%%%%%%%%%%%%%%%%%%%%%%%%%%%%%%%%
\subsection{Computation}

With this background notation in hand, the variance of $\mapthree$
can be written as
\begin{align}
\sigma^2\left[\MapStatEst{3}\right] = 
\left \langle 
\left(\pi \theta^2\right)^3 \frac{\sum_{\neq} w_{i_1}w_{i_2}w_{i_3}x_{i_1}x_{i_2}x_{i_3}}{\sum_{\neq} w_{i_1}w_{i_2}w_{i_3}} 
\cdot
\left(\pi \theta^2\right)^3 \frac{\sum_{\neq} w_{j_1}w_{j_2}w_{j_3}x_{j_1}x_{j_2}x_{j_3}}{\sum_{\neq} w_{j_1}w_{j_2}w_{j_3}}
\right \rangle
-\mapthree^2
\end{align}
We proceed as always by averaging over the source galaxies. For the
third order variance we then expect four structurally identical terms
each corresponding to various permutations of contractions. The
prefactor can be found by considering the following scheme. We
represent the two groups of indices in a similar shape to a six point
correlator and count the number of different contractions that
contract\footnote{In this note contraction of two indices means that
they are set equal to each other.} an index of the $i$ set with one of
the $j$ set. Let us do an example to count all double contractions
(see illustration below): For a single contraction we have $9$
possibilities, whereas for two contractions we can effectively do all
single contractions ($9$ terms) and delete the contracted indices,
leaving us with just four remaining indices. Connecting those gives 4
more possibilities. Finally we divide by the factorial of the number
of contractions, as those are interchangeable.
$$
\left|\contraction{}{i_1}{\ i_2\  i_3\  | \ }{j_1}
 \contraction[2ex]{i_1 \ }{i_2}{\ i_3\  j_1\  | \ }{j_2}
i_1 \ i_2 \ i_3 \ | \ j_1 \ j_2 \ j_3 \right| 
= \frac{9}{2!} \times 
\left| \contraction{}{i_2}{ \  i_3 \  | \ }{j_2}
i_2 \ i_3 \ | \ j_2 \ j_3 \right|
=\frac{9\cdot4}{2!} = 18
$$
Generalizing to $a$ contractions for two $n$th order index sets this
gives $C(n,a)=\frac{n^2 (n-1)^2 \cdots (n-(a-1))^2}{a!}$. Now we find
for the source galaxy averaging
\begin{align}
A(\epsts{i}{1}\epsts{i}{2}\epsts{i}{3}\epsts{j}{1}\epsts{j}{2}\epsts{j}{3})
&=
\gamts{i}{1}\gamts{i}{2}\gamts{i}{3}\gamts{j}{1}\gamts{j}{2}\gamts{j}{3} 
&\circled{1} \nonumber \\ 
\contraction{}{i_1}{\ i_2\  i_3\  | \ }{j_1}
i_1 \ i_2 \ i_3 \ | \ j_1 \ j_2 \ j_3 \hspace{1cm}
&+
\sigsqone \left( Q_{i_1}^2 \gamts{i}{2}\gamts{i}{3}\gamts{j}{2}\gamts{j}{3} \deltaij{1}{1} + \text{8 perm.} \right)
&\circled{2}\nonumber \\ 
\contraction{}{i_1}{\ i_2\  i_3\  | \ }{j_1}
\contraction[2ex]{i_1 \ }{i_2}{\ i_3\  j_1\  | \ }{j_2}
i_1 \ i_2 \ i_3 \ | \ j_1 \ j_2 \ j_3 \hspace{1cm}
&+
\sigsq{2} \left( Q_{i_1}^2 Q_{i_2}^2 \gamts{i}{3}\gamts{j}{3} \deltaij{1}{1}\deltaij{2}{2} + \text{17 perm.} \right)
&\circled{3}\nonumber \\ 
\contraction{}{i_1}{\ i_2\  i_3\  | \ }{j_1}
\contraction[2ex]{i_1 \ }{i_2}{\ i_3 \ | \  j_1\ }{j_2}
\bcontraction{i_1 \ i_2 \ }{i_3}{ \ | \ j_1 \ j_2 \ }{j_3}
i_1 \ i_2 \ i_3 \ | \ j_1 \ j_2 \ j_3 \hspace{1cm}
&+
\sigsq{3} \left( Q_{i_1}^2Q_{i_2}^2Q_{i_3}^2 \deltaij{1}{1}\deltaij{2}{2}\deltaij{3}{3} + \text{5 perm.} \right)
&\circled{4}
\end{align}
We now perform the positional averaging over those terms $\circled{1}
- \circled{4}$ individually. In order to shorten similar calculations
we note the following identity for the position average corresponding
to an $m$ point contraction of an $n$th order variance:
\begin{align}
\left\langle P\left(
\sum_{\substack{\neq \\i_1=j_1 \\ \cdots \\ i_m=j_m}}
w_{i_1} \cdots w_{i_n} w_{j_1} \cdots w_{j_n} \ x_{i_1} \cdots x_{i_n} x_{j_1} \cdots x_{j_n}
\right)\right\rangle
=
\sum_{\neq}
w_{i_1}^2 \cdots w_{i_m}^2 w_{i_{m+1}} \cdots w_{i_n} w_{j_{m+1}} \cdots w_{j_n} \ \msmap{m}{2(n-m)}
\end{align}
For the term $\circled{1}$ the index structure in the summation symbol
has not changed at all, so we can simply recycle the reasoning to get
to the ellipticity averaging calculation. Also adding in the ensemble
average we get
\begin{align}
\left\langle P\left(
\sum_{\neq} w_{i_1}w_{i_2}w_{i_3}x_{i_1}x_{i_2}x_{i_3}
\sum_{\neq} w_{j_1}w_{j_2}w_{j_3}x_{j_1}x_{j_2}x_{j_3}
\right)\right\rangle
\nonumber \\
&\hspace*{-4cm}= \sum_{\neq} w_{i_1}w_{i_2}w_{i_3}w_{j_1}w_{j_2}w_{j_3}\mapn{6}
+9\sum_{\neq} w_{i_1}^2w_{i_2}w_{i_3}w_{j_2}w_{j_3}\msmap{1}{4}
\nonumber \\
&\hspace*{-4cm}+18\sum_{\neq} w_{i_1}^2w_{i_2}^2w_{i_3}w_{j_3}\msmap{2}{2}
+6\sum_{\neq} w_{i_1}^2w_{i_2}^2w_{i_3}^2\msn{3}
\end{align}
Note that in this case we can pull out the prefactor from the
summation symbols as the ensemble average quantities are theory
values.

For the second set of terms $\circled{2}$ we have one Kronecker delta
in place such that we can only contract over the remaining five
indices. For example, the first term with the matching weights
yields\footnote{For an explicit computation of this term, see \App{sec:ExplicitTerm}}:
\begin{align}
\left\langle P\left(
\sum_{i_1} \sum_{i_2 \neq i_1} \sum_{i_3 \neq i_2 \neq i_1} \sum_{j_2 \neq i_1} \sum_{j_3 \neq j_2 \neq i_1}
w_{i_1}^2w_{i_2}w_{i_3}w_{j_2}w_{j_3}Q_{i_1}^2x_{i_2}x_{i_3}x_{j_2}x_{j_3}
\right)\right\rangle
\nonumber \\
&\hspace*{-6cm}= 
M_{g,2}\left(\sum_{\neq} w_{i_1}^2w_{i_2}w_{i_3}w_{j_2}w_{j_3}\mapn{4} + 4\sum_{\neq} w_{i_1}^2w_{i_2}^2w_{i_3}w_{j_3}\msmap{1}{2} + 2\sum_{\neq} w_{i_1}^2w_{i_2}^2w_{i_3}^3\msn{2}\right)
\end{align}
All the other permutations simply shift the squares in one of the
$w_i$s around, but does not change the result - hence we can simply
multiply by $9$.

Continuing with the terms in $\circled{3}$ the two Kronecker deltas
force us to do either one or no contraction. For the first term the
result looks like:
\begin{align}
\left\langle P\left(
\sum_{i_1} \sum_{i_2 \neq i_1} \sum_{i_3 \neq i_2 \neq i_1}  \sum_{j_3 \neq i_2 \neq i_1}
w_{i_1}^2w_{i_2}^2w_{i_3}w_{j_3}Q_{i_1}^2Q_{i_2}^2x_{i_3}x_{j_3}
\right)\right\rangle
\nonumber \\
&\hspace*{-4cm}= 
M_{g,2}^2\left(\sum_{\neq} w_{i_1}^2w_{i_2}^2w_{i_3}w_{j_3}\mapn{2} + \sum_{\neq} w_{i_1}^2w_{i_2}^2w_{i_3}^2\msone \right)
\end{align}
Again, all the other permutations yield the same result, so we can
multiply by $18$.

For the final \circled{4} term no further contractions can be done
and, again, all permutations give equivalent answers. For the first
term we have
\begin{align}
\left\langle P\left(
\sum_{i_1} \sum_{i_2 \neq i_1} \sum_{i_3 \neq i_2 \neq i_1} 
w_{i_1}^2w_{i_2}^2w_{i_3}^2Q_{i_1}^2Q_{i_2}^2Q_{i_3}^2
\right)\right\rangle
= 
M_{g,2}^3\sum_{\neq} w_{i_1}^2w_{i_2}^2w_{i_3}^3
\end{align}
Collecting together all the terms we find the weighted variance of the $\mapn{3}$ to be
\begin{align}
\sigma^2\left[\MapStatEst{3}\right] 
=
\frac{1}{\left( \sum_{\neq} w_{i_1}w_{i_2}w_{i_3} \right)^2} \left\{ \vphantom{18\sum_{\neq} w_{i_1}^2w_{i_2}^2w_{i_3}w_{j_3}\msmap{2}{2}}\right.
\nonumber \\ &\hspace*{-2cm}
\sum_{\neq} w_{i_1} w_{i_2}w_{i_3}w_{j_1}w_{j_2}w_{j_3}\mapn{6}
+9\sum_{\neq} w_{i_1}^2w_{i_2}w_{i_3}w_{j_1}w_{j_2}\msmap{1}{4}
\nonumber \\ &\hspace*{-2.8cm}
+18\sum_{\neq} w_{i_1}^2w_{i_2}^2w_{i_3}w_{j_1}\msmap{2}{2}
+6\sum_{\neq} w_{i_1}^2 w_{i_2}^2w_{i_3}^2\msn{3}
\nonumber \\ &\hspace*{-2.8cm}
+9M_{g,2}\sigsqone \left[\sum_{\neq} w_{i_1}^2w_{i_2}w_{i_3}w_{j_1}w_{j_2}\mapn{4} + 4\sum_{\neq} w_{i_1}^2w_{i_2}^2w_{i_3}w_{j_1}\msmap{1}{2} + 2\sum_{\neq} w_{i_1}^2w_{i_2}^2w_{i_3}^3\msn{2} \right]
\nonumber \\ &\hspace*{-2.8cm}
+18M_{g,2}^2\sigsq{2} \left[\sum_{\neq} w_{i_1}^2w_{i_2}^2w_{i_3}w_{j_1}\mapn{2} + \sum_{\neq} w_{i_1}^2w_{i_2}^2w_{i_3}^2 \msone  \right]
\nonumber \\ &\hspace*{-2.8cm}
\left. +6M_{g,2}^3\sigsq{3}\sum_{\neq} w_{i_1}^2w_{i_2}^2w_{i_3}^2\right\}
-\mapn{3}^2
\end{align}
%

%%%%%%%%%%%%%%%%%%%%%%%%%%%%%%%%%%%%%%%%%%%%%%%%%%%%%%%

\subsection{Explicit computation of one third order term}\label{sec:ExplicitTerm}

We now compute one contraction term explicitly and show that all
permutations and higher order contractions can be computed in a
similar fashion. As a first step let us write down all the
possibilities that the six indices can take. In here the $a$th element
of each tuple shows which index the $j_a$ is, it is either $i_1$,
$i_2$, $i_3$, or none of those which is labelled $\neq$. The
horizontal lines separate sets of tuples which have the same number of
unequal indices. Note that counting through the tuples we get the
numbers from the contractions. 
\begin{center}
\begin{tabular}{*{3}{c}}
& $(\neq \ \neq \ \neq)$ & \\\\
\hline \\
$(i_1 \ \neq \ \neq)$ & $(\neq \ i_1 \ \neq)$ & $(\neq \ \neq \ i_1)$ \\
$(i_2 \ \neq \ \neq)$ & $(\neq \ i_2 \ \neq)$ & $(\neq \ \neq \ i_2)$ \\
$(i_3 \ \neq \ \neq)$ & $(\neq \ i_3 \ \neq)$ & $(\neq \ \neq \ i_3)$ \\\\
\hline \\
$(i_1 \ i_2 \ \neq)$ & $(i_1 \ \neq \ i_2)$ & $(\neq \ i_1 \ i_2)$ \\
$(i_1 \ i_3 \ \neq)$ & $(i_1 \ \neq \ i_3)$ & $(\neq \ i_1 \ i_3)$ \\
$(i_2 \ i_3 \ \neq)$ & $(i_2 \ \neq \ i_3)$ & $(\neq \ i_2 \ i_3)$ \\
$(i_2 \ i_1 \ \neq)$ & $(i_2 \ \neq \ i_1)$ & $(\neq \ i_2 \ i_1)$ \\
$(i_3 \ i_1 \ \neq)$ & $(i_3 \ \neq \ i_1)$ & $(\neq \ i_3 \ i_1)$ \\
$(i_3 \ i_2 \ \neq)$ & $(i_3 \ \neq \ i_2)$ & $(\neq \ i_3 \ i_2)$ \\\\
\hline \\
$(i_1 \ i_2 \ i_3)$ & $(i_2 \ i_1 \ i_3)$ & $(i_3 \ i_1 \ i_2)$ \\
$(i_1 \ i_3 \ i_2)$ & $(i_2 \ i_3 \ i_1)$ & $(i_3 \ i_2 \ i_1)$ \\
\end{tabular}
\end{center}
The term we deal with is the first one in $\circled{1}$ where $j_1$ is
set equal to $i_1$. In a first step we rewrite the six summation
symbols in terms of summations that solely consist of unequal
indices. For this we can only choose the tuples that have $i_1$ as a
first entry. We then find successively
\begin{align*}
\sum_{i_1} \sum_{i_2 \neq i_1} \sum_{i_3 \neq i_2 \neq i_1} \sum_{j_1} \sum_{j_2 \neq j_1} \sum_{j_3 \neq j_2 \neq j_1}
w_{i_1}w_{i_2}w_{i_3}w_{j_1}w_{j_2}w_{j_3}Q_{i_1}x_{i_2}x_{i_3}Q_{j_1}x_{j_2}x_{j_3}\deltaij{1}{1}
\\ &\hspace{-6cm} = 
\sum_{i_1} \sum_{i_2 \neq i_1} \sum_{i_3 \neq i_2 \neq i_1} \sum_{j_2 \neq i_1} \sum_{j_3 \neq j_2 \neq i_1}
w_{i_1}^2w_{i_2}w_{i_3}w_{j_2}w_{j_3}Q_{i_1}^2x_{i_2}x_{i_3}x_{j_2}x_{j_3}
\\ &\hspace{-6cm} = 
\sum_{\neq} w_{i_1}^2w_{i_2}w_{i_3}w_{j_2}w_{j_3}Q_{i_1}^2x_{i_2}x_{i_3}x_{j_2}x_{j_3}
\\ &\hspace{-6cm} +
\sum_{\neq}
\left( w_{i_1}^2w_{i_2}^2w_{i_3}w_{j_3}Q_{i_1}^2x_{i_2}^2x_{i_3}x_{j_3} + w_{i_1}^2w_{i_2}w_{i_3}^2w_{j_3}Q_{i_1}^2x_{i_2}x_{i_3}^2x_{j_3} \right)
\\ &\hspace{-6cm} +
\sum_{\neq}
\left( w_{i_1}^2w_{i_2}^2w_{i_3}w_{j_2}Q_{i_1}^2x_{i_2}^2x_{i_3}x_{j_2} + w_{i_1}^2w_{i_2}w_{i_3}^2w_{j_2}Q_{i_1}^2x_{i_2}x_{i_3}^2x_{j_2} \right)
\\ &\hspace{-6cm} +
\sum_{\neq} w_{i_1}^2w_{i_2}^2w_{i_3}^2\left( Q_{i_1}^2x_{i_2}^2x_{i_3}^2 +  Q_{i_1}^2x_{i_3}^2x_{i_2}^2 \right)
\\ &\hspace{-6cm} = 
\sum_{\neq} w_{i_1}^2w_{i_2}w_{i_3}w_{j_1}w_{j_2}Q_{i_1}^2x_{i_2}x_{i_3}x_{j_1}x_{j_2} 
+2\sum_{\neq} w_{i_1}^2w_{i_2}^2w_{i_3}^2\ Q_{i_1}^2x_{i_2}^2x_{i_3}^2
\\ &\hspace{-6cm} +
2\sum_{\neq} w_{i_1}^2 w_{i_2} w_{i_3} w_{j_1} Q_{i_1}^2 x_{i_2}x_{i_3}x_{j_1} \left( w_{i_2}x_{i_2} + w_{i_3}x_{i_3} \right)
\end{align*}
where in the first step we applied the delta, in the second one subbed
in all the relevant terms and in the third one renamed indices and
combined equal terms. Note that the number of tuples chosen for each
number of $\neq$ symbols does match the one from the contraction
formalism. Now we apply the position and ensemble averaging.
\begin{align*}
\left\langle P\left( 
\sum_{i_1} \sum_{i_2 \neq i_1} \sum_{i_3 \neq i_2 \neq i_1} \sum_{j_1} \sum_{j_2 \neq j_1} \sum_{j_3 \neq j_2 \neq j_1}
w_{i_1}w_{i_2}w_{i_3}w_{j_1}w_{j_2}w_{j_3}Q_{i_1}x_{i_2}x_{i_3}Q_{j_1}x_{j_2}x_{j_3}\deltaij{1}{1} 
\right) \right\rangle
\\ &\hspace{-10cm} = 
\sum_{\neq} w_{i_1}^2w_{i_2}w_{i_3}w_{j_1}w_{j_2}M_{g,2}\mapn{4}
+2\sum_{\neq} w_{i_1}^2w_{i_2}^2w_{i_3}^2 M_{g,2}\msn{2}
+2\sum_{\neq} w_{i_1}^2 w_{i_2} w_{i_3} w_{j_1} \left( w_{i_2} + w_{i_3} \right) M_{g,2}\msmap{1}{2}
\\ &\hspace{-10cm} = 
\sum_{\neq} w_{i_1}^2w_{i_2}w_{i_3}w_{j_1}w_{j_2}M_{g,2}\mapn{4}
+2\sum_{\neq} w_{i_1}^2w_{i_2}^2w_{i_3}^2 M_{g,2}\msn{2}
+4\sum_{\neq} w_{i_1}^2 w_{i_2}^2 w_{i_3} w_{j_1} M_{g,2}\msmap{1}{2}
\end{align*}
where in the last step we noted the argument of the sum with brackets
is symmetric, and hence the results for both terms are equal. This is
exactly the result we would have expected from the contractions on
the subset excluding $i_1$ and $j_1$. \\ Looking at the other eight
permutations, the only difference is that we choose different indices
at the start - however there will always be equally many and all the
steps are essentially mirrored, therefore we can just multiply the
result we got by 9.

% Don't change these lines
\bsp	% typesetting comment
\label{lastpage}
\end{document}